\documentclass[a4paper,12pt]{article}
\pdfoutput=1 % if your are submitting a pdflatex (i.e. if you have
\usepackage{a4wide}
\usepackage{cite}

\usepackage[T1]{fontenc} 
\usepackage[latin1]{inputenc}

\usepackage{empheq}
\usepackage{amssymb}
\usepackage{amsmath}
\usepackage{amsfonts}
\usepackage{xcolor}

\DeclareMathAlphabet{\mathpzc}{OT1}{pzc}{m}{it}

\numberwithin{equation}{section}

\usepackage[colorlinks]{hyperref}

\hypersetup{
 citecolor=blue,
 linkcolor=blue,
 urlcolor=blue}

\newcommand{\lp}{\left(}
\newcommand{\rp}{\right)}
\newcommand{\nn}{\nonumber}

\newcommand{\ep}{\epsilon}
\newcommand{\be}{\begin{equation}}
\newcommand{\ee}{\end{equation}}
\newcommand{\bes}{\begin{equation}\begin{split}}

\def \as {\alpha_s}
\def \asontwopi {\frac{\alpha_s}{2\pi}}
\def \asontwopimu {\frac{\alpha_s(\mu)}{2\pi}}
\def \Li {\text{Li}}
\def \gsb {g_{s,b}}

\def \Ca {C_A}
\def \Ci {C_i}
\def \Cf {C_F}
\def \tr {T_R}
\def \nf {n_f}

\def \Em {E_{\rm max}}
\def \d {{\rm d}}

\def \ren {{\rm ren}}
\def \RR {{\rm RR}}
\def \RV {{\rm RV}}
\def \VV {{\rm VV}}

\def \SS {S{\hspace{-5pt}}S}
\def \CC {C{\hspace{-6pt}C}}
\def \FLM {F_{\rm LM}}
\def \FLMqqb {F_{{\rm LM},q\bar q}}

\def \I {I}
\def \ONLO {\hat{\mathcal O}_{\rm NLO}}

\def \LO {{\rm LO}}
\def \NLO {{\rm NLO}}
\def \NNLO {{\rm NNLO}}

\newcommand{\dg}[1]{[df_{#1}]}
\newcommand{\df}[1]{[df_{#1}]}
\newcommand{\FLMij}[2]{F_{{\rm LM},{#1}{#2}}}
\newcommand{\FLVijfin}[2]{F_{{\rm LV},{#1}{#2}}^{\rm fin}}
\newcommand{\FLVVijfin}[2]{F_{{\rm LVV},{#1}{#2}}^{\rm fin}}
\newcommand{\FLVsqijfin}[2]{F_{{\rm LV}^2,{#1}{#2}}^{\rm fin}}
\newcommand{\FLVfin}{F_{{\rm LV}}^{\rm fin}}
\newcommand{\FLVVfin}{F_{{\rm LVV}}^{\rm fin}}
\newcommand{\FLVsqfin}{F_{{\rm LV}^2}^{\rm fin}}

\def \zb {\bar z}
\newcommand{\LMu}{\ln\lp\frac{\mu^2}{s}\rp}
\newcommand{\LMusq}{\ln^2\lp\frac{\mu^2}{s}\rp}

\def \H {{\rm H}}
\def \DY {{\rm DY}}
\def \qb {{\bar q}}
\def \dhs {{\d\hat\sigma}}

\def\dLips {{{\rm dLips}}}

\newcommand{\fa}{{f_{a}}}
\newcommand{\fb}{{f_{b}}}
\newcommand{\fx}{{f_{x}}}
\newcommand{\fy}{{f_{y}}}

\def \Obs {{\mathcal O}}
\def \muf {{\mu_F}}
\def \mur {{\mu_R}}

\def \DD {{\mathcal D}}

\begin{document}
\vspace{-5.0cm}
\begin{flushright}
OUTP-19-03P, 
CERN-TH-2019-011, TTP19-007, P3H-19-003
\end{flushright}

\vspace{2.0cm}

\begin{center}
{\Large \bf 
Analytic results for color-singlet production at NNLO QCD with 
the nested soft-collinear subtraction scheme}\\
\end{center}

\vspace{0.5cm}

\begin{center}
Fabrizio Caola$^{1,2}$, 
Kirill Melnikov$^{3}$, Raoul R\"ontsch$^{3,4}$.\\
\vspace{.3cm}
{\it
{}$^1$Rudolf Peierls Centre for Theoretical Physics, Clarendon Laboratory, Parks Road, Oxford OX1 3PU, UK \& 
Wadham College, Oxford OX1 3PN\\
{}$^2$IPPP, Durham University, Durham DH1 3LE, UK\\
{}$^3$Institute for Theoretical Particle Physics, KIT, Karlsruhe, Germany\\
{}$^4$Theoretical Physics Department, CERN, 1211 Geneva 23, Switzerland\\
}

\vspace{1.3cm}

{\bf \large Abstract}
\end{center}
We present  analytic formulas that describe fully-differential production
of color-singlet final states  in $q\bar q$ and $gg$ annihilation, including all the
relevant partonic channels,  through NNLO QCD.  We work within the
nested soft-collinear scheme which allows for fully local subtraction
of infrared divergences.  We demonstrate analytic cancellation of
soft and collinear poles and present formulas for finite parts of
all integrated subtraction terms.  These results
provide an important building block for calculating NNLO QCD
corrections to arbitrary processes at hadron colliders within the
nested soft-collinear subtraction scheme.

\thispagestyle{empty}

\clearpage
{\hypersetup{linkcolor=black}
\tableofcontents
}
\thispagestyle{empty}
\clearpage

\pagenumbering{arabic}

\allowdisplaybreaks

\section{Introduction}
\label{sec:intro}

Perturbative computations play an important role in the contemporary
exploration of particle physics at the Large Hadron Collider (LHC). In particular, the lack of
direct evidence for physics beyond the Standard Model suggests that further progress in
particle physics will require a better understanding of hard hadron
collisions  and a confrontation of precise
theoretical predictions with  experimental results.

The quality of theoretical predictions for hard processes at the LHC
has increased dramatically in recent years thanks to the advent of
flexible methods for handling infrared singularities 
that led to the calculation of
next-to-next-to-leading-order (NNLO) QCD corrections for sufficiently
complex processes~\cite{ant,czakonsub,czakonsub4d,Boughezal:2011jf,
  Cacciari:2015jma,qt1,qt2,njet1,njet2,colorful}.  In spite of these
successes, there are ongoing efforts to either simplify and improve
existing methods or to devise ``better'' ones. For example, it
is believed that higher efficiency of the slicing formalism of
Refs.~\cite{njet1,njet2,qt1,qt2} may come from a deeper understanding
of power corrections to soft and collinear
limits~\cite{tackmann:nj:pow,frank:nj:pow,Ebert:2018gsn}. Similarly,
improved control of the interplay between soft and collinear dynamics
may lead to the formulation of simple ``minimal'' local subtraction
schemes~\cite{magnea,herzog}. 
Although there is no guarantee that any of these efforts will result
in an absolute breakthrough in fixed-order calculations for LHC
processes, it is plausible that these developments will lead to more
efficient computational frameworks and enable precise phenomenological
descriptions of complex multiparticle final states.

In Ref.~\cite{Caola:2017dug} we have attempted to simplify the
residue-improved subtraction scheme proposed in
Ref.~\cite{czakonsub}.  This subtraction scheme is very attractive because 
it is fully local,  completely   general  and perfectly  modular, so that the  subtractions for a generic
process are built from a relatively small set of basic ingredients.
Its main disadvantages include a lack of physical transparency and a certain redundancy,  as well
as the numerical integration of the subtraction terms that may inadvertently impact its  efficiency.

 We have argued in Ref.~\cite{Caola:2017dug} that QCD color coherence
 removes an interplay between angles and energies of soft and
 collinear particles in gauge-invariant QCD amplitudes, thus leading
 to a minimal number of subtraction terms that need to be considered.
 Perhaps more importantly, since soft and collinear singularities are
 not intertwined, it is possible to separate them cleanly, removing
 unnecessary redundancies of the subtraction procedure presented in
 Ref.~\cite{czakonsub}.  As shown in Ref.~\cite{Caola:2017dug}, it
 appears to be advantageous to first subtract the double-soft
 singularities from the full amplitude and then, iteratively, remove
 the remaining ones.  Once this is done, a transparent and physically
 appealing subtraction scheme is obtained. Moreover, 
 this scheme appears to be very efficient, at least in the
 color-singlet production case that we have studied up to
 now~\cite{Caola:2017dug,Caola:2017xuq}.

Given the improved efficiency and inherent simplicity of the
subtraction scheme developed in~Ref.~\cite{Caola:2017dug}, it is
natural to investigate whether one could obtain analytic results for
the integrated counterterms. A successful completion of this task
would lead to the formulation of the {\it first} subtraction scheme
applicable at the LHC that is both fully local and under complete
analytic control. Although it is hard to say to what extent these nice
features are actually important in practice, 
we do hope that they will lead to a
very efficient subtraction framework for completely generic processes.

It is easy to identify the major obstacles to obtaining fully
analytic subtraction schemes.  Indeed, any NNLO subtraction scheme
involves three ``double-unresolved'' contributions whose integration
is highly non-trivial.
They are $1)$ the double-soft emission of two partons with
energies $E_{f_1} \sim E_{f_2} \ll \sqrt{s} $, where $\sqrt{s}$ is the
center-of-mass energy of the partonic process; $2)$ the emission of
two partons collinear to one of the incoming legs and $3)$ the
emission of two partons collinear to one of the hard final-state legs.
  We note that as far as the numerical implementation of NNLO QCD
  corrections to a generic process is concerned, contributions 1) and
  2) are the most problematic.  Indeed, for any splitting process, the
  integrated contribution 3) is a number of the form $a_f/\ep + b_f$,
  so that it can be calculated numerically once and for all.  On the
  contrary, the integrated contributions 1) and 2) are functions of
  the relative angles between hard partons and the momentum fraction
  carried into the hard process, respectively.\footnote{Although the
    computation of NNLO QCD corrections to a generic process requires
    all three contributions, for sufficiently simple processes only
    a subset is needed.  For example, for color-singlet production
    only contributions 1) and 2) are needed, while for color singlet
    decay only 1) and 3) are required.}  Close to the end-points, these
  function may develop integrable singularities, which makes their
  numerical evaluation tedious.

In Ref.~\cite{max}, some of us presented analytic result for the
integrated double-soft subtraction term. In this paper, we will argue
that a minor modification of the subtraction procedure described in
Ref.~\cite{Caola:2017dug} greatly simplifies the analytic integration
of the triple-collinear subtraction terms. In fact, such an
integration of all relevant triple-collinear subtraction terms has
recently been  performed in Ref.~\cite{maxtc}.  Thanks to these results,
it is now possible to present a subtraction framework for the
production of color-singlet particles at hadron colliders that is both
fully analytic and fully local.

Although the production of color-singlets at NNLO QCD has been studied
many times, including the development of public computer codes, even
the simplest versions of these processes such as $pp \to Z$ and $pp\to H$ are
quite useful to us because NNLO QCD corrections to these processes
are known
analytically~\cite{Hamberg:1990np,Anastasiou:2002yz}.  This feature
allows us to check all the non-trivial ingredients of our
computational framework to a very high accuracy.  We believe such a
validation is important in view of its expected application to more
complex cases in the future.  Of course, it is also interesting to explore the
performance of our subtraction scheme by considering a well-known
process, where many benchmarks exist already.

Nevertheless, it should be clear that the goal of this paper is to present  analytic formulas relevant
for the production of  {\it generic}  color-singlet final states at a  hadron collider, 
written in a way that will allow us to move beyond color-singlet production.  For this reason we decided
to avoid using simplifications that are  particular to the cases of 
Drell-Yan or Higgs production.

The rest  of this  paper is organized as follows.  In
Section~\ref{sec:basics} we summarize  the main features of the
nested soft-collinear subtraction  scheme of Ref.~\cite{Caola:2017dug} and explain 
how we modified it to allow for an analytic  integration of the triple-collinear subtraction terms. 
 In Section~\ref{sec:dy} we provide formulas for the $q \bar q$-initiated
production of the color-singlet final state.  In Section~\ref{sec:h}
we provide formulas for the gluon annihilation into  a color-singlet
final state. We discuss the validation of our   results in Section~\ref{sec:validation} and 
conclude in Section~\ref{sec:conclusions}. 
A large number of useful formulas are collected in 
appendices, as well as in an ancillary file attached to this paper.

%%%

\section{Overview of the nested  subtraction scheme}
\label{sec:basics}

In this section we briefly review the method for computing
NNLO QCD corrections described in Ref.~\cite{Caola:2017dug}  and explain  how to modify it to simplify 
the  analytic integration of the triple-collinear subtraction terms.  
We consider the  production of a color-singlet state $V$ in hadronic collisions.
We write the (fiducial) cross section as
\be
\sigma_f = \sum\limits_{a,b \in [-n_f, n_f]}\int \d x_1 \d x_2 f_a(x_1,\muf)
f_b(x_2,\muf) \dhs_{\fa\fb}(x_1,x_2,\mur,\muf;\Obs),
\label{eq:fid_xsec}
\ee
where $\nf$ is the number of light flavors,  $\dhs_{\fa\fb}$ is the partonic 
cross section in the $\fa \fb$ channel,
and we employ the following notation for parton distributions functions: $f_0 = g$
and $f_{\pm 1, \pm 2, \pm 3, \pm 4, \pm 5} = \{ d/\bar d,u/\bar u,s/\bar s,c/\bar c,b/\bar b \}$.
Finally, $\Obs$ is a suitable infrared-safe observable that defines the  fiducial volume.

We consider the perturbative expansion of the partonic cross section and
write
\be
\dhs_{\fa\fb}(x_1,x_2,\mur,\muf,\Obs) = \dhs^{\LO}_{\fa\fb}
+ \dhs^{\NLO}_{\fa\fb}
+ \dhs^{\NNLO}_{\fa\fb} + \mathcal O\lp\alpha_s^{q+3}\rp.
\ee
Here $q=0$ for quark-initiated processes and $q = 2$ for gluon-initiated processes, and we have suppressed the arguments of the functions  on the right-hand side. We focus on the NNLO QCD contribution 
 $\dhs^{\NNLO}_{\fa\fb}$. 
 It can be written as
 \be
 \dhs^{\NNLO}_{\fa\fb} = \dhs^{\RR}_{\fa\fb}+\dhs^{\RV}_{\fa\fb}+
 \dhs^{\VV}_{\fa\fb}+\dhs^{\ren}_{\fa\fb},
 \label{eq:terms}
 \ee
where 
\bes
&
\dhs^{\RR}_{\fa\fb} = \frac{\mathcal N}{2s}\int \dLips(V) [d f_4][d f_5](2\pi)^d \delta_d(p_1+p_2-p_V-p_4-p_5)
\times\\
&\quad\quad\quad\quad\quad
|\mathcal M^{\rm tree}|^2({p_1,p_2,p_V,p_4,p_5}) \;\mathcal O(p_V,p_4,p_5),
\\
&\dhs^{\RV}_{\fa\fb} = \frac{\mathcal N}{2s}\int \dLips(V)[d f_4] (2\pi)^d \delta_d(p_1+p_2-p_V-p_4)
\times\\
&\quad\quad\quad\quad\quad
2{\rm Re}\big[\mathcal M^{\rm tree}
\mathcal M^{{\rm 1-loop},*}\big]({p_1,p_2, p_V,p_4}) \;\mathcal O(p_V,p_4),
\\
&
\dhs^{\VV}_{\fa\fb} = \frac{\mathcal N}{2s}\int \dLips(V) (2\pi)^d \delta_d(p_1+p_2-p_V)
\times\\
&\quad\quad\quad\quad\quad
\big[2{\rm Re}\big[\mathcal M^{\rm tree} \mathcal M^{{\rm 2-loop},*}\big]
+|\mathcal M^{\rm 1-loop}|^2\big]({p_1,p_2, p_V})\; \mathcal O(p_V),
\end{split}
\label{eq:cont}
\ee
and $\dhs^{\ren}$ contains all contributions that originate from the renormalization of input parameters,
such as the strong coupling constant  $\alpha_s$ and the parton distribution functions (PDFs). In Eq.~\eqref{eq:cont},
$\mathcal N$ is a normalization factor that takes into account color and spin averages, $s$ is the
partonic center-of-mass  energy squared, $\dLips(V)$ is the phase space for the
final state $V$, and 
\be
[df_i]  = \frac{\d^{d-1} p_i}{(2\pi)^{d-1} 2 E_i} \theta (\Em-E_i).
\ee
Here $d$ is the dimensionality of space-time that we use as the regularization parameter,  and 
$\Em$ is an arbitrary\footnote{The only requirement on $\Em$ is that it should be at least as large
as the maximum energy allowed by the momentum-conserving $\delta$-functions in 
Eq.~\eqref{eq:cont}. For simplicity, throughout this paper we use $\Em = \sqrt{s}/2$, where
$\sqrt{s}$ is the partonic center-of-mass  energy.} energy scale that is introduced for future 
convenience.

Each term in Eq.~\eqref{eq:terms} is individually divergent. These divergences can either appear explicitly as 
poles in $\ep = (4-d)/2$ or be hidden inside phase-space integrals. The goal of any subtraction scheme
is to extract these divergences and to arrive at the following representation of the NNLO contribution to the cross section 
\be
\dhs^{\NNLO}_{\fa\fb} = \dhs^{\rm NNLO}_{V+2,\fa\fb} + 
\dhs^{\rm NNLO}_{V+1,\fa\fb} + 
\dhs^{\rm NNLO}_{V,\fa\fb}, 
\label{eq:nnlo_split_fin}
\ee
where  $\dhs^{\NNLO}_{V+i}$ are  finite quantities that involve  contributions  with $V$ and up to $i$ partons 
in the final state. We will refer to $\dhs^{\NNLO}_{V+i}$, with $i=2,1,0$, 
as terms with NNLO, NLO and LO 
kinematics,   respectively. 

In Ref.~\cite{Caola:2017dug}, we proposed a framework to achieve the separation of the NNLO contributions to the cross section 
as in Eq.~\eqref{eq:nnlo_split_fin}. It is based on three ideas:

\begin{itemize}

\item a multiparticle phase space can be decomposed into a sum of
  elements (partitions) such that for each partition only a
  well-definite subset of particles gives rise to collinear
  singularities upon integration over the phase space of final state
  partons. An early discussion of this idea can be found in papers on
  NLO QCD subtractions~\cite{Frixione:1995ms,Frixione:1997np}; in the
  context of NNLO QCD calculations, it was reincarnated in
  Ref.\cite{czakonsub};
  
\item for each of these partitions there exists a  phase space
  parametrization that allows the extraction of both soft and collinear
  singularities in a fully factorized form~\cite{czakonsub};

\item thanks to gauge invariance and
  color coherence \cite{Ellis:1991qj}, soft and collinear singularities are not entangled in QCD
  amplitudes, in contrast to  individual
  diagrams \cite{Caola:2017dug}.

\end{itemize}
We argued in Ref.~\cite{Caola:2017dug} that these three points allow
us to follow the so-called FKS subtraction scheme
\cite{Frixione:1995ms,Frixione:1997np}, developed for NLO QCD
computations, and to perform the required soft and collinear
subtractions in a nested way.  As a consequence,
the computational framework becomes very transparent physically  and quite 
efficient numerically.

We will illustrate the main ideas of Ref.~\cite{Caola:2017dug} by
considering the double-real emission corrections to the Drell-Yan process
$q\bar q \to V$ as an example, focusing on the most complicated
$q(p_1) \bar q(p_2) \to V +g(p_4) g(p_5)$ channel. All other partonic channels 
can be dealt with  along the same lines although the
details can be somewhat  different.\footnote{ Results for all the relevant
channels are presented in the next sections.}  We write the corresponding
cross section as 
\be
2s\cdot \dhs_{q\bar q}^{\RR} = \int \dg4
\dg5 \theta(E_4-E_5) \FLMqqb(1,2,4,5) \equiv \langle
\FLMqqb(1,2,4,5) \rangle,
\label{eq2}
\ee 
where 
\bes
\FLMqqb(1,2,4,5) = {\cal N} \int {\rm dLips}(V)
(2\pi)^d \delta_d(p_1+p_2 - p_V - p_4 - p_5) \times
\\
|{\cal M}^{\rm tree}_{q\bar q \to V + gg}|^2(p_1,p_2,p_V,p_4,p_5) \;{\cal
  O}(p_V,p_4,p_5), 
 \end{split}
 \label{eq:defFLM1245}
\ee  
see Eq.~\eqref{eq:cont}. All energies in these
formulas are to be understood in the center-of-mass frame of the
colliding partons. Note that we have introduced the energy ordering $E_4>E_5$
in Eq.~\eqref{eq2}. Since the matrix
element is symmetric with respect to the permutations of the gluons
$g_4$ and $g_5$, we can remove the $1/2!$ symmetry factor from $\cal{N}$.

Our goal is to extract singularities from Eq.~\eqref{eq2}.  These
singularities have different origins. There
exist
\begin{itemize} 
\item a double-soft singularity that occurs  when energies of the two gluons
  vanish in such a way that their ratio $E_5/E_4$ is fixed;
  
\item a single-soft singularity that arises when  $E_5$ vanishes at fixed
  $E_4$. Note that due to the energy ordering in Eq.~\eqref{eq2} the
  opposite limit, $E_4 \to 0$ at fixed $E_5$, cannot occur;
  
\item many different collinear singularities that appear when one or both 
   gluons are emitted along the direction of the incoming 
  quark or the incoming anti-quark, or when the momenta of the two gluons become  parallel
  to each other. 
\end{itemize} 
We need to extract all these singularities in an unambiguous way.  We explain how to 
do this in the next two subsections. 

\subsection{Extraction of soft singularities}
As we explained in Ref.~\cite{Caola:2017dug}, it is convenient to begin by  
regularizing the double-soft singularity $E_4\sim E_5 \sim \lambda \sqrt{s} \to 0$.  We
write
\be
 \langle \FLMqqb(1,2,4,5) \rangle 
= 
\langle \SS \FLMqqb(1,2,4,5) \rangle 
+
\langle ( I - \SS) \FLMqqb(1,2,4,5) \rangle ,
\label{eq5}
\ee 
where $\SS$ is an operator that extracts the
double-soft $\lambda \to 0$ singular limit 
from 
$\FLMqqb$. To make this statement precise,
when the operator $\SS$ acts on $F_{LM}$, it removes the
four-momenta of the gluons from both the energy-momentum conserving
$\delta$-function and the observable, and extracts the leading
singular behavior from the matrix element squared. The result is 
\be
\SS \FLMqqb(1,2,4,5) = \gsb^4\; {\rm Eik}(1,2,4,5) \; \FLMqqb(1,2),
\label{eq6}
\ee
where $\gsb$ is the bare strong coupling and ${\rm Eik}(1,2,4,5)$ is the square of the eikonal factor derived e.g.
in Ref.~\cite{Catani:1999ss}. It is also given in Ref.~\cite{Caola:2017dug}
using notation that is identical to what we use in this paper.
Also, $\FLMqqb(1,2)$ is defined analogously to Eq.~\eqref{eq:defFLM1245}; it reads 
\be
\langle \FLMqqb(1,2) \rangle = \mathcal N \int {\rm dLips}(V)
(2\pi)^d \delta_d(p_1+p_2 - p_V) |{\cal M}^{\rm tree}_{q\bar{q} \to V}|^2(p_1,p_2,p_V) {\cal O}(p_V).
\ee
This tree-level matrix element squared integrated over the Born 
phase-space obviously provides the 
leading order result  for the observable $\cal{O}$.

We deal with the two terms on the right-hand side of Eq.~\eqref{eq5} in very 
different ways. In the first term, thanks to Eq.~\eqref{eq6}, the hard
matrix element decouples and only the eikonal factor needs to be
integrated over the two-gluon phase space.  In our original
paper~\cite{Caola:2017dug} this integral was calculated  numerically
but, since then, an analytic computation of this integral has become
available \cite{max}.  The result reads\footnote{We note
  that the result in Eq.~\eqref{eqmy1} also includes the $n_f$-part
  which originates from the radiation of a $q \bar q$ pair into the
  final state. We include it here for completeness.}  
\be
\frac{\langle \SS \FLMqqb(1,2,4,5) \rangle}{ a_{s,b}^2
  e^{-2\ep L} \langle \FLMqqb(1,2) \rangle } = C_F^2 \; D_S^{\Cf} +
C_F C_A D_S^{C_A} + C_F T_R n_f D_S^{n_f},
\label{eqmy1}
\ee
where we have defined
\be
a_{s,b} = \frac{\gsb^2}{8\pi^2} \frac{\left(4\pi\right)^{\ep}}{\Gamma(1-\ep)},
\label{defasb}
\ee
and
\be
L = \log\left(\frac{s}{\mu^2}\right).
\label{defL}
\ee
In Eq.~\eqref{eqmy1}, the abelian part is known  in a closed form 
\be
D_S^{C_F} = \frac{2}{\ep^4}\frac{ \Gamma^4(1-\ep)}{\Gamma^2(1-2\ep)}
= \frac{2}{\ep^4} - \frac{ 2 \pi^2}{3 \ep^2}  - \frac{8 \zeta_3}{\ep}  - \frac{2 \pi^4}{45} + {\cal O}(\ep), 
\ee
and the other two contributions are computed as an expansion in $\ep$
\be
\begin{split} 
& D_S^{C_A} = \frac{1}{2\ep^4} + \frac{11}{12\ep^3} 
+ \frac{1}{\ep^2} \left ( -\frac{16}{9} - \frac{\pi^2}{4} + \frac{11}{3}\ln 2\right ) \\
& + \frac{1}{\ep} \left ( \frac{217}{54} - \frac{11 \pi^2}{36} - 
\frac{137}{18}\ln 2
 - \frac{11}{3} \ln^2 2 
- \frac{21}{4}\zeta_3 \right ) \\
& -\frac{649}{81} + \frac{125\pi^2}{216} - \frac{11 \pi^4}{80} 
+\frac{434}{27}\ln 2 - \frac{11}{6}\pi^2 \ln 2
+\frac{137}{18}\ln^2 2 + \frac{22}{9} \ln^3 2 -\frac{275}{12}\zeta_3,
\\
& D_S^{n_f} = -\frac{1}{3\ep^3} + \frac{1}{\ep}\lp\frac{13}{18 } -\frac{4}{3}\ln 2\rp
+ \frac{1}{\ep } 
\left ( -\frac{125}{54}+ \frac{\pi^2}{9}
+ \frac{35}{9}\ln 2 + \frac{4}{3} \ln^2 2
\right ) \\
& + \frac{601}{162} - \frac{23 \pi^2}{108} -\frac{223}{27} \ln2 + \frac{2\pi^2}{3} \ln 2
-\frac{35}{9}\ln^2 2 -\frac{8}{9}\ln^3 2+ \frac{25}{3}\zeta_3 .
\end{split}
\label{eq:dsresults}
\ee
The equivalent results for gluon-initiated color singlet production can be obtained by
simply replacing  $\Cf \to  \Ca$ in Eq.~\eqref{eqmy1}.

We now turn to the second term in Eq.~\eqref{eq5} where the  double-soft divergencies are already 
regularized but  both the $E_5 \to 0$ divergence at fixed $E_4$
and collinear divergences  are still present.  To
extract  them, we repeat the above procedure and subtract the
$E_5 \to 0$ singularities at fixed $E_4$.  We call the corresponding operator
$S_5$ and write
\bes
\langle ( I - \SS) \FLMqqb(1,2,4,5) \rangle  &= 
 \langle (I - \SS) (I - S_5) \FLMqqb(1,2,4,5) \rangle\\
&+\langle S_5 (I - \SS) \FLMqqb(1,2,4,5) \rangle .
\end{split}
\label{eq7}
\ee
When the  operator $S_5$ acts   on $\FLMqqb(1,2,4,5)$, it   removes the gluon 
$g_5$ from the phase space and the observable,  and extracts the leading 
singularity 
\bes
S_5 \FLMqqb(1,2,4,5)  = &
\frac{\gsb^2}{E_5^2}
\left [ (2 C_F - C_A) \frac{\rho_{12}}{\rho_{15} \rho_{25}}
+ C_A \left ( \frac{\rho_{14}}{\rho_{15} \rho_{45}} + \frac{\rho_{24}}{
  \rho_{25} \rho_{45} } \right ) \right ] \\
&\times \FLMqqb(1,2,4).
\label{eq8}
\end{split}
\ee
We use the notation  $\rho_{ij} = 1 - \cos \theta_{ij}$ 
in Eq.~\eqref{eq8}, where $\theta_{ij}$ is the relative 
angle between partons $i$ and $j$. We have also introduced
\bes
\langle \FLMqqb(1,2,4) \rangle = \mathcal N 
\int  {\rm dLips}(V)\dg4 (2\pi)^d \delta_d(p_1+p_2 - p_V  - p_4)\times
\\
|{\cal M}^{\rm tree}_{q\bar{q} \to V+g}|^2(p_1,p_2,p_V,p_4) \;{\cal O}(p_V,p_4),
\end{split}
\ee
see Eq.~\eqref{eq:defFLM1245}.
From here on, we will omit the subscript on $\mathcal{M}$ indicating the partonic process.
It is clear that the second term in Eq.~\eqref{eq7} has a simplified
(i.e. independent of $g_5$) matrix element.
The integration over the energy and angles of the gluon $g_5$ can therefore be performed,
and the remaining infrared divergences in
the matrix element for the process $q \bar q \to V + g_4$ can be
dealt with in a way that is similarly to what is usually done in
next-to-leading-order computations. 
On the other hand, the first term in Eq.~\eqref{eq7} is now free of soft divergences
but it still contains collinear singularities. We explain how to extract them in the next subsection.

\subsection{Extraction of collinear singularities}
In the previous subsection we extracted soft singularities from the double-real emission contribution by writing it as  
\bes
\left\langle \FLMqqb(1,2,4,5)\right\rangle \to \left\langle(\I-\SS)(\I-S_5)\FLMqqb(1,2,4,5)\right\rangle
\\
+{\rm simpler~terms~with~reduced~kinematics.}
\end{split}
\ee
The procedure continues with the extraction of collinear singularities. 
This requires  an additional step, similar to the
energy ordering in Eq.~\eqref{eq2}. Indeed, we need to
split the phase space into regions such that in each region only a limited 
subset of momentum configurations can lead to collinear singularities. 

Doing that involves the first two  points in the itemized list that we
presented after Eq.~\eqref{eq:nnlo_split_fin}.  The first point  is  the 
phase space partitioning; our goal is to split the 
phase space so that collinear singularities are localized  in a clean 
and physical way. For
example, we may want to focus on the collinear emissions off the incoming quark 
{\it or} the collinear emissions off the incoming anti-quark, {\it or} on the
collinear emission of the gluon $g_4$ off the quark and the gluon $g_5$ off the
anti-quark etc.

We can do that by introducing a partition of unity and
using it to split the phase space. We write 
\be 1 = \omega^{41,51} +
\omega^{42,52} + \omega^{41,52} + \omega^{42,51}.
\label{part1}
\ee 
For the {\it double-collinear} partitions $\{4i,5j\},~i\ne j$, the damping factor
$\omega^{4i,5j}$ is engineered  in such a way that
 collinear
 singularities in  $\omega^{4i,5j} \FLMqqb(1,2,4,5)$  arise  only if  momentum $p_4$ is parallel to $p_i$ and/or the
momentum $p_5$ is parallel to $p_j$. Conversely, in the \emph{triple-collinear}
partitions $\{4i,5i\}, ~i=1,2$,  the damping factor $w^{4i,5i}$ is designed in such a way that
only the $p_4||p_i$, $p_5||p_i$ and $p_4||p_5$ momentum configurations lead to a singularity. 
Apart from these conditions, there is significant freedom in choosing the partition functions; 
we will present a possible choice in the forthcoming sections.\footnote{We note that
if one is only interested in color-singlet production, partitions can 
easily be avoided.
Nevertheless, we stress that here we use this class of processes to present
results that can be used as building blocks for NNLO calculations for \emph{generic}
processes. In the general case, partitions are crucial for the formalism presented here.}

Contributions from the double-collinear partitions $\omega^{41,52},
\omega^{42,51}$ can be computed right away since the singular limits
are easy to establish and no overlapping singularities are
present.  For example, in case of  $\omega^{41,52}$,  it is sufficient to use the angle between  the
three-momenta $p_4$ and  $p_1$
and the angle between the three-momenta $p_5$ and $p_2$ 
as independent variables to describe the collinear
singularities in this partition.

The situation is more complex for the triple-collinear partitions,
where overlapping singularities are present. The complexity stems from the fact that 
different hierarchies between  $\rho_{4i},\rho_{5i}$ and $\rho_{45}$ lead
to inequivalent limits in this case. 
To identify these limits and extract them in a unique way, we
further partition the phase space into four sectors.  Taking as an
example the $w^{41,51}$ partition, we introduce four sectors
as follows
\be
\begin{split}
1 &=  \theta \left (\rho_{51} < \frac{\rho_{41}}{2} \right ) 
+ \theta \left (\frac{\rho_{41}}{2} < \rho_{51} < \rho_{41}  \right ) 
+ \theta \left (\rho_{41} <  \frac{\rho_{51}}{2}  \right ) 
+ \theta \left ( \frac{\rho_{51}}{2} <  \rho_{41} < \rho_{51} \right ) \\
& = \theta^{(\mathpzc{a})}+\theta^{(\mathpzc{b})}
+\theta^{(\mathpzc{c})}+\theta^{(\mathpzc{d})}.
\end{split}
\label{eq:defsectors}
\ee
The four sectors in the  partition $w^{42,52}$ are constructed analogously.
It is clear  that Eq.~(\ref{eq:defsectors})
acts in such a way  that in each of the four sectors only a small 
number of singular collinear limits occurs. We then expect that by  choosing an appropriate
parametrization for {\it each of the four sectors}, these singularities can be isolated and extracted. A
convenient phase space parametrization for each of the four sectors
can be found in Ref.~\cite{czakonsub}.

In each of the four sectors shown in Eq.~\eqref{eq:defsectors}, the
nested subtraction of these collinear limits can then be performed
similar to what we discussed in connection with the soft limits. 
We sketch how to do this by considering sector $(\mathpzc{a})$ of the $w^{41,51}$ partition.
Because of the angular ordering Eq.~\eqref{eq:defsectors},  a double-collinear singularity in this sector 
can only occur if $p_5 || p_1$. Similar to  the soft case, we isolate it by writing
\bes
&\left\langle \theta^{(\mathpzc{a})} w^{41,51}
(\I-S_5)(\I-\SS) \FLMqqb(1,2,4,5)\right\rangle=
\\
&\quad\quad+
\left\langle \theta^{(\mathpzc{a})} C_{51} w^{41,51}
(\I-S_5)(\I-\SS)\FLMqqb(1,2,4,5)\right\rangle
\\
&\quad\quad+\left\langle \theta^{(\mathpzc{a})} (\I-C_{51}) w^{41,51}
(\I-S_5)(\I-\SS)\FLMqqb(1,2,4,5)\right\rangle,
\end{split}
\label{eq:sing_coll}
\ee
where $C_{51}$ is an operator that extracts the most singular contribution
in the collinear $5||1$ limit from the quantity on the left-hand side of Eq.~(\ref{eq:sing_coll}) 
and enforces this  collinear limit on the damping
factor $w^{41,51}$, the reduced matrix element, the momentum-conserving $\delta$-function
and the observable $\mathcal O$. This amounts to the replacements
$\rho_{51}\to 0$ and  $p_1 \to p'_1 = p_1 (E_1-E_5)/E_1$ in these quantities.  The result reads 
\be
C_{51} w^{41,51}\FLMqqb(1,2,4,5) = -\tilde w_{5||1}^{41,51}\frac{\gsb^2}{p_1\cdot p_5}
P_{qq}\lp\frac{E_1}{E_1-E_5}\rp \FLMqqb(1',2,4),
\ee
where $P_{qq}$ is the Altarelli-Parisi splitting function
$P_{qq}(z) = \Cf(1+z^2)/(1-z)$,
\be
\tilde w^{4i,5j}_{k||l} = \lim_{\rho_{kl}\to 0} w^{4i,5j},
\ee
 and the ``$1'$ '' notation in $\FLMqqb$ refers to  the $p_1\to p'_1$ substitution that 
 we just described. Compared to soft limits, there is an additional subtlety. 
 Indeed, in our construction
the angular part of the phase space is non-trivial. To unambiguously define the $C_{51}$ operator,
we must specify its action on the gluons' phase space $\dg4\dg5$. A convenient choice, adopted already in Ref.~\cite{Caola:2017dug}, 
is to let $C_{51}$ act on it, i.e. to take the $\rho_{51} \to 0$ limit of 
the measure $\dg4 \dg5$.

The  right hand side of Eq.~\eqref{eq:sing_coll} includes  a term with reduced kinematics,
which can be dealt with using methods similar to the ones used in NLO computations, and another  term that  only contains
a triple-collinear singularity. The latter occurs whenever $4||5||1$, without further 
hierarchy between $\rho_{51},\rho_{41}$ and $\rho_{45}$. 
To regulate this last singularity, we introduce a triple-collinear operator $\CC_1$
and write
\bes
&\left\langle \theta^{(\mathpzc{a})} (\I-C_{51}) w^{41,51}
(\I-S_5)(\I-\SS)\FLMqqb(1,2,4,5)\right\rangle = 
\\
&
\quad\quad+\left\langle \theta^{(\mathpzc{a})} \CC_1(\I-C_{51}) 
(\I-S_5)(\I-\SS)\FLMqqb(1,2,4,5)\right\rangle 
\\
&
\quad\quad+\left\langle \theta^{(\mathpzc{a})} (\I-\CC_1) (\I-C_{51}) w^{41,51}
(\I-S_5)(\I-\SS)\FLMqqb(1,2,4,5)\right\rangle,
\end{split}
\label{eq:last_sing}
\ee
where we used $\CC_1 w^{41,51}=\lim_{\rho_{51}\to0,\rho_{41}\to 0,\rho_{45}\to 0}=1$, which immediately follows from the definition
of $w^{41,51}$. Similar to the single-collinear case, 
the operator $\CC_1$ extracts the most singular behavior from the matrix element 
in the limit $\rho_{41}\sim\rho_{51}\sim\rho_{45}\to 0$
and sets
$p_1 \to p'_1 = p_1 (E_1-E_4-E_5)/E_1 $ in the reduced matrix element, momentum-conserving 
$\delta$-function and observable ${\cal O}$. We obtain 
\be
\CC_1 \FLMqqb(1,2,4,5) = \gsb^4\lp\frac{2}{s_{145}}\rp^2 P_{ggq}(s_{45},-s_{14},-s_{15}; z_4,z_5)
\FLMqqb(1',2),
\label{eq:tcsplit}
\ee
where $s_{145}= s_{45}-s_{14}-s_{15}$ and  $P_{ggq}$ is a triple-collinear splitting function \cite{Catani:1999ss} that depends on the  invariants
$s_{ij} = 2 E_i E_i \rho_{ij}$ and the momentum fractions $z_i = E_i/(E_4+E_5-E_1)$. 

Note that in the triple-collinear
limit the only effect of the gluon emission on the reduced matrix element and 
the kinematics of the initial state is the 
boost $p_1 \to p'_1 = (E_1-E_4-E_5)/E_1 \equiv z p_1$. We can then schematically write 
Eq.~\eqref{eq:last_sing} as
\bes
&\left\langle \theta^{(\mathpzc{a})} (\I-C_{51}) w^{41,51}
(\I-S_5)(\I-\SS)\FLMqqb(1,2,4,5)\right\rangle = 
\int \d z  \left\langle P(z)\right\rangle \left\langle\FLMqqb(z,2)\right\rangle
\\
&
\quad\quad+\left\langle \theta^{(\mathpzc{a})} (\I-\CC_1) (\I-C_{51}) w^{41,51}
(\I-S_5)(\I-\SS)\FLMqqb(1,2,4,5)\right\rangle,
\end{split}
\label{eq:gen_final}
\ee
where $\left\langle P(z)\right\rangle$ is the integral of the (soft-regulated) splitting function
over the phase space of the unresolved gluons, with the constraint $E_4+E_5 = (1-z)E_1$ 
and $E_4<E_5$. We note that the second term in Eq.~\eqref{eq:gen_final} is   free of all singularities, and
can be integrated in four dimensions using  standard Monte-Carlo techniques. 

Although this discussion  is valid for any  triple-collinear operator $\CC_1$ that extracts the corresponding
triple-collinear singularity from the matrix element squared, we must specify the
action of $\CC_1$ on $\dg4\dg5$ and on the 
$P_{ggq}$ function itself to unambiguously define the subtraction framework.
In Ref.~\cite{Caola:2017dug}, we let $\CC_1$ act  on both $\dg4\dg5$ and on the splitting function, 
i.e. we evaluated all the $s_{ij}$ invariants in Eq.~\eqref{eq:tcsplit} and the angular factors in
the $\dg4\dg5$ phase-space in the triple-collinear limit. While this is a valid option, it is not the only one.
In fact, this choice  makes the analytic integration over angles of the unresolved partons rather complicated, since 
it  constrains  the internal rotational symmetry of the unresolved phase space and does not allow for  simple reparametrizations.

To overcome these issues, we now define the operator 
$\CC_1$ in such a way that it \emph{does not} act on either $\dg4 \dg5$ or on $P_{ggq}$.
Rather, $\CC_1$ acts on the momentum-conserving $\delta$-function and on the observable, and extracts
the leading triple-collinear singularity from the matrix element according to Eq.~\eqref{eq:tcsplit},
but it leaves the angular factors in the $\dg4 \dg5$ phase space and all the $s_{ij}$ invariants in 
Eq.~\eqref{eq:tcsplit} untouched.
This modification of the subtraction scheme  leads to a simpler integration
of the triple-collinear splitting function over the unresolved phase space. Indeed, such a calculation  has  recently been performed for all  relevant
triple-collinear splitting functions in Ref.~\cite{maxtc}.

The results of Ref.~\cite{maxtc},  combined with integrated double-soft subtraction terms  presented earlier in Ref.~\cite{max},
 allow us to promote the fully local subtraction framework or 
Ref.~\cite{Caola:2017dug} to a fully
 analytic scheme. This implies that we can now check the
cancellation of all infrared poles analytically  and achieve faster and more stable 
physical predictions by using
analytic formulas for all the integrated subtraction terms. 

We will present analytic formulas required for the computation of NNLO QCD corrections to the production of color-singlet final states 
in the remaining parts of this paper.    However, before we do that,  a general comment is in order.
Indeed, as  should be clear from the discussion in this  section, our framework is highly modular;  we believe
that this   modularity ensures  that its generalization beyond color-singlet production  will proceed seamlessly. 
Indeed, the only
differences between the color-singlet production and the  general case with colored partons in the final state 
are:
\begin{enumerate}
\item compared to color-single production, a generic process has a more complicated color structure and requires double-soft integrals that are functions 
  of  relative angles of pairs of  hard emittors, rather than pure numbers as in Eq.~\eqref{eq:dsresults}.
  The results relevant for this case have been presented 
in Ref.~\cite{max};
\item a generic process also involves triple-collinear final state splitting. While the analytic
integration  of the relevant splitting functions over the unresolved triple-collinear phase space has not been
performed for all possible splittings, in Ref.~\cite{maxtc} it was shown  that techniques used
to deal with initial state splittings can be successfully applied to final splittings as well.
\end{enumerate}

It follows that the most general ingredients required for computing NNLO QCD corrections to generic partonic processes at the LHC
can be obtained.  From that perspective, the analytic formulas presented in this paper provide important  building blocks
for such a generic computation and give an excellent  starting point for its generalization that will be addressed in the future. For now, we will proceed
with presenting analytic formulas for all partonic channels that may contribute to the production of color-singlet
final states at a hadron colliders. 

%%%

\section{Quark-initiated color-singlet production}
\label{sec:dy}
In this section, we consider the production of a color-singlet final state
\be
p p \to V + X,
\ee
to NNLO QCD accuracy for 
reactions that are quark-initiated at leading order. We refer to these
processes as ``Drell-Yan processes'', however, we
emphasize that the results presented in this section are
applicable to  \emph{any} color-singlet production process
which is quark-initiated at LO. Typical examples include $pp \to Z,W^+,\gamma^*, ZZ, W^+W^-, WZ, WH, ZH$ and so on. 

Starting from Eq.~\eqref{eq:fid_xsec}, we find it convenient to group the different partonic channels in three
categories
\bes
\d\sigma_f^\DY &= \int \d x_1 \d x_2 
\sum\limits_{\substack{a,b \in [-\nf,\nf]\\a,b\ne0}}
\fa(x_1) \fb(x_2) \d\hat\sigma_{\fa\fb}^\DY(x_1,x_2)
\\
&+
\int \d x_1 \d x_2 \sum\limits_{\substack{a\in [-\nf,\nf]\\a\ne0}}
\bigg[\fa(x_1) g(x_2) \d\hat\sigma_{\fa g}^\DY(x_1,x_2)+
g(x_1) \fa(x_2) \d\hat\sigma_{g \fa}^\DY(x_1,x_2)\bigg]
\\
&+
\int \d x_1 \d x_2 ~ g(x_1) g(x_2) \d\hat\sigma_{gg}^\DY(x_1,x_2).
\end{split}
\label{eq:dy_channels}
\ee
We omit the dependence of ${\rm d} \sigma_f^{\rm DY}$ on the renormalization and factorization scales
$\mur,\muf$ and the observable $\mathcal O$ to shorten
the notation. 
The first term in Eq.~\eqref{eq:dy_channels}, $\d\hat\sigma^\DY_{\fa\fb}$, receives contributions
from quark channels and is present at LO.
The terms on the second line, $\d\hat\sigma_{\fa g}^\DY$ 
and $\d\hat\sigma^\DY_{g\fa}$, start
contributing at NLO, and the last term
$\d\hat\sigma_{gg}^\DY$  appears for the first time at NNLO.
In what follows, we will consider the LO, NLO and NNLO contributions in turn. 
To simplify the notation, we will omit the superscript
$\DY$ for the rest of this section.

For the NNLO contribution, we will consider the different channels defined in 
Eq.~\eqref{eq:dy_channels} separately. We also find it convenient to split
each of these channels further,
according to the highest final state multiplicity that they involve,
(cf. Eq.~\eqref{eq:nnlo_split_fin})
\be
\dhs^{\NNLO}_{\fa\fb} = \dhs^{\NNLO}_{V+2,\fa\fb}+\dhs^{\NNLO}_{V+1,\fa\fb}+
\dhs^{\NNLO}_{V,\fa\fb}. 
\ee
Finally, we separate the above terms into those involving only tree-level matrix elements and those terms involving
loop corrections, by writing\footnote{We note that certain partonic channels only contain a subset of these terms.}
\be
\begin{split}
&\dhs^{\NNLO}_{V+2,\fa\fb} = \dhs^{\NNLO}_{1245,\fa\fb},\\
&\dhs^{\NNLO}_{V+1,\fa\fb} = \dhs^{\NNLO}_{124,\fa\fb} + 
\dhs^{\NNLO}_{{\rm virt}_{124},\fa\fb},\\
&\dhs^{\NNLO}_{V,\fa\fb} \; \;\;\; = 
\dhs^{\NNLO}_{12,\fa\fb} + \dhs^{\NNLO}_{{\rm virt}_{12},\fa\fb}.
\end{split} \label{eq:nnlo_split_fin_dy}
\ee
The term $\dhs^{\NNLO}_{1245,\fa\fb}$ receives contributions from processes with NNLO-like kinematics
(i.e. with two additional resolved partons in the final state),
and corresponds to the fully subtracted real-real contribution.
The remaining terms arise from integrated subtraction terms, $\alpha_s$ and parton distribution
function renormalizations, and  real-virtual and purely virtual corrections. The terms
$\dhs^{\NNLO}_{124,\fa\fb}$ and $\dhs^{\NNLO}_{12,\fa\fb}$ only involve tree-level
matrix elements squared, while $\dhs^{\NNLO}_{{\rm virt}_{124},\fa\fb}$ and
$\dhs^{\NNLO}_{{\rm virt}_{12},\fa\fb}$ also involve finite remainders of virtual amplitudes. 
It is important to
emphasize that \emph{all} of the different terms in Eq.~\eqref{eq:nnlo_split_fin_dy} are
separately finite, so that we can discuss them separately. 
In what follows, we will present results for each of these terms.

\subsection{LO and NLO}
\label{sec:dy_lonlo}
We start by discussing the quark channel $\dhs_{\fa\fb}$, with $a,b\ne0$, which 
is the only channel contributing at leading order. 
The LO cross section reads 
\be
2s\cdot \d\hat\sigma^{\rm LO}_{\fa\fb} = 
\left\langle\FLMij{\fa}{\fb}(1,2)\right\rangle.
\ee
NLO QCD corrections to the quark channel can be written as
\bes
&
2s\cdot\d\hat\sigma^{\rm NLO}_{\fa\fb} = 
\bigg\langle \FLVijfin{\fa}{\fb}(1,2) 
+ \asontwopimu \left[\frac{2\pi^2}{3}\Cf-2\gamma_q\LMu\right]
\FLMij{\fa}{\fb}(1,2) \bigg\rangle
\\
&
+\asontwopimu\int\limits_0^1\d z 
\left[ 
\mathcal P'_{qq}(z) - \hat P^{(0)}_{qq,R}(z) \LMu 
\right]
\left\langle 
\frac{\FLMij{\fa}{\fb}(z\cdot1,2)+\FLMij{\fa}{\fb}(1,z\cdot 2)}{z}
\right\rangle
\\
&
+ \left\langle \ONLO \FLMij{\fa}{\fb}(1,2,4)\right\rangle,
\end{split}
\label{eq:dy_nlo_qqb}
\ee
with $a,b\ne 0$.
The $\ONLO$ operator that appears in this formula renders the
contribution of the single-gluon emission process finite. It reads
\be
\ONLO = (I-C_{41}-C_{42}) ( I - S_4).
\label{eq:onlo}
\ee
The other quantity in Eq.~\eqref{eq:dy_nlo_qqb}, $\left \langle
\FLVijfin{\fa}{\fb}(1,2)\right \rangle $, refers to the finite remainder
of the (UV-renormalized) one-loop virtual correction. Its definition
is given in Appendix~\ref{app:virtual}. Finally,
$\hat P^{(0)}_{qq,R}$ and  $\mathcal P_{qq}'$
are related to the 
LO Altarelli-Parisi splitting function,
and $\gamma_q = 3\Cf/2$, see Appendix~\ref{app:split} for 
explicit formulas.

The $qg$ and $gq$ channels start contributing at NLO. They read
\bes
2s\cdot\d\hat\sigma^{\rm NLO}_{\fa g} &= 
\asontwopimu\int\limits_0^1\d z \sum_x
\left\langle \frac{\FLMij{\fa}{\fx}(1,z\cdot 2)}{z}\right\rangle
\left[
\mathcal P'_{qg}(z) - \hat P^{(0)}_{qg,R}(z) \LMu  
\right]
\\
&+ \left\langle \ONLO \FLMij{\fa}{g}(1,2,4)\right\rangle,
\end{split}
\ee
and analogously
\bes
2s\cdot\d\hat\sigma^{\rm NLO}_{g\fa} &= 
\asontwopimu\int\limits_0^1\d z \sum_x
\left[
\mathcal P'_{qg}(z)-\hat P^{(0)}_{qg,R}(z) \LMu 
\right]
\left\langle \frac{\FLMij{\fx}{\fa}(z\cdot 1,2)}{z}\right\rangle
\\
&+ \left\langle \ONLO \FLMij{g}{\fa}(1,2,4)\right\rangle,
\end{split}
\ee
with $a\ne0$ and the various splitting functions defined in App.~\ref{app:split}. 
Note that in this case  only a subset of soft/collinear 
singularities are present in $\FLM(1,2,4)$; for example
\bes
\ONLO \FLMij{q}{g}(1,2,4) &= 
(I-C_{41}-C_{42}) ( I - S_4) \FLMij{q}{g}(1,2,4) 
\\
&=
(\I-C_{42})\FLMij{q}{g}(1,2,4).
\end{split}
\ee

\subsection{NNLO: quark channels}
\label{sec:dy_nnlo_quarks}

In this section we consider the NNLO corrections to $\dhs_{\fa\fb}$,
with $a,b\ne0$.  This includes the partonic processes $q_i\qb_j \to
V+gg$, $q_i\qb_j \to V+q_k\qb_l$ and $q_i q_j \to V+ q_k q_l$.  Of
these partonic processes, the $q_i\qb_j \to V+gg$ has the most
complicated singularity structure; it was discussed in detail in
Ref.~\cite{Caola:2017dug} and reviewed in
Section~\ref{sec:basics}. Recall that we introduced an energy ordering
$E_4 > E_5$ (cf. Eq.~\eqref{eq2}), which is natural since the
amplitude is symmetric under the exchange of the two final state
gluons.

The singularity structure is much simpler for final state quarks,
where one could use only two sectors to separate the collinear singularities.
Nevertheless, we find it convenient to treat the gluon and quark final states
on an equal footing.
We therefore need to symmetrize the amplitudes involving the final state quarks
explicitly,
since they are not symmetric in general; we do this by writing
\be
\begin{split}
\int  \df4\df5 &\FLMqqb(1,2,4,5) 
\\
&= \int  \df4\df5 \FLMqqb(1,2,4,5) \big[ \theta(E_4 > E_5)+ \theta(E_4 < E_5) \big]\\
&= \int \df4\df5 \theta(E_4 > E_5) \big[ \FLMqqb(1,2,4,5)  + \FLMqqb(1,2,5,4) \big]] \\
 &\equiv \left \langle \FLM(1,2,4,5) \right \rangle.
\end{split}
\ee
If one wishes to consider the final state gluons and quarks separately,
one could do away with the energy ordering and the symmetrization of the quark amplitudes.
We emphasize that in this case, the formulas in the forthcoming sections
would require modifications. 

As mentioned in Section~\ref{sec:basics}, an important part of the subtraction scheme
is the separation of the phase space into partitions such that in
each partition, only a limited number of kinematic configurations leads to
collinear divergences, cf. Eq.~\eqref{part1}.
Throughout this paper, we choose the partition  functions       to be
\be
\begin{split} 
  & w^{41,51} = \eta_{42} \eta_{52} \left (  1 + \frac{\eta_{41}}{\eta_{45} + \eta_{42} + \eta_{51} }
  + \frac{\eta_{51}}{\eta_{45} + \eta_{41} + \eta_{52}} \right ),
  \\
  & w^{42,52} = \eta_{41} \eta_{51} \left (  1 + \frac{\eta_{42}}{\eta_{45} + \eta_{41} + \eta_{52} }
  + \frac{\eta_{52}}{\eta_{45} + \eta_{42} + \eta_{51}} \right ),
  \\
  & w^{41,52} = \frac{\eta_{42} \eta_{51} \eta_{45}}{\eta_{45} + \eta_{41} + \eta_{52}},
  \;\;\;
  w^{42,51} = \frac{\eta_{41} \eta_{52} \eta_{45}}{\eta_{45} + \eta_{42} + \eta_{51}},
\end{split}
\label{eq:partitions}
\ee
where we have used $\eta_{ij} = \rho_{ij}/2$.
It is straightforward to check that these functions restrict the collinear singularities as discussed in Section~\ref{sec:basics},
and also that they sum up to one, cf. Eq.~\eqref{part1}.

We now present results for the different terms in Eq.~\eqref{eq:nnlo_split_fin_dy} that arise in the quark channel.

\subsubsection{Terms with NNLO kinematics}
\label{sec:dy_nnlo_quarks_nnlo}
This (hard) regularized contribution is the only one that involves the 
full  matrix element for $\fa \fb \to V+f_4 f_5$.  
It reads~\cite{Caola:2017dug}
\begin{align}
&  \d \hat \sigma^{\rm NNLO}_{1245,\fa\fb} =
\sum_{(ij)\in dc}\bigg\langle
\bigg[(\I- C_{5j})(\I-C_{4i})\bigg]
\big[\I-\SS\big]\big[\I-S_5\big]\times
\nonumber
\\
&\quad\quad\quad\quad\quad\quad
\times\df4\df5 w^{4i,5j}\FLMij{\fa}{\fb}(1,2,4,5)
\bigg\rangle
\nonumber
\\
&\quad\quad
+\sum_{i\in tc} 
\bigg\langle
\bigg[
\theta^{(\mathpzc{a})} \big[\I-\CC_i\big]\big[\I-C_{5i}\big] + 
\theta^{(\mathpzc{b})} \big[\I-\CC_i\big]\big[\I-C_{45}\big] 
\label{eq:qqhard}
\\
&\quad\quad\quad\quad~~
 + \theta^{(\mathpzc{c})} \big[\I-\CC_i\big]\big[\I-C_{4i}\big]+ 
\theta^{(\mathpzc{d})} \big[\I-\CC_i\big]\big[\I-C_{45}\big]
\bigg]
\nonumber
\\
&\quad\quad\quad\quad~~
\times \big[\I-\SS\big]\big[\I-S_5\big]
\df4 \df5 w^{4i,5i}\FLMij{\fa}{\fb}(1,2,4,5)
\bigg\rangle.\nonumber
\end{align}
In this equation, $dc = \{ (1,2), (2,1)\}$ and $tc = \{1,2\}$ refer to
the double- and triple-collinear partitions, respectively, while the
sectors $(\mathpzc{a})$--$(\mathpzc{d})$ are defined by the angular ordering conditions in
Eq.~\eqref{eq:defsectors}.  The operators $S_5$, $\SS$, $C_{ij}$ and $\CC_i$
have been discussed in great detail in Ref.~\cite{Caola:2017dug}, and in Sec.~\ref{sec:basics}.

We note that $ \d \hat \sigma^{\rm NNLO}_{1245\,\fa\fb}$ is computed
numerically in four dimensions.  In order to do so, we must provide
the explicit parametrization of the phase space for the {\it complete}
final state which includes two radiated partons and a vector boson (or
its decay products).  It is clear that there are many different ways
to do so.  We find it useful to describe the phase space using
tree-level variables, i.e. the invariant mass $M_V^2$ and the rapidity
$Y$ of the vector boson, but other choices are possible.  In addition,
we have to choose the energies of the two final state partons and the
relative angles between them and the hard emittor\footnote{The
identity of the ``hard emittor'' depends on the partition.} as
independent variables, in order to extract singularities in the same
way as in the computation of the integrated subtraction terms, which
are presented in the forthcoming subsections.  For this reason, there
is less freedom in  choosing how to parametrize the momenta  of the radiated
partons. We have discussed this point   in some detail in Appendix B of
Ref.~\cite{Caola:2017dug}.  We will not repeat this discussion,
instead, our goal here is to provide a guide for a numerical implementation of
Eq.~\eqref{eq:qqhard}.

We work in the center-of-mass frame of the colliding partons.
As the first step, we determine  the center-of-mass collision energy squared $s$.
To do so, we parametrize the energies of the radiated partons as\footnote{We remind the reader that we have chosen $\Em =\sqrt{s}/2$.}
\be
E_4 = \frac{\sqrt{s}}{2} x_1,\;\;\;\; E_5 = \frac{\sqrt{s}}{2} x_1 x_2,
\ee
where $x_1, x_2 \in [0:1]$, 
and use momentum conservation $(p_1 + p_2 - p_4 - p_5)^2 = M_V^2$ to find 
\be
s = \frac{M_V^2}{1 - x_1(1+x_2) + x_1^2 x_2 \eta_{45}}.
\ee
There is an obvious constraint $s < S_h$, where
$S_h$ is the center-of-mass energy squared of the colliding hadrons, 
that we have to impose while generating the events. 

The choice of angular variables depends on the partition and the
sector; for the sake of definiteness, we will discuss the sector 
``$\mathpzc{a}$''
of the partition $w^{41,51}$.  In this case, the scalar products
$\eta_{ij}=(1 - \cos\theta_{ij}) /2$ may be parametrized by
the           variables $x_3, x_4, \lambda \in [0:1]$ according to
\be
\eta_{41} = x_3,\;\;\; \eta_{51} = 
x_3 \frac{x_4}{2},
\;\;\; \eta_{45} = \frac{ x_3(1-x_4/2)^2}{N_F(x_3,x_4/2,\lambda)},
\label{eq316}
\ee
where
\be
N_F(x_3,x_4,\lambda) = 1 + x_4(1-2 x_3) - 2(1-2\lambda) \sqrt{x_4(1-x_3)(1-x_3 x_4)},
\ee
see~\cite{czakonsub}.
We note that, since $0 \le x_4 \le 1$, the angular ordering $\eta_{51} < \eta_{41}/2$ is assured.

In addition to the invariant mass, we also fix the rapidity of the vector boson in the laboratory
frame, $Y$. This allows
us to determine fractions of hadron energies carried by the  colliding partons, $\xi_{1,2}$. We find
\be
\xi_{1,2} = \sqrt{ \frac{s}{S_{\rm h}}} e^{\pm y}, 
\ee
where
\be
y = Y - \frac{1}{2} \ln \frac{ 1 - x_1 (1- \eta_{41}) - x_1 x_2 (1 - \eta_{51})  }{ 1 - x_1 (1- \eta_{42}) - x_1 x_2 (1 - \eta_{52})         }.
\label{eq:319}
\ee
We require  that $ 0 < \xi_{1,2} < 1$ and that both the numerator and the denominator in the argument
of the logarithm in Eq.~\eqref{eq:319} are positive definite.

We are now in position to write down the four-momenta of the QCD partons in an event. We do so in the
partonic center-of-mass frame. The knowledge of $\xi_{1,2}$ then allows us to boost momenta
to the laboratory frame where all kinematic constraints are defined.  The momenta read
\be
\begin{split}
  & p_{1,2} = \frac{\sqrt{s}}{2} \left ( 1, 0, 0, \pm 1 \right ), \\
  & p_4 = \frac{\sqrt{s}}{2} x_1 \left ( 1, \sin \theta_{41} \cos \varphi_4, \sin \theta_{41} \sin \varphi_4, \cos \theta_{41} \right ),  \\
  & p_5 = \frac{\sqrt{s}}{2} x_1 x_2 \left ( 1, \sin \theta_{51} \cos(\varphi_4 + \varphi_{45})  ,
   \sin \theta_{51} \sin(\varphi_4 + \varphi_{45}) , \cos \theta_{51} \right ),
  \end{split}
  \label{eq4.6}
\ee
where
$\cos\theta_{ij} = 1-\rho_{ij}$ and~\cite{czakonsub}
\be
\sin \varphi_{45} = \frac{ 2 \sqrt{\lambda (1-\lambda)} (1-x_4/2)}{N_F(x_3,x_4/2,\lambda)},
\;\;\;\;\;
\cos \varphi_{45} = \pm \sqrt{1 -  \sin^2 \varphi_{45}}.
\ee
The four-momentum of the vector boson is obtained by
momentum conservation $p_V= p_1 + p_2 - p_4 - p_5$. If needed, further details of the colorless final
state can be described. For example, in case of $V \to l^+l^-$, the phase space for leptonic decays is generated in the
$V$-rest frame and the lepton momenta are boosted back into the partonic
center-of-mass frame using the known $p_V$ and $M_V^2$. 

For the chosen partition and sector,  the  phase space weight reads
\be
w_h(\{x_i\},\lambda;\{y_i\}) =  
\frac{w_{\LO}(\{y_i\})}{(8\pi^2)^2}
\frac{s^3}{8 M_V^2}\; \frac{x_1^3 x_2 x_3 (1-x_4/2)}{N_F(x_3,x_4/2,\lambda)}
\; w^{41,51}(\eta_{41},\eta_{42},\eta_{51},\eta_{52},\eta_{45}),
\ee
where $w_{\LO}$ is the weight for the Born $\fa\fb\to V$ process, which 
depends in general on a set of variables $\{y_i\}$ that describe the 
$V$ final state. 
The contribution of the generated {\it hard} event to the phase-space integral is then
\be
 \I\FLMqqb(1,2,4,5)  \to 
\mathcal N f_q(\xi_1) f_{\bar q}(\xi_2) w_h(\{x_i\},\lambda;\{y_i\}) |{\cal M}(p_1,p_2,p_V,p_4,p_5)|^2.
\ee
The matrix element squared can be calculated either in the center-of-mass frame or in the laboratory frame since
all required boosts are defined at this point.

The contribution that we just described corresponds to the
product of identity operators in Eq.~\eqref{eq:qqhard}; below we
discuss how the subtraction terms in Eq.~\eqref{eq:qqhard} are to be
calculated.
To this end, we consider first the class of terms in Eq.~\eqref{eq:qqhard} where
the double-soft operator $\SS$ appears.  We will start with the term
$\SS F_{\rm LM}$ and describe the weight of the counter-event produced
by this term.  To compute the weight, we take the limit $x_1 \to 0$
everywhere; this corresponds to $E_{4,5} \to 0$ at $E_5/E_4$ held
fixed.  We obtain
\be
s_{\SS} = M_V^2,\;\; y_{\SS} = Y,\;\; \xi^{\SS}_{1,2} = \sqrt{\frac{M_V^2}{S_h}} e^{\pm Y}.
\ee
The four-vectors for $p_{1,2}$ and $p_{4,5}$ are the same as in
Eq.~\eqref{eq4.6} but the four-momentum of the vector boson reads $P_V
= p_1 + p_2$, i.e. the radiation of the two soft partons does not
impact the kinematics of the vector boson.  The phase space weight of
the counter-event reads
\be
w_{\SS}(\{x_i\},\lambda;\{y_i\})
= \frac{w_{\LO}(\{y_i\})}{(8\pi^2)^2}\frac{M_V^4}{8}\; \frac{x_1^3 x_2 x_3 (1-x_4/2)}{N_F(x_3,x_4/2,\lambda)} \; w^{41,51}(\eta_{41},\eta_{42},\eta_{51},\eta_{52},\eta_{45}), 
\ee
and its  contribution to the fiducial cross section becomes 
\be
\SS  \FLMqqb(1,2,3,4,5) \to \mathcal N f_q(\xi^{\SS}_1) f_{\bar q}(\xi^{\SS}_2) w_{\SS}(\{x_i\},\lambda;\{y_i\})
{\rm } {\rm Eik}(1,2,4,5) |{\cal M}(p_1,p_2)|^2.
\label{eq:ssweight}
\ee

Suppose we consider terms in Eq.~\eqref{eq:qqhard} where, in addition 
to the double-soft operator $\SS$, some other operator acts on $\FLMqqb$. In this
case, we should just set the relevant 
variables(s)
to zero in Eq.~\eqref{eq:ssweight}
and, if necessary, change the way the four-momenta  are generated.
For example, consider a term $C_{51} \SS \FLMqqb(1,2,4,5)$.  For this sector,
the operator $C_{51}$ implies that $x_4$ should be set to zero everywhere after the leading $1/x_4$
asymptotic is extracted.
The four-momenta are then unchanged, except for $p_5$ which becomes
\be
p_5 =  \frac{M_V}{2} x_1 x_2 \left ( 1, 0, 0, 1\right).
\ee
Computing the $5||1$ limit of the double-soft eikonal function,
we arrive at the contribution from the $C_{51} \SS \FLMqqb(1,2,4,5)$
kinematic configuration
\be
\begin{split} 
  C_{51} \SS  &\FLMqqb(1,2,3,4,5) \to \\
&  \mathcal N f_q(\xi^{\SS}_1) f_{\bar q}(\xi^{\SS}_2) \;
w_{\SS}(\{x_i\},\lambda;\{y_i\})|_{x_4\to 0}
   \;
  16 C_F^2 \frac{s_{12}}{s_{15} s_{25}} \frac{s_{12}}{s_{14}s_{24}} |{\cal M}(p_1,p_2)|^2,
  \end{split}
\ee
where
\be
\begin{split}
&s_{12} = 4 E_1 E_2, \;\;\;\;\;\;\; s_{15} = 2 E_1 E_5 x_3 x_4,
 \;\;\;\;\;\; s_{25} = 4 E_2 E_5, \;\;\;\;\;\; \\
&  s_{14} = 4 E_1 E_4 x_3, \;\;\;\;\; s_{24} = 4 E_2 E_4(1-x_3).
\end{split}
\ee
We note that, according to these equations, $s_{15} \ne 2 (p_1 \cdot p_5)$. This is so because  the $1/s_{15}$ term
describes the leading $x_4 \to 0$ singularity, so $x_4 \ne 0$ is kept there and set to zero everywhere else.  

As the last example, consider the triple-collinear limit, which  corresponds to $x_3 \to 0$, cf. 
Eqs.~(\ref{eq316},~\ref{eq4.6}).
In this case, the partonic center of mass collision energy squared is
\be
s_{\CC} = \frac{M_V^2}{1 - x_1(1+x_2)}.
\label{eq:C}
\ee
Similar to the case of hard event, we require $0 < s_{\CC} < S_h$.  The four-momenta of partons to be used in the matrix element and the momentum-conserving $\delta$-function are 
\be
\begin{split}
  & p_{1,2} = \frac{\sqrt{s_{\CC}}}{2} \left ( 1, 0, 0, \pm 1 \right ), \\
  & p_4 = \frac{\sqrt{s_{\CC}}}{2} x_1 \left ( 1, 0, 0, 1 \right ), \\
  & p_5 = \frac{\sqrt{s_{\CC}}}{2} x_1 x_2 \left ( 1,0,0,1 \right ),
\end{split}
\ee
and the vector boson four-momentum is
\be
p_V = (p_1 - p_4 - p_5) + p_2 = \frac{E_1-E_4-E_5}{E_1} p_1 + p_2
= (1-x_1(1+x_2)) p_1 + p_2.
\label{eq:pVC}
\ee
Combining Eq.~\eqref{eq:pVC}  with Eq.~\eqref{eq:C}, 
we easily check that $p_V^2 = M_V^2$ as, of course,
it should be. 

As we emphasized in Sec.~\ref{sec:basics}, we define $\CC_i$ in such a way that
it \emph{does not} act on the phase-space. The weight of the counter-event is then
identical to the one for the hard process
\be
w_{\CC}(\{x_i\},\lambda;\{y_i\})=w_h(\{x_i\},\lambda;\{y_i\})
\ee
and the parameter $y$ relevant for calculating  fractions of hadron energies carried by the incoming
partons reads
\be
y_{\CC} = Y - \frac{1}{2} \ln \left ( 1 - x_1 (1+x_2) \right ).
\ee
The momentum fractions themselves are then given by
\be
\xi_{1,2}^{\CC}= \sqrt{\frac{s_{\CC}}{S_h}} e^{\pm y_{\CC}}.
\ee
Combining the different ingredients, we derive  the weight of the triple-collinear counter-event
\bes
\CC_{51} \FLMqqb(1,2,4,5) 
\to \mathcal N f_q(\xi^{\CC}_1) f_{\bar q}(\xi^{\CC}_2) w_{\CC} (\{x_i\},\lambda;\{y_i\})
\times\\
\lp\frac{2}{s_{145}}\rp^2P_{ggq}(s_{45},-s_{14},-s_{15};z_4,z_5) |{\cal M}(p'_1,p_2)|^2,
\end{split}
\ee
where $p'_1 = (1-x_1(1+x_2)) p_1$.
The arguments of the triple-collinear splitting function $P_{ggq}$ are then computed
as
\be
s_{14} = 4 E_1 E_4 x_3, \;\;\; s_{15} = 2 E_1 E_5 x_3 x_4,
\;\;\;\; s_{45} = \frac{4 E_4 E_5 x_3 (1-x_4/2)^2}{N_{F}(x_3,x_4/2,\lambda)},
\ee
and $s_{145} = s_{45}-s_{14} -s_{15}$, $z_i = E_i/(E_4+E_5-E_1)$. We stress that the above scalar products in the splitting function are evaluated with $x_3\ne 0$, i.e.
\emph{not} in the triple-collinear limit. 

The remaining contributions to the fully-subtracted cross section $  \d  \hat\sigma^{\rm NNLO}_{1245,\fa\fb} $  are computed along the same lines. The important
thing is that we always take the leading singularity in the relevant variables 
and employ the  limiting behavior of amplitudes squared
to calculate weights of the subtraction terms. We also make sure that 
the subtraction counter-terms that make the hard matrix element finite
are identical to the  subtraction terms that have been analytically integrated.

\subsubsection{Tree-level terms with NLO kinematics}
In this section, we consider the term with NLO kinematics and
tree-level matrix elements, $\d\hat\sigma^{\rm NNLO}_{124,\fa\fb} $.
The general structure of this contribution is
\bes
\d\hat\sigma^{\rm NNLO}_{124,\fa\fb} &
= 
\asontwopimu
\sum_x\int\limits_0^1\d z 
\Bigg\{
\hat P^{(0)}_{\fx\fa,R}(z)
\left\langle
\ln\frac{\rho_{41}}{4}\ONLO
\left[\frac{\tilde w_{5||1}^{41,51} \FLMij{\fx}{\fb}(z\cdot1,2,4)}{z}\right]
\right\rangle
\\
&
+\left[\mathcal P'_{\fx\fa}(z) - \hat P^{(0)}_{\fx\fa,R}(z)\LMu\right]
\left\langle
\ONLO\left[\frac{\FLMij{\fx}{\fb}(z\cdot1,2,4)}{z}\right]
\right\rangle
\\
&
+\left\langle
\ln\frac{\rho_{42}}{4}\ONLO
\left[\frac{\tilde w_{5||2}^{42,52} \FLMij{\fa}{\fx}(1,z\cdot2,4)}{z}\right]
\right\rangle \hat P^{(0)}_{\fx\fb,R}(z)
\\
&
+\left\langle
\ONLO\left[\frac{\FLMij{\fa}{\fx}(1,z\cdot2,4)}{z}\right]
\right\rangle
\left[\mathcal P'_{\fx\fb}(z) - \hat P^{(0)}_{\fx\fb,R}(z)\LMu\right]
\Bigg\}
\\
&
+\asontwopimu \bigg\langle
\ONLO\bigg[
\Delta_q \cdot \FLMij{\fa}{\fb}(1,2,4)
+\Delta^r \cdot \big[r_\mu r_\nu \FLMij{\fa}{\fb}^{\mu\nu}(1,2,4)\big]\bigg]
\bigg\rangle,
\end{split}
\label{eq:FLM124}
\ee
where the 
various splitting functions are defined in Appendix~\ref{app:split}.
Although we only need $\Delta_q$ in Eq.(\ref{eq:FLM124}), it appears to be convenient to introduce  a more general object
$\Delta_{i \in \{q,g \}}$. It is defined as 
\allowdisplaybreaks
\begin{align}
\Delta_i & =
\Ci\left[
\frac{2}{3}\pi^2
-2\ln\frac{2 E_4}{\sqrt{s}}\lp
\tilde w_{5||1}^{41,51}\ln\frac{\eta_{41}}{2}
+
\tilde w_{5||2}^{42,52}\ln\frac{\eta_{42}}{2}
\rp
\right]
\nonumber
\\
&+
\Ca \left[
\ln\frac{2E_4}{\sqrt{s}}
\lp 
\ln\frac{2E_4}{\sqrt{s}} + 
\ln(\eta_{41}\eta_{42})
\rp
-\ln\eta_{41} \ln\eta_{42}
\right]
\nonumber
\\
&
+\lp\frac{137}{18}-\frac{7}{6}\pi^2\rp \Ca 
- \frac{13}{9}\nf
+\beta_0
\bigg[
\tilde w_{4||5}^{41,51} \ln\frac{\eta_{42}}{\eta_{41}}
+
\tilde w_{4||5}^{42,52} \ln\frac{\eta_{41}}{\eta_{42}}
+\frac{\ln (\eta_{41}\eta_{42})}{2} 
\label{eq:delta}
\\
&
- \ln \frac{2 E_4}{\sqrt{s}} 
+ 2\ln 2
\bigg]
+{\mathcal X}_i 
\bigg[\ln\frac{2E_4}{\sqrt{s}} + \frac{\ln(\eta_{41}\eta_{42})}{2}\bigg]
 - 2 \gamma_i \LMu,
\nonumber
\end{align}
In the case  $i = q$, we find $C_q = \Cf$, $\gamma_q = 3\Cf/2$ 
and $\mathcal X_q=3\Ca/2$, cf.
Appendix.~\ref{app:realvirtual}. In addition
\be
\Delta^r  = -\frac{\Ca}{3} + \frac{\nf}{3}.
\ee
Note that to obtain Eq.~\eqref{eq:delta}, we used $\eta_{41}+\eta_{42}=1$. 
In Eq.~\eqref{eq:delta}, $\FLMij{\fa}{\fb}^{\mu\nu}$ is analogous to
$\FLMij{\fa}{\fb}$ but with the polarization vector for gluon 4 removed
\bes
\langle \FLMij{\fa}{\fb}^{\mu\nu}(1,2,4) \rangle = \mathcal N 
\int  {\rm dLips}(V)\dg4 (2\pi)^d \delta_d(p_1+p_2 - p_V  - p_4)\times
\\
\big[{\cal M}^{{\rm tree},\mu}{\cal M}^{*,{\rm tree},\nu}\big]
(p_1,p_2,p_V,p_4) \;{\cal O}(p_V,p_4),~~~~~
{\cal M}^{\rm tree} = \epsilon_\mu(p_4) {\cal M}^{{\rm tree},\mu}.
\end{split}
\ee
$\FLM^{\mu\nu}$ is contracted with $r^\mu r^\nu$ where $r^\mu$ is a unit vector that spans the two-dimensional space orthogonal
to $p_4$, see Ref.~\cite{Caola:2017dug} for further details. If $p_4$ is 
parametrized as in Eq.~\eqref{eq4.6}, then 
\be
r^{\mu} = 
(0,-\cos\theta_{41}\cos\varphi_4,-\cos\theta_{41}\sin\varphi_4,\sin\theta_{41}).
\ee
Note that since $r \cdot p_4 = 0$ and $r^2 = - 1$, we can view $r^\mu$ as the polarization vector of the emitted gluon. 

As an illustration, we now explicitly write Eq.~\eqref{eq:FLM124} in the
case of $Z$ production.
For the same-flavor channel $(\fa,\fb)=(q,\qb)$, 
Eq.~\eqref{eq:FLM124} becomes
\bes
&\d\hat\sigma^{Z,\rm NNLO}_{124,q\bar q} 
= 
\asontwopimu
\int\limits_0^1\d z 
\Bigg\{
\hat P^{(0)}_{qq,R}(z)
\Bigg\langle
\ln\frac{\rho_{41}}{4}\ONLO
\bigg[
\frac{\tilde w_{5||1}^{41,51} 
\FLMij{q}{\qb}(z\cdot1,2,4)
}{z}\bigg]
\\
&
+
\ln\frac{\rho_{42}}{4}\ONLO
\bigg[
\frac{\tilde w_{5||2}^{42,52} 
\FLMij{q}{\qb}(1,z\cdot2,4)
}{z}\bigg]
\Bigg\rangle
+
\hat P^{(0)}_{gq,R}(z)
\Bigg\langle
\ln\frac{\rho_{41}}{4}\ONLO
\\
&\times
\bigg[
\frac{\tilde w_{5||1}^{41,51} 
\FLMij{g}{\qb}(z\cdot1,2,4)
}{z}\bigg]
+
\ln\frac{\rho_{42}}{4}\ONLO
\bigg[
\frac{\tilde w_{5||2}^{42,52} 
\FLMij{q}{g}(1,z\cdot2,4)
}{z}\bigg]
\Bigg\rangle+
\\
&
\left[\mathcal P'_{qq}(z) - \hat P^{(0)}_{qq,R}(z)\LMu\right]
\left\langle
\ONLO\left[\frac{\FLMij{q}{\qb}(z\cdot1,2,4)
+\FLMij{q}{\qb}(1,z\cdot2,4)}{z}\right]
\right\rangle + 
\\
&
\left[\mathcal P'_{gq}(z) - \hat P^{(0)}_{gq,R}(z)\LMu\right]
\left\langle
\ONLO\left[\frac{\FLMij{g}{\qb}(z\cdot1,2,4)
+\FLMij{q}{g}(1,z\cdot2,4)}{z}\right]
\right\rangle
\Bigg\}
\\
&
+\asontwopimu \bigg\langle
\ONLO\bigg[
\Delta_q \cdot \FLMij{q}{\qb}(1,2,4)
+\Delta^r \cdot \big[r_\mu r_\nu \FLMij{q}{\qb}^{\mu\nu}(1,2,4)\big]\bigg]
\bigg\rangle.
\end{split}
\ee
For different-quark channels $(\fa,\fb)=(q_i,q_j)$ with $q_i \ne \bar q_j$,
we find
\bes
&\d\hat\sigma^{Z,\rm NNLO}_{124,q_i q_j} 
= 
\asontwopimu
\int\limits_0^1\d z 
\Bigg\{
\hat P^{(0)}_{gq,R}(z)
\Bigg\langle
\ln\frac{\rho_{41}}{4}\ONLO
\bigg[
\frac{\tilde w_{5||1}^{41,51} 
\FLMij{g}{q_j}(z\cdot1,2,4)
}{z}\bigg]
\\
&
+
\ln\frac{\rho_{42}}{4}\ONLO
\bigg[
\frac{\tilde w_{5||2}^{42,52} 
\FLMij{q_i}{g}(1,z\cdot2,4)
}{z}\bigg]
\Bigg\rangle
+
\left[\mathcal P'_{gq}(z) - \hat P^{(0)}_{gq,R}(z)\LMu\right]
\\
&
\times
\left\langle
\ONLO\left[\frac{\FLMij{g}{q_j}(z\cdot1,2,4)
+\FLMij{q_i}{g}(1,z\cdot2,4)}{z}\right]
\right\rangle
\Bigg\}.
\end{split}
\ee

\subsubsection{Tree-level terms with LO kinematics}
\label{sec:dy_nnlo_quarks_lo}
We now turn to the contribution involving terms with 
LO kinematics and tree-level matrix elements,
$\d\hat\sigma^{\rm NNLO}_{12,\fa\fb}$.
Accounting for the boost of the initial state along the collision axis, 
it can naturally be split into

\be
\d\hat\sigma^{\rm NNLO}_{12,\fa\fb} = 
\d\hat\sigma^{\rm NNLO}_{(z,\zb),\fa\fb}
+\d\hat\sigma^{\rm NNLO}_{(z,2),\fa\fb}
+\d\hat\sigma^{\rm NNLO}_{(1,z),\fa\fb}
+\d\hat\sigma^{\rm NNLO}_{(1,2),\fa\fb},
\ee
$a,b\ne0$.
We now consider each of these terms separately.

\begin{enumerate}
\item \underline{{\bf Terms involving} $\FLM(z\cdot 1,\zb\cdot2)$:}
\bes
\d\hat\sigma^{\rm NNLO}_{(z,\zb),\fa\fb} &
= 
\lp\asontwopimu\rp^2\sum_{x,y}\int\limits_0^1\d z~\d\zb
\left[
\mathcal P'_{\fx\fa}(z)
-\LMu \hat P^{(0)}_{\fx\fa,R}(z)
\right]
\\
&\times
 \left\langle\frac{\FLMij{\fx}{\fy}(z\cdot1,\zb\cdot2)}{z\zb}\right\rangle
\left[
\mathcal P'_{\fy\fb}(\zb)
-\LMu \hat P^{(0)}_{\fy\fb,R}(\zb)
\right].
\end{split}
\label{eq:FLMzzb}
\ee

Once again, to illustrate this equation we consider the case of $Z$
production.  Here, this contribution is only relevant for the 
$(\fa,\fb)=(q,\qb)$
channel, where it reads
\bes
\d\hat\sigma^{Z,\rm NNLO}_{(z,\zb),q\qb} &
= 
\lp\asontwopimu\rp^2\int\limits_0^1\d z~\d\zb
\left[
\mathcal P'_{qq}(z)
-\LMu \hat P^{(0)}_{qq,R}(z)
\right]
\\
&\times
 \left\langle\frac{\FLMij{q}{\qb}(z\cdot1,\zb\cdot2)}{z\zb}\right\rangle
\left[
\mathcal P'_{qq}(\zb)
-\LMu \hat P^{(0)}_{qq,R}(\zb)
\right].
\end{split}
\ee

\item \underline{{\bf Terms involving} $\FLM(z\cdot1,2)$ {\bf and} 
$\FLM(1,z\cdot2)$:}
\bes
\d\hat\sigma^{\rm NNLO}_{(z,2),\fa\fb} + 
\d\hat\sigma^{\rm NNLO}_{(1,z),\fa\fb} &
=
\lp\asontwopimu\rp^2 \sum_x \int\limits_0^1\d z
\bigg[ \mathcal T_{\fx\fa}(z) 
\left\langle\frac{\FLMij{\fx}{\fb}(z\cdot1,2)}{z}\right\rangle
\\
&+
\left\langle\frac{\FLMij{\fa}{\fx}(1,z\cdot2)}{z}\right\rangle
\mathcal T_{\fx\fb}(z) \bigg].
\end{split}
\label{eq:FLMz}
\ee
This term has a non-trivial flavor structure. To simplify it,
we employ the notation used 
to describe the NNLO QCD contributions to the 
Altarelli-Parisi splitting functions, and write the functions $\mathcal{T}$ in terms of nonsinglet, singlet, and vector functions
\bes
\mathcal T_{q_i q_j} & = \delta_{ij} \mathcal T_{qq}^{\rm NS} 
+ \mathcal T_{qq}^{\rm S},\\
\mathcal T_{q_i \qb_j} & = \delta_{ij} \mathcal T_{q\qb}^{\rm V} + 
\mathcal T_{q\qb}^{\rm S}. 
\end{split}
\ee
Similar to the Altarelli-Parisi splitting functions, we have 
$\mathcal T_{q\qb}^{\rm S} = \mathcal T_{qq}^{\rm S}$ through NNLO,
and we will always use the latter in what follows.
Again, we consider the example of $Z$ production. For the
$(\fa,\fb)=(q\qb)$ channel, Eq.~\eqref{eq:FLMz} becomes
\bes
\d\hat\sigma^{Z,\rm NNLO}_{(z,2),q\qb} + 
\d\hat\sigma^{Z,\rm NNLO}_{(1,z),q\qb} &
=
\lp\asontwopimu\rp^2 \int\limits_0^1\d z
\big[\mathcal T_{qq}^{\rm NS}(z) + \mathcal T_{qq}^{\rm S}(z)\big]
\\
&\times \left\langle\frac{\FLMij{q}{\qb}(z\cdot1,2)+
\FLMij{q}{\qb}(1,z\cdot2)}{z}\right\rangle,
\end{split}
\ee
while for the $q_i q_j$ with $i\ne -j$ it reads
\bes
\d\hat\sigma^{Z,\rm NNLO}_{(z,2),q_i q_j} + 
\d\hat\sigma^{Z,\rm NNLO}_{(1,z),q_i q_j}
&=
\lp\asontwopimu\rp^2 \int\limits_0^1\d z
~\big[\mathcal \delta_{ij} \mathcal T_{q\qb}^{\rm V}(z) 
+ \mathcal T_{qq}^{\rm S}(z)\big]
\\
&\times \left\langle\frac{\FLMij{\qb_j}{q_j}(z\cdot1,2)
+\FLMij{q_i}{\qb_i}(1,z\cdot2)}{z}\right\rangle.
\end{split}
\ee
The transition function $\mathcal T_{qq}^{\rm NS}$ is explicitly shown
in Appendix~\ref{app:transition}. All other $\mathcal T_{ij}$ functions
are presented in an ancillary file.

\item \underline{{\bf Terms involving} $\FLM(1,2)$:}
\begin{align}
&\d\hat\sigma^{\rm NNLO}_{(1,2),\fa\fb} = 
 \biggl\langle\FLMij{\fa}{\fb}(1,2)\biggr\rangle \times
\lp\asontwopimu\rp^2\Bigg\{
\Cf^2\bigg[\frac{8\pi^4}{45} 
-\big(2\pi^2 + 16 \zeta_3\big)\LMu
\nonumber\\
&\quad
+ \lp \frac{9}{2} - \frac{2\pi^2}{3}\rp
\LMusq\bigg]
+\Ca\Cf\bigg[\frac{739}{81} + \frac{209\pi^2}{72} 
- \frac{7\pi^4}{80}
+\ln2\times
\nonumber\\
&\quad
\lp\frac{4}{3}+\frac{11\pi^2}{9}-\frac{7}{2}\zeta_3\rp
+(\zeta_2-2)\ln^2{2} - \frac{\ln^4{2}}{6} 
-\frac{407}{36}\zeta_3 - 4\Li_4\lp\frac{1}{2}\rp
\label{eq:FLM12}
\\
&\quad
+\LMu\lp -\frac{199}{54}+\frac{23\pi^2}{24} - 7\zeta_3\rp - \frac{11}{4}\LMusq
\bigg]
+\Cf\nf\bigg[-\frac{214}{81}
\nonumber\\
&\quad
-\frac{7\pi^2}{18}-\ln2\lp\frac{4}{3}+
\frac{2\pi^2}{9}\rp + 2 \ln^2{2} + \frac{37}{18}\zeta_3
+\LMu\lp\frac{17}{27}-\frac{\pi^2}{12}\rp 
\nonumber\\
&\quad
+\frac{1}{2}\LMusq\bigg]
+\Theta_{bd}\bigg[\frac{23}{36}\Cf\nf + \Ca\Cf\lp \frac{\pi^2}{3}
-\frac{131}{36}
\rp + (2\ln2) \Cf\beta_0\bigg]\Bigg\}.
\nonumber
\end{align}
The $\Theta_{bd}$ term in Eq.~\eqref{eq:FLM12} depends on the choice of 
partition functions. It is defined as follows
\be
\Theta_{bd} \equiv 
-\bigg\langle
\left[\I-C_{41}-C_{42}\right]\bigg[\frac{\rho_{12}}{\rho_{41}\rho_{42}}
\lp \tilde w^{41,51}_{4||5} \ln\frac{\eta_{41}}{1-\eta_{41}}
+ \tilde w^{42,52}_{4||5} \ln\frac{\eta_{42}}{1-\eta_{42}}\rp\bigg]
\bigg\rangle.
\label{eq:thetabcdef}
\ee
If the partition functions are chosen as in Eq.~\eqref{eq:partitions},
it is immediate to obtain
\be
\Theta_{bd} = 2-\frac{\pi^2}{3}.
\label{eq:thetabc}
\ee
\end{enumerate}

\subsubsection{Terms involving virtual corrections}
Finally, we consider the two terms in Eq.~\eqref{eq:nnlo_split_fin_dy}
which involve virtual corrections, $\dhs^{\NNLO}_{{\rm virt}_{124},\fa\fb}$ and
$\dhs^{\NNLO}_{{\rm virt}_{12},\fa\fb}$.  The former corresponds to the
real-virtual corrections, which have NLO kinematics.  As such, they
have singularities that appear when the radiated parton becomes
unresolved.  These singularities can be subtracted as at NLO, so that
this term reads
\be
\d\hat\sigma^{\rm NNLO}_{{\rm virt}_{124},\fa\fb} = 
\big\langle \ONLO \FLVijfin{\fa}{\fb}(1,2,4)
\big\rangle,
\label{eq:rv}
\ee
where $\FLVijfin{\fa}{\fb}(1,2,4)$ is a finite remainder of the one-loop
amplitude, see Appendix~\ref{app:realvirtual}.
The other term corresponds to virtual contributions with LO kinematics.
It reads
\bes
\d\hat\sigma^{\rm NNLO}_{{\rm virt}_{12},\fa\fb} = &
\bigg\langle
\FLVVijfin{\fa}{\fb}(1,2)
+\FLVsqijfin{\fa}{\fb}(1,2)
 + \asontwopimu\left[\frac{2\pi^2}{3}\Cf-2\gamma_q\LMu\right] 
\times
\\
&
\quad
\FLVijfin{\fa}{\fb}(1,2)
\bigg\rangle
+\asontwopimu\int\limits_0^1 dz\bigg[\mathcal P'_{qq}(z) 
-\LMu \hat P^{(0)}_{qq,R}(z)\bigg]
\times
\\
&
\left\langle
\frac{\FLVijfin{\fa}{\fb}(z\cdot1,2)+\FLVijfin{\fa}{\fb}(1,z\cdot2)}{z}
\right\rangle,
\end{split}
\label{eq:vv}
\ee
where $\gamma_q$ is defined after Eq.~\eqref{eq:delta} and 
$\FLVVfin$, $\FLVsqfin$ and $\FLVfin$ are
defined in Appendix~\ref{app:virtual}.

\subsection{NNLO: quark-gluon channels}
\label{sec:dy_nnlo_qg}

In this section, we describe the NNLO contributions to the $qg$ channel, see Eq.~\eqref{eq:dy_channels}.
Similar results hold for the $gq$ channel. In principle, this channel could
be treated in the same fashion as the quark channels discussed in the 
previous section. However, its singularity structure is much simpler,
and so we need to consider a smaller number of limits.
Indeed, no double-soft singularities are present in this case. Because of
this, we find it convenient not to order the energies of partons 4 and 5.
We write
\be
\int \dg4\dg5\FLMij{q}{g}(1,2,4,5) \equiv 
\big\langle \FLMij{q}{g}(1,2,4,5) \big\rangle, 
\ee
and parametrize $E_{4,5} = x_{1,2} \Em$. However, the structure
of the collinear singularities is similar to that discussed in 
Sec.~\ref{sec:dy_nnlo_quarks}, so we use the same angular parametrization
and partitioning as defined there. 

There is another important difference compared to the
$ q \bar q$ channel  discussed in Sec.~\ref{sec:dy_nnlo_quarks},
namely that in the $qg$ channel spin correlations
appear in the collinear emissions off the incoming gluon. 
We postpone their discussion to Sec.~\ref{sec:h_spincorr}, where we
consider the most general  case of spin correlations. Apart from this,
the structure of the result is very similar to the one discussed 
previously, so we limit ourselves to reporting the relevant formulas.

We write
\be
\dhs^{\NNLO}_{\fa g} = \dhs^{\NNLO}_{V+2,\fa g} + 
\dhs^{\NNLO}_{V+1,\fa g}
+\dhs^{\NNLO}_{V,\fa g},
\ee
and
\bes
&\dhs^{\NNLO}_{V+2,\fa g} = \dhs^{\NNLO}_{1245,\fa g},
\\
&\dhs^{\NNLO}_{V+1,\fa g} = \dhs^{\NNLO}_{124,\fa g}
                   +\dhs^{\NNLO}_{{\rm virt}_{124},\fa g},
\\
&\dhs^{\NNLO}_{V,\fa g} ~~= \dhs^{\NNLO}_{(z,\zb),\fa g}
                   +\dhs^{\NNLO}_{(1,z),\fa g}
                   +\dhs^{\NNLO}_{{\rm virt}_{12},\fa g}.
\end{split}
\ee
We consider the case with $a\ne 0$, and discuss each term separately. 
Note that in this channel there are no terms proportional 
to $\FLM(1,2)$ and to $\FLM(z\cdot 1,2)$
since the
process $qg \to V$ at leading order is impossible. 
For the other terms, we obtain the following results.

\begin{enumerate}
\item \underline{\bf Tree-level terms with NNLO kinematics:}
\begin{align}
&  \d \hat \sigma^{\rm NNLO}_{1245,\fa g} =
\sum_{(ij)\in dc}\bigg\langle
\bigg[(\I- C_{5j})(\I-C_{4i})\bigg]\big[\I-S_5\big]
\times
\nonumber
\\
&\quad\quad\quad\quad\quad\quad
\times\dg4\dg5 w^{4i,5j}\FLMij{\fa}{g}(1,2,4,5)
\bigg\rangle
\nonumber
\\
&\quad\quad
+\sum_{i\in tc} 
\bigg\langle
\bigg[
\theta^{(\mathpzc{a})} \big[\I-\CC_i\big]\big[\I-C_{5i}\big] + 
\theta^{(\mathpzc{b})} \big[\I-\CC_i\big]\big[\I-C_{45}\big] 
\label{eq:qghard}
\\
&\quad\quad\quad\quad~~
 + \theta^{(\mathpzc{c})} \big[\I-\CC_i\big]\big[\I-C_{4i}\big]+ 
\theta^{(\mathpzc{d})} \big[\I-\CC_i\big]\big[\I-C_{45}\big]
\bigg]
\nonumber
\\
&\quad\quad\quad\quad~~
\times \big[\I-S_5\big]
\dg4 \dg5 w^{4i,5i}\FLMij{\fa}{g}(1,2,4,5)
\bigg\rangle.\nonumber
\end{align}
The discussion of the individual terms is identical to that 
of  the quark channels, cf. Sec.~\ref{sec:dy_nnlo_quarks_nnlo}, with the only difference that now energies
of final state partons are parametrized differently.

\item \underline{\bf Tree-level terms with NLO kinematics:}
\begin{align}
&\d\hat\sigma^{\rm NNLO}_{124,\fa g} 
= 
\asontwopimu
\int\limits_0^1\d z 
\Bigg\{
\hat P^{(0)}_{qq,R}(z)
\Bigg\langle
\ln\frac{\rho_{41}}{4}\ONLO
\bigg[
\frac{\tilde w_{5||1}^{41,51} 
\FLMij{\fa}{g}(z\cdot1,2,4)
}{z}\bigg]\Bigg\rangle
\nonumber
\\
&
+
\sum_{x\ne0}
\hat P^{(0)}_{qg,R}(z)
\Bigg\langle
\ln\frac{\rho_{42}}{4}\ONLO
\bigg[
\frac{\tilde w_{5||2}^{42,52} 
\FLMij{\fa}{\fx}(1,z\cdot2,4)
}{z}\bigg]\Bigg\rangle
\nonumber
\\
&
+
\hat P^{(0)}_{gg,R}(z)
\Bigg\langle
\ln\frac{\rho_{42}}{4}\ONLO
\bigg[
\frac{\tilde w_{5||2}^{42,52} 
\FLMij{\fa}{g}(1,z\cdot2,4)
}{z}\bigg]\Bigg\rangle
\nonumber
\\
&
+\bigg[\mathcal P'_{qq}(z) - \hat P^{(0)}_{qq,R}(z)\LMu\bigg]
\left\langle
\ONLO\left[\frac{\FLMij{\fa}{g}(z\cdot1,2,4)
}{z}\right]
\right\rangle 
\label{eq:FLM124qg}
\\
&
+\sum_{x\ne0}\bigg[\mathcal P'_{qg}(z) - \hat P^{(0)}_{qg,R}(z)\LMu\bigg]
\left\langle
\ONLO\left[\frac{\FLMij{\fa}{\fx}(1,z\cdot2,4)
}{z}\right]
\right\rangle
\nonumber
\\
&
+
\bigg[\mathcal P'_{gg}(z) - \hat P^{(0)}_{gg,R}(z)\LMu\bigg]
\left\langle
\ONLO\left[\frac{\FLMij{\fa}{g}(1,z\cdot2,4)
}{z}\right]
\right\rangle 
\Bigg\}
\nonumber
\\
&
+\asontwopimu \bigg\langle
\ONLO\bigg[
\Delta^{(qg)} \cdot \FLMij{\fa}{g}(1,2,4)
\bigg]
\bigg\rangle,
\nonumber
\end{align}
which is analogous to Eq.~\eqref{eq:FLM124}
for the quark channels.
The splitting functions
in Eq.~\eqref{eq:FLM124qg}
are defined in Appendix~\ref{app:split}, and
$\Delta^{(qg)}$ is given by
\bes
&\Delta^{(qg)} = 
\Cf\bigg[ \ln\frac{2 E_4}{\sqrt{s}}
\bigg(  \ln\frac{2 E_4}{\sqrt{s}} - 2 \ln\eta_{41} - 4 \ln2\bigg)
+\lp\frac{3}{2}-2\ln\frac{2 E_4}{\sqrt{s}}\rp\times
\\
&\quad\quad
\lp \tilde w^{41,51}_{4||5} \ln\frac{\eta_{42}}{\eta_{41}}
+ \tilde w^{42,52}_{4||5} \ln\frac{\eta_{41}}{\eta_{42}}\rp
+ \frac{13}{2} + 3\ln2 - \pi^2 + 3\ln\eta_{41} 
\\
&\quad\quad
+ 2\Li_2(\eta_{42}) - \frac{3}{2}\LMu\bigg]
+\beta_0\bigg[\frac{1}{2}\lp \ln\frac{2 E_4}{\sqrt{s}} + \ln\eta_{42}\rp
-\LMu\bigg]
\\
&\quad\quad
+\Ca\bigg[\frac{\pi^2}{3} - \frac{3}{4}\ln\eta_{42} + 
\lp\frac{3}{2}-\ln\frac{2 E_4}{\sqrt{s}}\rp
\ln\frac{\eta_{42}}{\eta_{41}} 
-\frac{3}{4}\ln\frac{2 E_4}{\sqrt{s}}
\\
&\quad\quad
-\Li_2(\eta_{42})+\Li_2(\eta_{41})\bigg].
\end{split}
\label{eq:deltaqg}
\ee 
\item
\underline{{\bf 
Tree-level terms with LO kinematics involving} $\FLM(z\cdot1,\zb\cdot2)$:}
\bes
\d\hat\sigma^{\rm NNLO}_{(z,\zb),\fa g} &
= 
\lp\asontwopimu\rp^2\sum_x\int\limits_0^1\d z~\d\zb
\left[
\mathcal P'_{qq}(z)
-\LMu \hat P^{(0)}_{qq,R}(z)
\right]
\\
&\times
 \left\langle\frac{\FLMij{\fa}{\fx}(z\cdot1,\zb\cdot2)}{z\zb}\right\rangle
\left[
\mathcal P'_{qg}(\zb)
-\LMu \hat P^{(0)}_{qg,R}(\zb)
\right],
\end{split}
\ee
which is analogous to Eq.~\eqref{eq:FLMzzb}.

\item
\underline{{\bf Terms with LO kinematics involving}
$\FLM(z\cdot1,2)$ {\bf and} $\FLM(1,z\cdot2)$:}
\bes
\d\hat\sigma^{\rm NNLO}_{(1,z),\fa g} 
=
\lp\asontwopimu\rp^2 \sum_x\int\limits_0^1\d z
~\mathcal T_{qg}(z)
\left\langle\frac{\FLMij{\fa}{\fx}(1,z\cdot2)}{z}\right\rangle,
\end{split}
\ee
analogous to Eq.~\eqref{eq:FLMz}.
The function $\mathcal T_{qg}$ can be found in the ancillary file.

\item \underline{{\bf Terms involving virtual corrections with NLO
kinematics:}}
\bes
\d\hat\sigma^{\rm NNLO}_{{\rm virt}_{124},\fa g} = 
\big\langle \ONLO \FLVijfin{\fa}{g}(1,2,4)\big\rangle,
\end{split}
\ee
analogous to Eq.~\eqref{eq:rv}. The finite remainder $\FLVijfin{\fa}{g}(1,2,4)$ is
defined in Appendix~\ref{app:realvirtual}.

\item \underline{{\bf Terms involving virtual corrections with LO
kinematics:}}
\bes
\d\hat\sigma^{\rm NNLO}_{{\rm virt}_{12},\fa g} = 
\asontwopimu\sum_x\int\limits_0^1
\bigg[\mathcal P'_{qg}(z) -\LMu \hat P^{(0)}_{qg,R}(z)\bigg]
\left\langle
\frac{\FLVijfin{\fa}{\fx}(1,z\cdot2)}{z}
\right\rangle,
\end{split}
\ee
analogous to Eq.~\eqref{eq:vv}.
The finite remainder $F_{LV}^{\rm fin}$ is defined in 
Appendix~\ref{app:virtual}.

\end{enumerate}
Results for the $gq$ channel can be  obtained from the above
formulas in a straightforward manner, by replacing  labels 
$1\leftrightarrow 2$.

\subsection{NNLO: gluon-gluon channel}
\label{sec:dy_nnlo_gg}

In this section, we describe the $gg$ channel, see Eq.~\eqref{eq:dy_channels}.
This case is particularly simple, since no soft or triple-collinear singularities are
present. As the result, only double-collinear configurations need to be considered.

As a consequence, it is not necessary to partition the phase space in any way
and the singularity structure can be dealt with as in  NLO computations described earlier. 
We parametrize the energies and angles of the emitted partons as
\be \label{eq:param_dy_gg}
E_{4,5}=x_{1,2}\Em,~~~~ \eta_{41} = x_3,~~~~ \eta_{51} = x_4.
\ee 
Singularities only appear when $x_{3,4}$ are equal to either zero or {\it one}. 
Because of the simplicity of the singularity structure, we treat the two
cases at once. To deal with them, we use a similar but simpler strategy
to the one discussed in the preceding sections.  
We write
\bes
& \FLMij{g}{g}(1,2,4,5) = 
\\
&\quad
\big[(\I-C_{41}-C_{42}) + C_{41} + C_{42}\big]
\big[(\I-C_{51}-C_{52}) + C_{51} + C_{52}\big]
\FLMij{g}{g}(1,2,4,5)=
\\
&\quad
\big[\I - C_{41} - C_{42} - C_{51} - C_{52} + C_{42} C_{51} + 
C_{41} C_{52}\big]\FLMij{g}{g}(1,2,4,5)
\\
&
\quad+
\big[C_{41}+C_{42}+C_{51}+C_{52} - C_{42} C_{51} - C_{41} C_{52}\big]
\FLMij{g}{g}(1,2,4,5),
\end{split}
\label{eq:Iggch}
\ee
where we used the fact that $C_{41} C_{51} = C_{42} C_{52} \to 0$ when it 
acts on the matrix element. The first term on the right hand side in 
Eq.~\eqref{eq:Iggch}
is
free of singularities and corresponds to what we called 
$\d\hat\sigma^{\rm NNLO}_{1245}$ in the previous sections, while
the second term contains the subtraction counterterms.
We combine Eq.~\eqref{eq:Iggch}    with contributions from the  PDFs renormalization and write 
\be
\d\hat\sigma^{\rm NNLO}_{gg} = 
\d\hat\sigma^{\rm NNLO}_{V+2,gg}
+\d\hat\sigma^{\rm NNLO}_{V+1,gg}
+\d\hat\sigma^{\rm NNLO}_{V,gg},
\ee
where
\be
\begin{split}
&  \d\hat\sigma^{\rm NNLO}_{V+2,gg} = \d\hat\sigma^{\rm NNLO}_{1245,gg}, \\
&  \d\hat\sigma^{\rm NNLO}_{V+1,gg} = \d\hat\sigma^{\rm NNLO}_{124,gg}, \\
&  \d\hat\sigma^{\rm NNLO}_{V,gg} ~= \d\hat\sigma^{\rm NNLO}_{(z,\bar{z}),gg}.
  \end{split}
\ee
At variance to the  cases discussed above, virtual corrections  do not contribute in this channel.
The individual terms read as follows.
\begin{enumerate}
\item \underline{\bf Tree-level terms with NNLO kinematics:}
\bes
  \d \hat \sigma^{\rm NNLO}_{1245,gg} = &
\bigg \langle \big[\I - C_{41} - C_{42} - C_{51} - C_{52} + C_{42} C_{51} + 
C_{41} C_{52}\big]  \\
& \times \dg4 \dg5 \FLMij{g}{g}(1,2,4,5) \bigg\rangle,
\end{split}
\end{equation}
where again the construction of each term follows the discussion in
Sec.~\ref{sec:dy_nnlo_quarks_nnlo}, but with the parametrization shown in
Eq.~\eqref{eq:param_dy_gg}.

\item \underline{\bf Tree-level terms with NLO kinematics:}
\bes
&\d\hat\sigma^{\rm NNLO}_{124,gg} 
= 
\asontwopimu
\sum_{x\ne0}\int\limits_0^1\d z 
\left[\mathcal P'_{qg}(z) - \hat P^{(0)}_{qg,R}(z)\LMu\right]
\\
&\times
\left\langle
\ONLO\left[\frac{\FLMij{\fx}{g}(z\cdot1,2,4)
+\FLMij{g}{\fx}(1,z\cdot2,4)}{z}\right]
\right\rangle 
.
\end{split}
\label{eq:FLM124gg}
\ee
\item
\underline{{\bf 
Tree-level terms with LO kinematics involving} $\FLM(z\cdot1,\zb\cdot2)$:}
\bes
\d\hat\sigma^{\rm NNLO}_{(z,\zb),gg} &
= 
\lp\asontwopimu\rp^2\sum_{x,y}\int\limits_0^1\d z~\d\zb
\left[
\mathcal P'_{qg}(z)
-\LMu \hat P^{(0)}_{qg}(z)
\right]
\\
&\times
 \left\langle\frac{\FLMij{\fx}{\fy}(z\cdot1,\zb\cdot2)}{z\zb}\right\rangle
\left[
\mathcal P'_{qg}(\zb)
-\LMu \hat P^{(0)}_{qg}(\zb)
\right].
\end{split}
\ee

\end{enumerate}

%%%

\section{Gluon-initiated color-singlet production}

\label{sec:h}
In this section, we consider the production of a color-singlet
final state $H$
in gluon fusion through  NNLO QCD.  We refer to this process
as ``Higgs production'', although, similarly to the ``Drell-Yan process''
discussed in the previous section,
these results are applicable to the production of any color-singlet final state 
which proceeds through gluon fusion at LO.
The procedure of extracting the infrared divergences is
identical to what has already been
discussed in the case of the $q \bar q$ annihilation and we do not repeat it.
However, in gluon-initiated processes, initial state radiation leads to spin correlations that we did not discuss up to now.
In the next section we show how to deal with this complication. 

\subsection{Spin correlations}
\label{sec:h_spincorr}

We have discussed spin correlations relevant to the computation of
NNLO QCD corrections to the $q \bar q\to V $ process in
Ref.~\cite{Caola:2017dug}. In that case, the spin correlations
appeared because of the splitting of a virtual gluon to two final
state partons, $g^* \to f_{4,5}$.  For the $gg\to H$ process, the situation
is different in that spin correlations also appear in the initial
state radiation, including its triple-collinear limit. In this section,
we discuss this case. 

We will begin with  the discussion of NLO QCD corrections to $gg \to H$. The computation proceeds exactly as for
 the $q \bar q\to V$ process, cf. Sec.~\ref{sec:dy_lonlo}. However, when the collinear operator acts on the
matrix element squared, we find
\be
\langle C_{41} F_{{\rm LM},gg}(1,2,4) \rangle =-
\left \langle \frac{g_{s,b}^2}{p_1 \cdot p_4 } \; P_{gg,\mu \nu} \left ( z^{-1} \right )
F_{{\rm LM}, gg}^{\mu \nu} \left( z \cdot 1, 2 \right ) \right \rangle,
\ee
where the splitting function reads
\be
P_{gg}^{\mu \nu}(z) = 2 C_A \left [ 
\left ( \frac{1-z}{z} + \frac{z}{1-z} \right ) 
\left [ -g_{\perp,d-2}^{\mu \nu}  \right ]
  - 2(1-\ep) z (1-z) \frac{k_\perp^\mu k_\perp^\nu}{k_\perp^2} \right ].
\ee
The transverse momentum $k_\perp$  is  defined using the Sudakov decomposition
\be
p_4 = \left ( 1 - z \right ) p_1 + \beta p_2 + k_\perp, 
\label{eq4.3}
\ee
where  $k_\perp \cdot p_1 = k_\perp \cdot p_2 = 0$.
The metric tensor of the transverse space $g_{\perp,d-2}$ satisfies 
$g_{\perp,d-2}^{\mu \nu}\; p_{1,\nu} =  g_{\perp,d-2}^{\mu \nu}\; p_{2,\nu} = 0$ and $g_{\perp,d-2}^{\mu \nu} k_{\perp,\nu}
= k_{\perp}^{\mu}$.

We write the four-momenta of the QCD partons as for  the $q \bar q\to V$ process.
We 
introduce $d-$dimensional vectors
$t^\mu = (1,\vec 0)$ and $e_3^\mu = (0,0,0,1,\vec 0)$,
and an additional vector $a^\mu$ that is orthogonal to both $t$ and $e_3$ and is normalized  $a^2 = -1$.
We write the four-momenta in terms of $t$, $e_3$ and $a$ and obtain
\be
p_{1,2} = E_1 ( t \pm e_3),\;\;\;\; p_4 = E_4 (t + \cos \theta_{41} e + \sin \theta_{41} a).
\ee
  By comparing the two parametrizations of the vector $p_4$, we find 
\be
\frac{k_\perp^\mu k_\perp^\nu}{k_\perp^2} = -a^\mu a^\nu.
\ee
Since the transverse components of the gluon four-momentum
decouple from the hard matrix element in the collinear limit, the vector $a^\mu$ only appears in the splitting
function. We can then integrate the splitting function over
the directions of  the vector $a^\mu$ using
\be
\int {\rm d} \Omega_{d-2}^{(a)}\; a^\mu a^\nu = -\frac{g_{\perp,d-2}^{\mu \nu} }{d-2} \Omega_{d-2}.
\ee
We find
\be
\int \frac{{\rm d} \Omega_{d-2}^{(a)}}{\Omega_{d-2}}\;
P_{\mu \nu, gg}(z^{-1})\;  F_{\rm LM}^{\mu \nu}( z \cdot 1, 2)  = \langle P_{gg}(z^{-1}) \rangle \; F_{\rm LM}(z \cdot 1,2), 
\ee
where
\be
\langle P_{gg}(z) \rangle = 2 C_A \left ( \frac{1-z}{z} + \frac{z}{1-z} + z(1-z) \right ) 
\ee
is the spin-averaged splitting function. It follows that averaging over the directions of the
transverse components of the gluon momentum naturally appears in our construction at NLO; 
as the consequence, the rest of the NLO QCD calculation is identical to the $q \bar q$ case. 

Before discussing spin correlations in the computation of NNLO QCD corrections,
we note that, in the  particular case of Higgs boson production,
 spin correlations are actually not needed. Indeed, the spin-correlated matrix element
for $gg \to H$ is proportional to $g_{\perp,d-2}^{\mu \nu}$. As the result, the spin-averaged
splitting function $P_{gg,\mu \nu} g_{\perp,d-2}^{\mu \nu}$ naturally appears in the calculation.
We emphasize, however, that this feature is particular to the process $gg \to H$, so that 
understanding spin correlations is necessary in a more general context.

We then consider the generic NNLO case. Here, the situation is more complex since we have to consider the momenta
of the two gluons becoming collinear to the direction of the incoming partons.  In the
double-collinear partitions the situation is identical to the NLO case
since the averaging over the transverse spaces of the 
two gluons is  performed independently. The triple-collinear partitions require
some discussion.
We consider the case when collinear singularities arise because of the emissions of gluons
$g_{4,5}$ off the gluon $g_1$.
We parametrize the four-momenta of the final-state gluons as  \cite{Caola:2017dug}
\be
\begin{split} 
& p_4 = E_4 \left(  t + \cos \theta_4 e_3 + \sin \theta_4 a \right ), \\
  & p_5 = E_5 \left ( t + \cos \theta_5 e_3 + \sin \theta_5 ( \cos \varphi_{45} \; a
  + \sin \varphi_{45} \; b ) \right ),
  \label{eq:stp}
\end{split} 
\ee
where the vectors $t,a,e_3$ have already been defined in the discussion after Eq.(\ref{eq4.3})  and the vector $b$
satisfies $t \cdot b =  e_3 \cdot b = a \cdot e_3 = 0$, as well as $b^2 = -1$. 

We begin with the double-collinear limits that develop spin
correlations. There are three possibilities: $g_4$ is collinear to
$g_5$, $g_5$ is collinear to $g_1$ and $g_4$ is collinear to $g_1$.
The first case is identical to the $q \bar q\to V$ process; it was
discussed in Ref.~\cite{Caola:2017dug} and we do not repeat this
discussion here.  The second  case, $p_5 || p_1$, relevant for
sector $(a)$, is discussed  below. After that we comment on the third case, relevant for sector $(c)$.

Starting from  the angular part of the phase space for sector $(a)$ and 
considering the limit
$x_4 \to 0$, corresponding to $\theta_{51} \to 0$, we find
\bes
\lim_{x_4\to 0} \d\Omega_{45}^{(\mathpzc{a})}=
\left[\frac{1}{8\pi^2}\frac{(4\pi)^\ep}{\Gamma(1-\ep)}\right]^2
\left[\frac{\Gamma^2(1-\ep)}{\Gamma(1-2\ep)}\right]
\left[\frac{\d\Omega_{d-2}^{(b)}}{\Omega_{d-2}}\frac{\d\Omega_{d-3}^{(a)}}{\Omega_{d-3}}\right]
\frac{\d x_3}{x_3^{1+2\ep}}
\frac{\d x_4}{x_4^{1+\ep}}
\times
\\
\frac{\d\lambda}{\pi\left[\lambda(1-\lambda)\right]^{1/2+\ep}}
\left[128(1-x_3)\right]^{-\ep}
2 x_3^2 x_4,
\end{split}
\ee
see Ref.~\cite{Caola:2017dug}.
Also, in this limit
\be
\sin\varphi = \sqrt{4\lambda(1-\lambda)}, ~~~~~\cos\varphi = -1+2\lambda,~~~~~\lambda\in[0,1].
\ee
This follows immediately from the definition of $\sin\varphi$ and by 
inverting the definition of $\lambda$ in terms of $\cos\varphi$. We then rewrite
\bes
\lim_{x_4\to 0} \d\Omega_{45}^{(\mathpzc{a})}=
\left[\frac{1}{8\pi^2}\frac{(4\pi)^\ep}{\Gamma(1-\ep)}\right]^2
\left[\frac{\d\Omega_{d-2}^{(b)}}{\Omega_{d-2}}\frac{\d\Omega_{d-3}^{(a)}}{\Omega_{d-3}}\right]
\frac{2\;\d x_3}{\left[4x_3(1-x_3)\right]^\ep}
\frac{2\;\d (x_3 x_4/2)}{\left[4 (x_3 x_4)/2\right]^{\ep}}
\d\widetilde\Lambda,
\end{split}
\ee
with
\be
\d\widetilde\Lambda=\left[16^{-\ep}\frac{\Gamma^2(1-\ep)}{\Gamma(1-2\ep)}
\frac{\d\lambda}{\pi\left[\lambda(1-\lambda)\right]^{1/2+\ep}}
\right],~~~~~~\int \d\widetilde\Lambda = 1.
\ee
We also have
\be
\begin{gathered}
\int\d\widetilde\Lambda \cos\varphi\sin\varphi = 0,
\\
\int\d\widetilde\Lambda \cos^2\varphi = \frac{1}{2(1-\ep)},~~~~~~
\int\d\widetilde\Lambda \sin^2\varphi = 1-\frac{1}{2(1-\ep)}.
\end{gathered}
\ee
These identities imply that in the $x_4\to 0$ limit
\begin{align}
\left\langle \kappa_{5,\perp}^\mu\kappa_{5,\perp}^\nu \right\rangle &= 
\int
\frac{\d\Omega_{d-3}^{(b)}}{\Omega_{d-3}}
\d\widetilde\Lambda 
\left[a^\mu a^\nu \cos^2\varphi_{45}  
+
b^\mu b^\nu \sin^2\varphi_{45}
+ ( b^\mu a^\nu + a^\mu b^\nu ) \cos \varphi_{45} \sin \varphi_{45} 
\right]
\nonumber\\
&
= \frac{a^{\mu}a^{\nu}}{2(1-\ep)}
+\frac{1-2\ep}{2(1-\ep)}\int \frac{\d\Omega_{d-3}^{(b)}}{\Omega_{d-3}} b^{\mu}b^{\nu} 
\nonumber\\
&
=
\frac{a^{\mu}a^{\nu}}{2(1-\ep)}
+\frac{1-2\ep}{2(1-\ep)}\frac{\left[-g_{\perp,d-3}^{\mu\nu}\right]}{d-3}
\\
&
=\frac{1}{2(1-\ep)}\bigg[a^{\mu}a^{\nu} - g_{\perp,d-3}^{\mu\nu}\bigg]=
\frac{\left[-g_{\perp,d-2}^{\mu\nu}\right]}{2(1-\ep)}.
\nonumber
\end{align}
Hence, in case of double-collinear limits with respect to the incoming partons,
integration over the transverse directions of the collinear gluons always leads to
spin-averaged splitting functions. This implies that subsequent computational
steps are conceptually identical to those of the  $q \bar q\to V$ process.
Finally, we note that the  above discussion can be repeated verbatim also in case $p_4 || p_1$, relevant  for sector $(c)$, 
if instead of Eq.~\eqref{eq:stp} we use
\be
\begin{split} 
& p_4 = E_4 \left(  t + \cos \theta_4 e_3 + \sin \theta_4 ( \cos \varphi_{45} \; a
  + \sin \varphi_{45} \; b ) \right ), \\
  & p_5 = E_5 \left ( t + \cos \theta_5 e_3 + \sin \theta_5 a  \right ).
\end{split} 
\ee

It remains to discuss the triple-collinear limit that corresponds to  the splitting $g_{1}  \to
g_{4} + g_{5} + g^*$. This splitting is described by the $P_{ggg}^{\mu \nu}$
splitting function  that 
 contains spin correlations, see e.g. Ref.~\cite{Catani:1999ss}. 
This splitting function is a symmetric rank-two tensor constructed from    $g_{\perp,{d-2}}^{\mu \nu}$, and the
vectors $k_{4(5),\perp}^{\mu}$. 
These vectors read, c.f. Eq.~\eqref{eq:stp},
\be
k_{4,\perp}^\mu = E_4 \sin \theta_4 a^\mu,\;\;\; k_{5,\perp}^\mu = E_5 \sin \theta_5
\left ( a^\mu \cos \varphi_{45} + b^\mu \sin \varphi_{45} \right ) .
\ee
In the triple-collinear limit, we need to integrate over the directions of the vectors $a$ and $b$.
Note that the integration over the angle $\varphi_{45}$ is non-trivial since 
$2 (p_4 \cdot p_5) = s_{45}$ depends on it.
To describe the integration of  different tensor structures
over the directions of $a$ and $b$, we introduce the notation
\be
\int \frac{{\rm d} \Omega_{d-2}^{(a)}}{\Omega_{d-2}} 
\frac{{\rm d} \Omega_{d-3}^{(b)}}{\Omega_{d-3}} X
= \langle X \rangle_{a,b},\;\;\;
\int \frac{{\rm d} \Omega_{d-2}^{(a)} }{\Omega_{d-2}} X = \langle X \rangle_{a},\;\;\;
\int \frac{{\rm d} \Omega_{d-3}^{(b)} }{\Omega_{d-3}} X = \langle X \rangle_{b}.
\ee
We find the following results for the four  tensor structures that
contribute to $P_{ggg}^{\mu \nu}$
\be
\begin{split}
& \langle g_{\perp,d-2}^{\mu \nu} \rangle = g_{\perp,d-2}^{\mu \nu},\;\;\;\;
 \langle   k_{4 \perp}^{\mu} k_{4 \perp}^{\nu} \rangle_{a,b} = k_{4,\perp}^2 \frac{g_{\perp,d-2}^{\mu \nu}}{d-2},
 \;\;\;\; \langle  k_{5 \perp}^{\mu} k_{5 \perp}^{\nu} \rangle_{a,b} = k_{5,\perp}^2 \frac{g_{\perp,d-2}^{\mu \nu}}{d-2},
 \\
 &   \langle   k_{4 \perp}^{\mu} k_{5 \perp}^{\nu} \rangle_{a,b}
= E_4 \sin \theta_4 E_5 \sin \theta_5
\langle a^\mu \left ( a^\nu \cos \varphi_{45} + b^\mu \sin \varphi_{45} \right ) \rangle_{a,b}
=  k_{4,\perp} \cdot k_{5,\perp} \;\frac{g_{\perp,d-2}^{\mu \nu} }{d-2} .
\end{split} \label{eq:418}
\ee
We write the spin-correlated splitting function as 
\be
P_{ggg}^{\mu \nu} = A_1 g_{\perp,d-2}^{\mu \nu}
+ A_2 (k_{4,\perp}^\mu k_{4,\perp}^\nu + k_{5,\perp}^\mu k_{5,\perp}^\nu )
+ A_3 (k_{4,\perp}^\mu k_{5,\perp}^\nu + k_{4,\perp}^\nu k_{5,\perp}^\mu ), 
\ee
and observe that Eq.~\eqref{eq:418} leads to 
\be
\langle P_{ggg}^{\mu \nu} \rangle_{a,b}  =
g_{\perp,d-2}^{\mu \nu} \left ( A_1 
+ \frac{A_2}{d-2}  ( k_{4,\perp}^2 + k_{5,\perp}^2 )
+ \frac{A_3}{d-2} ( 2 k_{4,\perp} k_{5,\perp} )
\right ).
\ee
The same result is obtained upon replacing
the spin-correlated splitting function with its spin-averaged version
\be
P_{ggg}^{\mu \nu} \to \frac{P_{ggg,\alpha \beta} g_{\perp,d-2}^{\alpha \beta}}{d-2}\; g_{\perp,d-2}^{\mu \nu}. 
 \ee
 Once this is done, the triple-collinear splittings in the $gg\to H$ 
process  can be treated in exactly the same way as in the Drell-Yan 
$q \bar q\to V$ case.

\subsection{Definition of partonic channels}
After discussinng spin correlations, we proceed with setting up the
NNLO QCD calculation for color-singlet production in gluon fusion. 
Starting from Eq.~\eqref{eq:fid_xsec}, we find it convenient to 
write the cross section for the generic process
\be
p p \to H + X
\ee
as
\bes
\d\sigma_f^\H &
= \int\d x_1 \d x_2~ g(x_1) g(x_2) \d\hat\sigma^\H_{gg} \\
&
+
\int\d x_1 \d x_2\sum\limits_{\substack{a\in [-\nf,\nf]\\a\ne0}} 
\left[
f_a(x_1) g(x_2) \d\hat\sigma^\H_{qg}+
g(x_1)f_a(x_2)  \d\hat\sigma^\H_{gq}\right] \\
&
+\int\d x_1 \d x_2\sum\limits_{\substack{a\in [-\nf,\nf]\\a\ne0}} 
f_a(x_1) f_{-a}(x_2) \d\hat\sigma^\H_{q\qb}
\\
&
+
\int\d x_1 \d x_2
\sum\limits_{\substack{a,b\in [-\nf,\nf]\\a,b\ne0,~a\ne -b}} 
f_a(x_1) f_b(x_2) \d\hat\sigma^\H_{q_i q_j}
.
\end{split}
\ee
The first term is the $gg$ channel which is the only partonic channel
contributing at LO.  The terms on the second and third line correspond
to the quark-gluon channels and quark-antiquark channel respectively and enter
at NLO. The $q_i q_j$ channel, where $q_i$ and $q_j$ can be both identical
or different (anti)quarks first appears at NNLO.  We will discuss each of
these channels separately in the following subsections. For simplicity,
we will omit the ``H'' superscript from now on. We express our results in 
terms of fully renormalized amplitudes for the $p p \to H+X$ process, where
$H$ is a generic color-singlet state. In the case of Higgs production in the 
$m_t\to \infty$ approximation, this implies that our results include both the
divergent \emph{and} the finite renormalization of the $Hgg$
Wilson coefficient, see App.~\ref{app:virtual} for more details. 

\subsection{LO and NLO}
\label{sec:h_lonlo}
At leading order, only the $gg$ channel contributes. We write 
\be
2s\cdot \d\hat\sigma^{\rm LO}_{gg} = 
\left\langle\FLMij{g}{g}(1,2)\right\rangle.
\ee
NLO corrections have a similar structure to those discussed in Sec.~\ref{sec:dy_lonlo}.
The result for the $gg$ channel reads
\bes
&
2s\cdot\d\hat\sigma^{\rm NLO}_{gg} = 
\bigg\langle \FLVijfin{g}{g}(1,2) 
+ \asontwopimu \left[\frac{2\pi^2}{3}\Ca
-2\gamma_g \LMu \right]\FLMij{g}{g}(1,2) \bigg\rangle
\\
&
+\asontwopimu\int\limits_0^1\d z 
\left[ 
\mathcal P'_{gg}(z) - \hat P^{(0)}_{gg,R} \ln \left(\frac{\mu^2}{s}\right) 
\right]
\left\langle 
\frac{\FLMij{g}{g}(z\cdot1,2)+\FLMij{g}{g}(1,z\cdot 2)}{z}
\right\rangle
\\
&
+ \left\langle \ONLO \FLMij{g}{g}(1,2,4)\right\rangle,
\end{split}
\label{eq:h_nlo_qqb}
\ee
with $\gamma_g=\beta_0=11\Ca/6-\nf/3$, and 
$\FLVijfin{g}{g}$ and the various splitting functions
defined in Appendix~\ref{app:virtual} and Appendix~\ref{app:split},
respectively. 

NLO corrections to the $gq$ and $qg$ channel read
\bes
2s\cdot\d\hat\sigma^{\rm NLO}_{gq} &= 
\asontwopimu\int\limits_0^1\d z 
\left\langle \frac{\FLMij{g}{g}(1,z\cdot2)}{z}\right\rangle
\left[
 \mathcal P'_{gq}(z)
-\hat P^{(0)}_{gq,R} \LMu
\right]
\\
&+ \left\langle \ONLO \FLMij{g}{q}(1,2,4)\right\rangle,
\end{split}
\ee
and 
\bes
2s\cdot\d\hat\sigma^{\rm NLO}_{qg} &= 
\asontwopimu\int\limits_0^1\d z 
\left[
 \mathcal P'_{gq}(z)
-\hat P^{(0)}_{gq,R} \LMu
\right]
\left\langle \frac{\FLMij{g}{g}(z\cdot1,2)}{z}\right\rangle
\\
&+ \left\langle \ONLO \FLMij{q}{g}(1,2,4)\right\rangle,
\end{split}
\ee
respectively. 
Finally, the $q\qb$ channel starts contributing at NLO but it is finite 
finite at this order and can simply be written as 
\be
2s\cdot\d\hat\sigma^{\rm NLO}_{q\qb} = 
\big\langle \ONLO\FLMij{q}{\qb}(1,2,4)\big\rangle=
\big\langle \FLMij{q}{\qb}(1,2,4)\big\rangle.
\ee

\subsection{NNLO: gluon channel}
\label{sec:h_nnlo_gg}
This channel has the same singularity structure as 
the Drell-Yan quark channels,
cf. Sec.~\ref{sec:dy_nnlo_quarks}, and we  use the same phase-space
parametrization and partitioning described there. This means that the
structure of the result is identical to what was discussed in Sec.~\ref{sec:dy_nnlo_quarks},
and we can write it as 
\be
\dhs^{\NNLO}_{gg} = \dhs^{\NNLO}_{H+2,gg}+\dhs^{\NNLO}_{H+1,gg}+
\dhs^{\NNLO}_{H,gg},
\ee
and
\bes
&
\dhs^{\NNLO}_{H+2,gg} = \dhs^{\NNLO}_{1245,gg},\\
&
\dhs^{\NNLO}_{H+1,gg} = \dhs^{\NNLO}_{124,gg} + \dhs^{\NNLO}_{{\rm virt}_{124},gg},\\
&
\dhs^{\NNLO}_{H,gg} ~= 
\dhs^{\NNLO}_{(z,\zb),gg}
+
\dhs^{\NNLO}_{(1,z),gg}
+
\dhs^{\NNLO}_{(z,2),gg}
+
\dhs^{\NNLO}_{(1,2),gg}
+
\dhs^{\NNLO}_{{\rm virt}_{12},gg}.
\end{split}
\ee
We now list all these terms separately. 
\begin{enumerate}
\item \underline{\bf Tree-level terms with NNLO kinematics:}
\begin{align}
&  \d \hat \sigma^{\rm NNLO}_{1245,gg} =
\sum_{(ij)\in dc}\bigg\langle
\bigg[(\I- C_{5j})(\I-C_{4i})\bigg]
\big[\I-\SS\big]\big[\I-S_5\big]
\times
\nonumber
\\
&\quad\quad\quad\quad\quad\quad
\times\dg4\dg5 w^{4i,5j}\FLMij{g}{g}(1,2,4,5)
\bigg\rangle
\nonumber
\\
&\quad\quad
+\sum_{i\in tc} 
\bigg\langle
\bigg[
\theta^{(a)} \big[\I-\CC_i\big]\big[\I-C_{5i}\big] + 
\theta^{(b)} \big[\I-\CC_i\big]\big[\I-C_{45}\big] 
\label{eq:gghard}
\\
&\quad\quad\quad\quad~~
 + \theta^{(c)} \big[\I-\CC_i\big]\big[\I-C_{4i}\big]+ 
\theta^{(d)} \big[\I-\CC_i\big]\big[\I-C_{45}\big]
\bigg]
\nonumber
\\
&\quad\quad\quad\quad~~
\times\ \big[\I-\SS\big]\big[\I-S_5\big]
\dg4 \dg5 w^{4i,5i}\FLMij{g}{g}(1,2,4,5)
\bigg\rangle,\nonumber
\end{align}
which is analogous to Eq.~\eqref{eq:qqhard}.

\item \underline{\bf Tree-level terms with NLO kinematics:}
\begin{align}
&\d\hat\sigma^{\rm NNLO}_{124,gg} 
= 
\asontwopimu
\int\limits_0^1\d z 
\Bigg\{
\hat P^{(0)}_{gg,R}(z)
\Bigg\langle
\ln\frac{\rho_{41}}{4}\ONLO
\bigg[
\frac{\tilde w_{5||1}^{41,51} 
\FLMij{g}{g}(z\cdot1,2,4)
}{z}\bigg]
\nonumber
\\
&
+
\ln\frac{\rho_{42}}{4}\ONLO
\bigg[
\frac{\tilde w_{5||2}^{42,52} 
\FLMij{g}{g}(1,z\cdot2,4)
}{z}\bigg]
\Bigg\rangle
+
2\nf\hat P^{(0)}_{qg,R}(z)
\Bigg\langle
\ln\frac{\rho_{41}}{4}\ONLO\times
\nonumber
\\
&
\bigg[
\frac{\tilde w_{5||1}^{41,51} 
\FLMij{q}{g}(z\cdot1,2,4)
}{z}\bigg]
+
\ln\frac{\rho_{42}}{4}\ONLO
\bigg[
\frac{\tilde w_{5||2}^{42,52} 
\FLMij{g}{q}(1,z\cdot2,4)
}{z}\bigg]
\Bigg\rangle
\nonumber
\\
&
+\left[\mathcal P'_{gg}(z) - \hat P^{(0)}_{gg,R}(z)\LMu\right]
\left\langle
\ONLO\left[\frac{\FLMij{g}{g}(z\cdot1,2,4)
+\FLMij{g}{g}(1,z\cdot2,4)}{z}\right]
\right\rangle 
\nonumber
\\
&
+2\nf\left[\mathcal P'_{qg}(z) - \hat P^{(0)}_{qg,R}(z)\LMu\right]
\times
\label{eq:hFLM124gg}
\\
&
\left\langle
\ONLO\left[\frac{\FLMij{q}{g}(z\cdot1,2,4)
+\FLMij{g}{q}(1,z\cdot2,4)}{z}\right]
\right\rangle
\Bigg\}
\nonumber
\\
&
+\asontwopimu \bigg\langle
\ONLO\bigg[
\Delta_g \cdot \FLMij{g}{g}(1,2,4)
+\Delta^r \cdot \big[r_\mu r_\nu \FLMij{g}{g}^{\mu\nu}(1,2,4)\big]\bigg]
\bigg\rangle,
\nonumber
\end{align}
which is analogous to Eq.~\eqref{eq:FLM124}. 
The splitting functions in Eq.~\eqref{eq:hFLM124gg}
are defined in Appendix~\ref{app:split}, and
$\Delta_g$ is given in Eq.~\eqref{eq:delta} with $C_g = \Ca$ and
$\gamma_g = \mathcal{X}_g = \beta_0$.

\item
\underline{{\bf 
Tree-level terms with LO kinematics involving} $\FLM(z\cdot1,\zb\cdot2)$:}
\bes
\d\hat\sigma^{\rm NNLO}_{(z,\zb),gg} &
= 
\lp\asontwopimu\rp^2\int\limits_0^1\d z~\d\zb
\left[
\mathcal P'_{gg}(z)
-\LMu \hat P^{(0)}_{gg,R}(z)
\right]
\\
&\times
 \left\langle\frac{\FLMij{g}{g}(z\cdot1,\zb\cdot2)}{z\zb}\right\rangle
\left[
\mathcal P'_{gg}(\zb)
-\LMu \hat P^{(0)}_{gg,R}(\zb)
\right].
\end{split}
\ee

\item
\underline{{\bf Tree-level terms with LO kinematics involving}
$\FLM(z\cdot1,2)$ {\bf and} $\FLM(1,z\cdot2)$:}
\bes
\d\hat\sigma^{\rm NNLO}_{(1,z),gg} + 
\d\hat\sigma^{\rm NNLO}_{(z,2),gg} 
=
\lp\asontwopimu\rp^2 \int\limits_0^1\d z
~\mathcal T_{gg}(z)\times
\\
\left\langle\frac{\FLMij{g}{g}(z\cdot1,2)+\FLMij{g}{g}(1,z\cdot2)}{z}\right\rangle.
\end{split}
\ee

\item \underline{{\bf Tree-level terms with LO kinematics involving}
$\FLM(1,2)$:}
\bes
&\d\hat\sigma^{\rm NNLO}_{(1,2),gg} = \biggl\langle\FLM(1,2)\biggr\rangle
\times\lp\asontwopimu\rp^2 \Bigg\{
\Ca^2\bigg[
\frac{739}{81}+\frac{4}{3}\ln2 + \frac{187\pi^2}{54} - 2 \ln^2(2)
\\
&\quad
+\frac{11}{9}\pi^2\ln2 
-\frac{407}{36}\zeta_3
+\frac{13\pi^4}{144}
+\frac{\pi^2}{6}\ln^2{2}
-\frac{\ln^4{2}}{6}
-\frac{7}{2}\zeta_3\ln2
\\
&\quad
-4\Li_4\lp\frac{1}{2}\rp
-\LMu\lp\frac{37}{6} + \frac{11\pi^2}{8}+23\zeta_3\rp
+\LMusq\lp\frac{121}{36}-\frac{2\pi^2}{3}\rp\bigg]
\\
&\quad
+ \Ca\nf \bigg[
  -\frac{214}{81} - \frac{227}{216}\pi^2 - \frac{4}{3}\ln2 - \frac{2}{9}\pi^2\ln2 + 
  2 \ln^2 2 + \frac{37}{18}\zeta_3
  \\
  & \quad
  +  \LMu \lp \frac{94}{27}  + \frac{\pi^2}{4} \rp - \frac{11}{9}\LMu^2  
  \bigg]
\\
&\quad
+\nf^2\bigg[
\frac{11\pi^2}{108} - \frac{10}{27}\LMu + \frac{1}{9}\LMusq\bigg]
\\
&\quad
+\Theta_{bd}\bigg[
\Ca^2\lp -\frac{131}{36}+\frac{11}{3}\ln2 + \frac{\pi^2}{3}\rp
+\Ca\nf\lp \frac{23}{36} - \frac{2}{3}\ln2\rp\bigg]\Bigg\}.
\end{split}
\ee

\item \underline{{\bf  terms involving virtual corrections with NLO
kinematics}}
\bes
\d\hat\sigma^{\rm NNLO}_{{\rm virt}_{124},gg} = 
\big\langle \ONLO \FLVijfin{g}{g}(1,2,4)\big\rangle.
\end{split}
\ee

\item \underline{{\bf Terms involving virtual corrections with LO 
kinematics}:}
\bes
\d\hat\sigma^{\rm NNLO}_{{\rm virt}_{12},gg} = &
\bigg\langle
\FLVVijfin{g}{g}(1,2) + \FLVsqijfin{g}{g}(1,2)
+
\asontwopimu\left[\frac{2\pi^2}{3}\Ca-2\gamma_g\LMu\right] \times
\\
&\quad
\FLVijfin{g}{g}(1,2)
\bigg\rangle
+\asontwopimu\int\limits_0^1 dz\bigg[\mathcal P'_{gg}(z) 
-\LMu \hat P^{(0)}_{gg,R}(z)\bigg]\times
\\
&
\left\langle
\frac{\FLVijfin{g}{g}(z\cdot1,2)+\FLVijfin{g}{g}(1,z\cdot2)}{z}
\right\rangle.
\end{split}
\ee
\end{enumerate}
The above equations are analogous to
Eqs.~\eqref{eq:FLMzzb},~\eqref{eq:FLMz},~\eqref{eq:FLM12},~\eqref{eq:rv} and \eqref{eq:vv}
in Drell-Yan production, respectively.
The finite remainders $\FLVVfin$, 
$\FLVsqfin$ and $\FLVfin$ are defined in 
Appendix~\ref{app:virtual}, the finite remainder $\FLVfin(1,2,4)$ is defined in Appendix~\ref{app:realvirtual},
the function $\mathcal T_{gg}$ are given in the ancillary file 
and $\Theta_{bd}$ is given in Eqs.~(\ref{eq:thetabcdef}, \ref{eq:thetabc}).

\subsection{NNLO: quark-gluon channels}
\label{sec:h_nnlo_qg}
The structure of this channel is analogous to the $qg$ channel for 
the Drell-Yan process, discussed in Sec.~\ref{sec:dy_nnlo_qg}.
We don't repeat the discussion here, and limit ourselves to presenting
final results. To closely follow Sec.~\ref{sec:dy_nnlo_qg}, we focus on the
$g\qb$ channel. We write
\be
\dhs^{\NNLO}_{g\qb} = \dhs^{\NNLO}_{H+2,g\qb}+
\dhs^{\NNLO}_{H+1,g\qb}+\dhs^{\NNLO}_{H,g\qb},
\ee
with
\bes
&\dhs^{\NNLO}_{H+2,g\qb} = \dhs^{\NNLO}_{1245,g\qb},
\\
&\dhs^{\NNLO}_{H+1,g\qb} = \dhs^{\NNLO}_{124,g\qb}
                   +\dhs^{\NNLO}_{{\rm virt}_{124},g\qb},
\\
&\dhs^{\NNLO}_{H,g\qb} ~~= \dhs^{\NNLO}_{(z,\zb),g\qb}
                   +\dhs^{\NNLO}_{(1,z),g\qb}
                   +\dhs^{\NNLO}_{{\rm virt}_{12},g\qb}.
\end{split}
\ee
We display the individual contributions below.

\begin{enumerate}
\item \underline{\bf Tree-level terms with NNLO kinematics:}
\begin{align}
&  \d \hat \sigma^{\rm NNLO}_{1245,g\qb} =
\sum_{(ij)\in dc}\bigg\langle
\bigg[(\I- C_{5j})(\I-C_{4i})\bigg]\big[\I-S_5\big]
\times
\nonumber
\\
&\quad\quad\quad\quad\quad\quad
\times\dg4\dg5 w^{4i,5j}\FLMij{g}{\qb}(1,2,4,5)
\bigg\rangle
\nonumber
\\
&\quad\quad
+\sum_{i\in tc} 
\bigg\langle
\bigg[
\theta^{(\mathpzc{a})} \big[\I-\CC_i\big]\big[\I-C_{5i}\big] + 
\theta^{(\mathpzc{b})} \big[\I-\CC_i\big]\big[\I-C_{45}\big] 
\label{eq:h_qghard}
\\
&\quad\quad\quad\quad~~
 + \theta^{(\mathpzc{c})} \big[\I-\CC_i\big]\big[\I-C_{4i}\big]+ 
\theta^{(\mathpzc{d})} \big[\I-\CC_i\big]\big[\I-C_{45}\big]
\bigg]
\nonumber
\\
&\quad\quad\quad\quad~~
\times \big[\I-S_5\big]
\dg4 \dg5 w^{4i,5i}\FLMij{g}{\qb}(1,2,4,5)
\bigg\rangle.\nonumber
\end{align}

\item \underline{\bf Tree-level terms with NLO kinematics:}
\begin{align}
&\d\hat\sigma^{\rm NNLO}_{124, g\qb} 
= 
\asontwopimu
\int\limits_0^1\d z 
\Bigg\{
\Bigg\langle
\ln\frac{\rho_{41}}{4}\ONLO
\bigg[
\tilde w_{5||1}^{41,51} 
\times
\nonumber
\\
&
\frac{\hat P^{(0)}_{gg,R}(z)
\FLMij{g}{\qb}(z\cdot1,2,4)
+\hat P^{(0)}_{qg,R} \FLMij{q}{\qb}(z\cdot1,2,4)
}{z}\bigg]\Bigg\rangle
\nonumber
+
\Bigg\langle
\ln\frac{\rho_{42}}{4}\ONLO
\times
\nonumber\\
&
\bigg[
\tilde w_{5||2}^{42,52} 
\frac{
\FLMij{g}{\qb}(1,z\cdot2,4)
\hat P^{(0)}_{qq,R}(z)
+
\FLMij{g}{g}(1,z\cdot2,4)
\hat P^{(0)}_{gq,R}(z)
}{z}\bigg]\Bigg\rangle
\nonumber
\\
&
+\bigg[\mathcal P'_{gg}(z) - \hat P^{(0)}_{gg,R}(z)\LMu\bigg]
\left\langle
\ONLO\left[\frac{\FLMij{g}{\qb}(z\cdot1,2,4)
}{z}\right]
\right\rangle 
\label{eq:FLM124qg_h}
\\
&
+\bigg[\mathcal P'_{qg}(z) - \hat P^{(0)}_{qg,R}(z)\LMu\bigg]
\left\langle
\ONLO\left[\frac{\FLMij{q}{\qb}(z\cdot1,2,4)
}{z}\right]
\right\rangle 
\nonumber
\\
&
+
\left\langle
\ONLO\left[\frac{\FLMij{g}{\qb}(1,z\cdot 2,4)
}{z}\right]
\bigg[\mathcal P'_{qq}(z) - \hat P^{(0)}_{qq,R}(z)\LMu\bigg]
\right\rangle 
\nonumber
\\
&
+
\left\langle
\ONLO\left[\frac{\FLMij{g}{g}(1,z\cdot 2,4)
}{z}\right]
\bigg[\mathcal P'_{gq}(z) - \hat P^{(0)}_{gq,R}(z)\LMu\bigg]
\right\rangle \Bigg\}
\nonumber
\\
&
+\asontwopimu \bigg\langle
\ONLO\bigg[
\Delta^{(qg)} \cdot \FLMij{g}{q}(1,2,4)
\bigg]
\bigg\rangle.
\nonumber
\end{align}

\item
\underline{{\bf 
Tree-level terms with LO kinematics involving} $\FLM(z\cdot1,\zb\cdot2)$:}
\bes
\d\hat\sigma^{\rm NNLO}_{(z,\zb),g\qb} &
= 
\lp\asontwopimu\rp^2\int\limits_0^1\d z~\d\zb
\left[
\mathcal P'_{gg}(z)
-\LMu \hat P^{(0)}_{gg,R}(z)
\right]
\\
&\times
 \left\langle\frac{\FLMij{g}{g}(z\cdot1,\zb\cdot2)}{z\zb}\right\rangle
\left[
\mathcal P'_{gq}(\zb)
-\LMu \hat P^{(0)}_{gq,R}(\zb)
\right].
\end{split}
\ee

\item
\underline{{\bf Terms with LO kinematics involving}
$\FLM(1,z\cdot2)$:}
\bes
\d\hat\sigma^{\rm NNLO}_{(1,z),g\qb} 
=
\lp\asontwopimu\rp^2 \int\limits_0^1\d z
~\mathcal T_{gq}(z)
\left\langle\frac{\FLMij{g}{g}(1,z\cdot2)}{z}\right\rangle.
\end{split}
\ee

\item \underline{{\bf Terms involving virtual corrections with NLO
kinematics}:}
\bes
\d\hat\sigma^{\rm NNLO}_{{\rm virt}_{124},g\qb} = 
\big\langle \ONLO \FLVijfin{g}{\qb}(1,2,4)\big\rangle.
\end{split}
\ee

\item \underline{{\bf Terms involving virtual corrections with LO
kinematics}:}
\bes
\d\hat\sigma^{\rm NNLO}_{{\rm virt}_{12}, g\qb} = 
\asontwopimu\int\limits_0^1
\bigg[\mathcal P'_{gq}(z) -\LMu \hat P^{(0)}_{gq,R}(z)\bigg]
\left\langle
\frac{\FLVijfin{g}{g}(1,z\cdot2)}{z}
\right\rangle.
\end{split}
\ee

\end{enumerate}
The splitting functions used in these equations are defined in 
App.~\ref{app:split}, $\Delta^{(qg)}$ is given in 
Eq.~\eqref{eq:deltaqg}, $\mathcal T_{gq}$ is given in the ancillary file
 and the $\FLVfin$ finite remainders are defined in 
App.~\ref{app:virtual} and \ref{app:realvirtual}.
Results for the $qg$ channel can be trivially obtained from the above
formulas under $1\leftrightarrow 2$ replacement.

\subsection{NNLO: $q\qb$ channel}
\label{sec:h_nnlo_qqb}

The singularity structure of this channel can be organized as follows. 
There are purely collinear singularities, coming from configurations 
where the $q$ 
and the $\qb$ emit two $t$-channel gluons that produce a Higgs. 
These have the same structure as those appearing in the 
 $gg$ channel for the Drell-Yan process. 
Apart from these, there are other singular contributions, 
that don't have a Drell-Yan equivalent. They stem from
extra gluon emission from the $s$-channel $q\qb \to H + g$ process. These
are of NLO origin, so they don't pose any particular challenge. 

Because of its simple singularity structure, we don't discuss this channel
in detail. For completeness, we present final formulas that are obtained
using the same setup that we employed for the $gg$ channel. We write
\be
\dhs^{\NNLO}_{q\qb} = \dhs^{\NNLO}_{H+2,q\qb}+
\dhs^{\NNLO}_{H+1,q\qb}+\dhs^{\NNLO}_{H,q\qb},
\ee
and
\bes
&\dhs^{\NNLO}_{H+2,q\qb} = \dhs^{\NNLO}_{1245,q\qb},\\
&\dhs^{\NNLO}_{H+1,q\qb} = \dhs^{\NNLO}_{124,q\qb} + 
\dhs^{\NNLO}_{{\rm virt}_{124},q\qb},\\
&\dhs^{\NNLO}_{H,q\qb} ~= \dhs^{\NNLO}_{(z,\zb),q\qb}.
\end{split}
\ee
We now present the individual contributions:
\begin{enumerate}
\item \underline{\bf Tree-level terms with NNLO kinematics:}
\begin{align}
&  \d \hat \sigma^{\rm NNLO}_{1245,q\qb} =
\sum_{(ij)\in dc}\bigg\langle
\bigg[(\I- C_{5j})(\I-C_{4i})\bigg]
\big[\I-\SS\big]\big[\I-S_5\big]
\times
\nonumber
\\
&\quad\quad\quad\quad\quad\quad
\times\dg4\dg5 w^{4i,5j}\FLMij{q}{\qb}(1,2,4,5)
\bigg\rangle
\nonumber
\\
&\quad\quad
+\sum_{i\in tc} 
\bigg\langle
\bigg[
\theta^{(a)} \big[\I-\CC_i\big]\big[\I-C_{5i}\big] + 
\theta^{(b)} \big[\I-\CC_i\big]\big[\I-C_{45}\big] 
\label{eq:gghard_qqb}
\\
&\quad\quad\quad\quad~~
 + \theta^{(c)} \big[\I-\CC_i\big]\big[\I-C_{4i}\big]+ 
\theta^{(d)} \big[\I-\CC_i\big]\big[\I-C_{45}\big]
\bigg]
\nonumber
\\
&\quad\quad\quad\quad~~
\times\ \big[\I-\SS\big]\big[\I-S_5\big]
\dg4 \dg5 w^{4i,5i}\FLMij{q}{\qb}(1,2,4,5)
\bigg\rangle.\nonumber
\end{align}

\item \underline{\bf Tree-level terms with NLO kinematics:}
\begin{align}
&\d\hat\sigma^{\rm NNLO}_{124,q\qb} 
= 
\asontwopimu
\int\limits_0^1\d z 
\Bigg\{
\hat P^{(0)}_{qq,R}(z)
\Bigg\langle
\ln\frac{\rho_{41}}{4}\ONLO
\bigg[
\frac{\tilde w_{5||1}^{41,51} 
\FLMij{q}{\qb}(z\cdot1,2,4)
}{z}\bigg]
\nonumber
\\
&
+
\ln\frac{\rho_{42}}{4}\ONLO
\bigg[
\frac{\tilde w_{5||2}^{42,52} 
\FLMij{q}{\qb}(1,z\cdot2,4)
}{z}\bigg]
\Bigg\rangle
+
\hat P^{(0)}_{gq,R}(z)
\Bigg\langle
\ln\frac{\rho_{41}}{4}\ONLO\times
\nonumber
\\
&
\bigg[
\frac{\tilde w_{5||1}^{41,51} 
\FLMij{g}{\qb}(z\cdot1,2,4)
}{z}\bigg]
+
\ln\frac{\rho_{42}}{4}\ONLO
\bigg[
\frac{\tilde w_{5||2}^{42,52} 
\FLMij{q}{g}(1,z\cdot2,4)
}{z}\bigg]
\Bigg\rangle
\nonumber
\\
&
+\left[\mathcal P'_{qq}(z) - \hat P^{(0)}_{qq,R}(z)\LMu\right]
\left\langle
\ONLO\left[\frac{\FLMij{q}{\qb}(z\cdot1,2,4)
+\FLMij{q}{\qb}(1,z\cdot2,4)}{z}\right]
\right\rangle 
\nonumber
\\
&
+\left[\mathcal P'_{gq}(z) - \hat P^{(0)}_{gq,R}(z)\LMu\right]
\times
\label{eq:hFLM124gg_qqb}
\\
&
\left\langle
\ONLO\left[\frac{\FLMij{g}{\qb}(z\cdot1,2,4)
+\FLMij{q}{g}(1,z\cdot2,4)}{z}\right]
\right\rangle
\Bigg\}
\nonumber
\\
&
+\asontwopimu \bigg\langle
\ONLO\bigg[
\Delta_q \cdot \FLMij{q}{\qb}(1,2,4)
+\Delta^r \cdot \big[r_\mu r_\nu \FLMij{q}{\qb}^{\mu\nu}(1,2,4)\big]\bigg]
\bigg\rangle.
\nonumber
\end{align}

\item
\underline{{\bf 
Tree-level terms with LO kinematics involving} $\FLM(z\cdot1,\zb\cdot2)$:}
\bes
\d\hat\sigma^{\rm NNLO}_{(z,\zb),q\qb} &
= 
\lp\asontwopimu\rp^2\int\limits_0^1\d z~\d\zb
\left[
\mathcal P'_{gq}(z)
-\LMu \hat P^{(0)}_{gq,R}(z)
\right]
\\
&\times
 \left\langle\frac{\FLMij{g}{g}(z\cdot1,\zb\cdot2)}{z\zb}\right\rangle
\left[
\mathcal P'_{gq}(\zb)
-\LMu \hat P^{(0)}_{gq,R}(\zb)
\right].
\end{split}
\ee

\item \underline{{\bf  Terms involving virtual corrections with NLO
kinematics}:}
\bes
\d\hat\sigma^{\rm NNLO}_{{\rm virt}_{124},q\qb} = 
\big\langle \FLVijfin{q}{\bar{q}}(1,2,4)\big\rangle.
\end{split}
\ee
\end{enumerate}
The splitting functions used in these equations are defined in 
App.~\ref{app:split}, $\Delta_q$ is given in 
Eq.~\eqref{eq:delta} with $C_i = \Cf$, $\gamma_q=3\Cf/2$, 
$\mathcal{X}_q = 3\Ca/2$, and the $\FLVfin$ finite remainder is defined in 
App.~\ref{app:realvirtual}. Note that, contrary to all the cases
discussed so far, $\FLVijfin{q}{\qb}$ does not  require any additional
regularization.

\subsection{NNLO: quark channels}
\label{sec:h_nnlo_qq}
The singularity structure of this channel is the same as the $gg$ channel
for the Drell-Yan process. Because of this, we use the same parametrization 
described in Sec.~\ref{sec:dy_nnlo_gg}. We write 
\be
\dhs^{\NNLO}_{q_i q_j} = \dhs^{\NNLO}_{H+2,q_i q_j}
+\dhs^{\NNLO}_{H+1,q_i q_j}
+\dhs^{\NNLO}_{H,q_i q_j},
\ee
with
\bes
&\dhs^{\NNLO}_{H+2,q_i q_j} = \dhs^{\NNLO}_{1245,q_i q_j},\\
&\dhs^{\NNLO}_{H+1,q_i q_j} = \dhs^{\NNLO}_{124,q_i q_j},\\
&\dhs^{\NNLO}_{H,q_i q_j} ~~~= \dhs^{\NNLO}_{(z,\zb),q_i q_j}.
\end{split}
\ee
Repeating the same steps discussed in Sec.~\ref{sec:dy_nnlo_gg}
we obtain the following results. 
\begin{enumerate}
\item \underline{\bf Tree-level terms with NNLO kinematics:}
\bes
  \d \hat \sigma^{\rm NNLO}_{1245,q_i q_j} = &
\bigg \langle \big[\I - C_{41} - C_{42} - C_{51} - C_{52} + C_{42} C_{51} + 
C_{41} C_{52}\big]  \\
& \times \dg4 \dg5 \FLMij{q_i}{q_j}(1,2,4,5) \bigg\rangle.
\end{split}
\end{equation}

\item \underline{\bf Tree-level terms with NLO kinematics:}
\bes
&\d\hat\sigma^{\rm NNLO}_{124,q_i q_j} 
= 
\asontwopimu
\int\limits_0^1\d z 
\left[\mathcal P'_{gq}(z) - \hat P^{(0)}_{gq,R}(z)\LMu\right]
\\
&\times
\left\langle
\ONLO\left[\frac{\FLMij{g}{q_j}(z\cdot1,2,4)
+\FLMij{q_i}{g}(1,z\cdot2,4)}{z}\right]
\right\rangle 
.
\end{split}
\ee
\item
\underline{{\bf 
Tree-level terms with LO kinematics involving} $\FLM(z\cdot1,\zb\cdot2)$:}
\bes
\d\hat\sigma^{\rm NNLO}_{(z,\zb),q_i q_j} &
= 
\lp\asontwopimu\rp^2 \int\limits_0^1\d z~\d\zb
\left[
\mathcal P'_{gq}(z)
-\LMu \hat P^{(0)}_{gq}(z)
\right]
\\
&\times
 \left\langle\frac{\FLMij{g}{g}(z\cdot1,\zb\cdot2)}{z\zb}\right\rangle
\left[
\mathcal P'_{gq}(\zb)
-\LMu \hat P^{(0)}_{gq}(\zb)
\right].
\end{split}
\ee

\end{enumerate}

\section{Validation of results}
\label{sec:validation}
In this section, we describe  the numerical checks that have been used to 
validate  results described in the preceding
sections.  We use the processes $pp \to Z$ and $pp \to H$ as test
cases, since for both of  these processes 
NNLO QCD corrections to the inclusive cross sections  are 
known analytically~\cite{Hamberg:1990np,Anastasiou:2002yz}. This
allows us to perform a high-precision check of our formulas. 

We begin by describing our setup. We consider proton-proton collisions with
$13$~TeV center-of-mass energy. We use  $m_Z = 91.1876~{\rm GeV}$,
$m_H = 125~{\rm GeV}$ and $m_t=173.2~{\rm GeV}$ for the $Z$, 
the Higgs and the top quark masses, respectively.  We 
derive the weak coupling constant from
 $g_W^2 = 4 \sqrt{2} m_W^2 G_F $, with $m_W = 80.398~{\rm GeV}$ and
$G_F = 1.16639 \times 10^{-5}~{\rm GeV}^{-2}$. The weak mixing 
angle is computed
from $\sin^2\theta_W = 1- m_W^2/m_Z^2$. We will consider
both on-shell $pp \to Z$ production and $p p \to e^+ e^-$ production. 
In the latter
case, we include both the $Z$ and the $\gamma^*$ contributions, and we
use $\Gamma_Z = 2.4952~{\rm GeV}$.

For the Higgs case, we consider $pp \to H$ production
in the 
$m_t\to\infty$ approximation and describe Higgs coupling to gluons 
using the  effective Lagrangian $\mathcal L_I = 
-\lambda_{Hgg} H G^{(a)}_{\mu\nu} G^{\mu\nu,(a)}$, see App.~\ref{app:virtual}
for details. The coupling $\lambda_{Hgg}$ depends on the Higgs vacuum
expectation value $v$. For our results, we use $v^2 = (G_F\sqrt{2})^{-1}$.
All computations are done using the NNPDF3.0 parton distribution 
set~\cite{nnpdf}, with 5 active flavors.
We employ LO/NLO/NNLO sets for LO/NLO/NNLO predictions, respectively. We
use the value of the strong coupling and its evolution 
provided by the PDFs sets, with 
$\as(m_Z)=0.118$ at (N)NLO, and $\as(m_Z) = 0.130$ at LO. 

We first consider fully inclusive on-shell $Z$ production. 
We compare results obtained within our framework 
to the analytic results of Ref.~\cite{Hamberg:1990np} that we
implemented in {\tt HOPPET}~\cite{Salam:2008qg}. 
We study each partonic channel individually, 
and additionally split some of them into contributions with different
color factors
in order to validate all the different singularity structures independently.
We show results for a single fixed factorization and
renormalization scale $\mu=2m_Z$, although we have performed the same check
for different scales as well. We note  that
these inputs are not chosen for their phenomenological relevance, but
rather to provide stringent checks on our results.
\begin{table}
\centering
\begin{tabular}{|c|c|c|c|}
\hline
Channel & Color structures & Numerical result (nb) & Analytic result (nb)\\
\hline
\hline
$q_i\qb_i\to gg$ & -- & 8.351(1) & 8.3516 \\
\hline
$q_i\qb_i \to q_j \qb_j $ & $\Cf \tr n_{\rm up},~\Cf \tr n_{\rm dn}$ & -2.1378(5) & -2.1382 \\
& $\Cf(\Ca-2\Cf)$ & $-4.8048(3)\cdot 10^{-2}$ &  $-4.8048\cdot 10^{-2}$ \\
& $\Cf\tr$ & 5.441(7)$\cdot 10^{-2}$ & 5.438$\cdot 10^{-2}$ \\
\hline
$q_i q_j \to q_i q_j\;\; (i\ne -j) $ & $\Cf \tr $ & 0.4182(5) & 0.4180 \\
& $\Cf(\Ca-2\Cf)$ & $-9.26(1)\cdot 10^{-4}$ & $-9.26\cdot 10^{-4}$ \\
\hline
\hline
$q_i g + g q_i$ & -- & -9.002(9) & -8.999\\
\hline
\hline
$gg$ & -- & 1.0772(1) & 1.0773\\
\hline
\hline
\end{tabular}
\caption{Different contributions to the NNLO \emph{coefficient} for
on-shell $Z$ production at the 13 TeV LHC with $\mu_R=\mu_F = 2m_Z$. 
All the color factors are included in the numerical
results. The residual Monte-Carlo integration error is shown in
brackets. See text for details.
\label{tab:z}
}
\end{table}

We summarize
our findings  in Table~\ref{tab:z}.
It  shows that our framework allows for extremely
high precision results, with numerical errors at the per mille level or 
better\footnote{The larger error in some channels is caused by non-negligible
cancellations between different contributions to the final result.}.
These results are always fully compatible with the analytic ones
within the numerical uncertainties. 
We remind the reader that these numbers refer to the NNLO \emph{coefficients},
which implies absolute precision on the physical cross section. 
Analogous results for the case of Higgs production 
for equal renormalization and factorization scales $\mu_R=\mu_F=m_H/2$ 
are shown in
Tab.~\ref{tab:h}.
Again, we find it convenient to perform numerical checks for different
partonic channels independently. 
Also in this case, our numerical results have tiny uncertainties 
and are in perfect agreement with the analytic values
obtained from Ref.~\cite{Anastasiou:2002yz}.

\begin{table}
\centering
\begin{tabular}{|c|c|c|c|}
\hline
Channel & Numerical result (pb) & Analytic result (pb)\\
\hline
\hline
$ gg \to gg$ & 9.397(1) & 9.398 \\
\hline
$ gg \to q\bar{q}$ &-1.243(2) & -1.243 \\
\hline
\hline
$qg + gq$ & 0.7865(8)  & 0.7861 \\
\hline
\hline
$q\bar{q}$ & $1.145(1) \cdot 10^{-2}$  &  $ 1.146 \cdot 10^{-2}$ \\
\hline
\hline
$q q$ & $2.139(3) \cdot 10^{-2}$  & $2.140  \cdot 10^{-2}$\\
\hline
$q q'$ & $5.967(5) \cdot 10^{-2}$    & $5.970 \cdot 10^{-2}$ \\
\hline

\hline
\hline
\end{tabular}
\caption{Different contributions to the NNLO \emph{coefficient} for
on-shell $H$ production at the 13 TeV LHC with $\mu_R=\mu_F = m_H/2$. 
The residual Monte-Carlo integration error is shown in
brackets. The labels $qq$ and $qq'$ 
refer to quark initial states with identical
and different flavors, respectively. See text for details.
\label{tab:h}
}
\end{table}

Having fully validated our results, we now briefly investigate the performance
of the framework when applied to the computation of physically relevant predictions.
Specifically, we explore the computational effort required to obtain
predictions for physical quantities at the per mille level. 
We start by considering inclusive
Higgs production, at the 13 TeV LHC. 
For this study, we set $\mu_R=\mu_F = m_H$. 
Running for less than an hour on a single core of a standard laptop, we obtain
\be
\sigma_{\H}^{\rm LO} = 15.42(1)~{\rm pb}; \;\;\;\;\;\;\; \sigma_{\H}^{\rm NLO} = 30.25(1)~\rm{pb}; \;\;\;\;\;\;\; \sigma_{\H}^{\rm NNLO} = 39.96(2)~{\rm pb}.
\label{eq:5.1}
\ee
As one can see from Eq.~\eqref{eq:5.1} the 
numerical uncertainty on the full NNLO cross section 
is below one per mille.
The result is in full agreement\footnote{
The different LO cross section w.r.t. Ref.~\cite{Grazzini:2017mhc} is due
to a different choice of LO PDFs.} with the benchmark predictions 
reported in Ref.~\cite{Grazzini:2017mhc}.

We now move to fiducial cross sections. 
We  consider $pp \to Z/\gamma^* \to e^-e^+$ production in
the fiducial volume defined by symmetric lepton cuts 
studied in Ref.~\cite{Grazzini:2017mhc}.
Specifically, 
we require that the transverse momentum and rapidity of each lepton satisfy
\be \label{eq:Higgs_xsecs}
p_{T,\ell} > 25~\rm{GeV} \;\;\;\;\;\;\;\ |\eta_{\ell}| < 2.47,
\ee
and that the invariant mass of the lepton pair is in a window $66~{\rm GeV} < m_{e^-e^+} < 116~{\rm GeV}$.
In this case, we use $\mu_R=\mu_F=m_Z$. Running on a single core of a standard
laptop for about an hour, we obtain
\be\label{eq:DY_xsecs}
\sigma_{{\rm DY}}^{\rm LO} = 650.4 \pm 0.1~{\rm pb}; \;\;\;\;\;\;\; \sigma_{{\rm DY}}^{\rm NLO} = 700.2\pm 0.3~\rm{pb}; \;\;\;\;\;\;\; \sigma_{{\rm DY}}^{\rm NNLO} = 734.8 \pm 1.4 ~{\rm pb}.
\ee
We note that in this case the error is at the few per mille level. 
We compared the NNLO $K$-factor against the benchmark result presented in Ref.~\cite{Grazzini:2017mhc},
and found agreement within the numerical precision. 

As a final comment, we note that although
the processes studied here are very simple, which makes it 
difficult to
predict how the framework will perform for more complicated ones,
these results are very encouraging. 

\section{Conclusions}
\label{sec:conclusions}

In this paper, we presented compact analytic formulas that describe fully-differential production of color-singlet final
states in hadron collisions. We studied final states that, at leading order, can be produced either in $q \bar q$ or in $gg$ annihilation. 

Our calculation employs the nested soft-collinear subtraction scheme that we 
developed earlier in Ref.~\cite{Caola:2017dug}.
However, compared to its original formulation, we found it useful to modify it 
 to allow for a simpler analytic integration of the triple-collinear limits. We explained the required changes in Section~\ref{sec:basics}.

 We validated our results  by using them to numerically compute
   NNLO QCD
   contributions  to total cross sections of $Z$ and $H$ production in proton collisions
   and comparing them with  results based on the convolution of known analytic results for partonic cross sections relevant
   for these two processes with parton distribution functions.  
 In both cases, we found an agreement at a better-than-per-mille level for NNLO {\it coefficient
functions}. We also showed that our computation can deal with fiducial cuts quite efficiently.

 As far as we know, the results presented in this paper  provide the 
first implementation of a
 fully local and fully analytic NNLO QCD subtraction
scheme. Although  -- as we already emphasized in the introduction -- it is difficult  to say to what extent these nice features of the
subtraction scheme will help with its efficiency,  we hope that they will be helpful in that respect. However, we also
believe that,  regardless of the performance issues,  physical clarity and overall  transparency of the obtained formulas  gives us hope that
analytic, local  NNLO QCD subtractions for arbitrary complex hadron collider processes are within reach.

\vspace*{1cm}
{\bf Acknowledgments} 
We are grateful to Mainz Institute for Theoretical Physics for  hospitality extended to us during the work
on this paper. R.R. is also grateful to the Galileo Galilei Institute for hospitality and support.
We thank B.~Mistlberger for providing computer-readable results for the NNLO corrections to the Higgs inclusive
cross section.   The research of K.M. is supported by BMBF grant 05H18VKCC1 and by the DFG Collaborative Research Center
TRR 257 ``Particle Physics Phenomenology after the Higgs Discovery''.

\newpage
\appendix

\section{Purely virtual contribution: definitions}
\label{app:virtual}
We consider the UV-renormalized amplitude for the process $p_1+p_2\to V$
\be
\mathcal A(p_1,p_2;\{p_V\}) = \mathcal A_{0}(1,2) + \asontwopimu \mathcal A_{1}(1,2) + 
\lp\asontwopimu\rp^2 \mathcal A_{2}(1,2) + \ldots .
\ee
Following Ref.~\cite{Catani:1998bh}, we write
\bes
&\mathcal A_1(1,2) = \mathcal I_{1}(\ep) \mathcal A_0(1,2) + \mathcal A_{1,\rm fin}(1,2),
\\
&\mathcal A_2(1,2) = \mathcal I_{2}(\ep) \mathcal A_0(1,2) + 
\mathcal I_{1}(\ep) \mathcal A_{1}(1,2) + \mathcal A_{2,\rm fin}(1,2),
\end{split}
\label{eq:catani}
\ee
with $\mathcal A_{i,\rm fin}$ finite in four dimensions. The explicit form of the 
$\mathcal I_i$ operators read~\cite{Catani:1998bh}
\bes
&\mathcal I_{1}(\ep) = -\frac{e^{\ep\gamma_E}}{\Gamma(1-\ep)}
\lp\frac{\Ci}{\ep^2} + \frac{\gamma_i}{\ep}\rp e^{i \pi \ep} \lp \frac{\mu^2}{s_{12}}\rp^\ep,
\\
&\mathcal I_2(\ep) = -\frac{1}{2}\mathcal I_1^2(\ep) - \frac{\beta_0}{\ep} \mathcal I_1(\ep) 
+\frac{e^{-\ep \gamma_E}\Gamma(1-2\ep)}{\Gamma(1-\ep)}\lp\frac{\beta_0}{\ep} + K\rp
\mathcal I_1(2\ep) + \frac{e^{\ep\gamma_E}}{\Gamma(1-\ep)}\frac{H_i}{2\ep},
\end{split}
\ee
with 
\be
\beta_0 = \frac{11}{6}\Ca - \frac{2}{3}\tr \nf,~~~
K=\lp\frac{67}{18}-\frac{\pi^2}{6}\rp\Ca-\frac{10}{9}\tr\nf.
\label{eq:b0}
\ee
For $q\bar q\to V$ reactions, $i=q$ and
\bes
C_q &= \Cf,~~~~\gamma_q = \frac{3}{2}\Cf,
\\
H_q &= \Cf^2\lp\frac{\pi^2}{2}-6\zeta_3-\frac{3}{8}\rp + 
\Ca\Cf\lp\frac{245}{216}-\frac{23}{48}\pi^2+\frac{13}{2}\zeta_3\rp
+\Cf\nf\lp\frac{\pi^2}{24}-\frac{25}{108}\rp,
\end{split}
\ee
while for $gg\to V$ reactions $i=g$ and
\bes
& C_g = \Ca,~~~~ \gamma_g = \beta_0,
\\
& H_g = \Ca^2\lp\frac{5}{12} + \frac{11}{144}\pi^2+\frac{\zeta_3}{2}\rp
+\Ca\nf \lp-\frac{29}{27}-\frac{\pi^2}{72}\rp
+\frac{\Cf\nf}{2}+\frac{5}{27}\nf^2.
\end{split}
\ee

To express the virtual contribution to the cross-section, it is also useful to define
\be
I_{12}(\ep) = 2{\rm Re}[\mathcal I_1(\ep)] = -2\cos(\ep\pi)\frac{e^{\ep \gamma_E}}{\Gamma(1-\ep)}
\lp\frac{\mu^2}{s_{12}}\rp^\ep
\left[\frac{\Ci}{\ep^2} + \frac{\gamma_i}{\ep}\right].
\ee
The NLO virtual correction can then be written as
\be
2s\cdot \dhs^{\rm NLO,V} = \lp\asontwopimu\rp I_{12}(\ep) \left\langle F_{LM}(1,2)\right\rangle
+ \left\langle F_{LV, \rm fin}(1,2)\right\rangle,
\ee
with $F_{LV,\rm fin}$ finite and proportional to $2{\rm Re}\big[\mathcal A_0 \mathcal A_{1,\rm fin}^*\big]$.
Similarly, we can write the purely virtual corrections at NNLO as
\bes
2s\cdot \d\hat\sigma^{\rm NNLO,VV} & = 
\bigg[
\frac{I_{12}^2(\ep)}{2} - \frac{\beta_0}{\ep} I_{12}(\ep)+
\frac{e^{-\ep\gamma_E}\Gamma(1-2\ep)}{\Gamma(1-\ep)}\lp\frac{\beta_0}{\ep}+K\rp
I_{12}(2\ep)
\\
&
+\frac{e^{\ep\gamma_E}}{\Gamma(1-\ep)}\frac{H_i}{\ep}\bigg]
\lp\asontwopimu\rp^2\left\langle\FLM(1,2)\right\rangle
\\
&
+ I_{12}(\ep) \lp\asontwopimu\rp 
\left\langle \FLVfin(1,2)\right\rangle 
+
\left\langle \FLVVfin(1,2)\right\rangle + \left\langle \FLVsqfin(1,2)\right\rangle,
\end{split}
\ee
with $\FLVVfin$ and $\FLVsqfin$ finite and proportional to
$2{\rm Re}\big[\mathcal A_0 \mathcal A_{2,\rm fin}^*\big]$ and 
$|\mathcal A_{1,\rm fin}|^2$, respectively. 

\allowdisplaybreaks
\subsection{Finite remainder: Drell-Yan}
In this section, we report the finite remainders for the Drell-Yan process, see e.g.~\cite{Gehrmann:2005pd}.
We obtain
\begin{align}
&\left\langle \FLVijfin{\fa}{\fb}(1,2)\right\rangle|_{\mu^2=Q^2} = 
-8\Cf\lp\asontwopi\rp\left\langle\FLMij{\fa}{\fb}(1,2)\right\rangle + \mathcal O(\ep),
\nn\\
&\left\langle \FLVsqijfin{\fa}{\fb}(1,2)\right\rangle|_{\mu^2=Q^2} = 
16\Cf^2 \lp\asontwopi\rp^2\left\langle\FLMij{\fa}{\fb}(1,2)\right\rangle + \mathcal O(\ep),
\nn\\
&\left\langle \FLVVijfin{\fa}{\fb}(1,2)\right\rangle|_{\mu^2=Q^2} = 
\lp\asontwopi\rp^2 \Bigg[
\Cf^2\lp
\frac{255}{16}+\frac{29 \pi^2}{12}-15 \zeta_3-\frac{11 \pi^4}{90}
\rp
\\
&\quad
+\Cf\Ca
\lp
-\frac{51157}{1296}-\frac{107 \pi^2}{72}+\frac{659 \zeta_3}{36}+\frac{31 \pi^4}{240}
\rp
+\Cf\nf
\lp
\frac{4085}{648}+\frac{7 \pi^2}{36}-\frac{\zeta_3}{18}
\rp
\Bigg]
\nn\\
&\quad
\times \left\langle \FLMij{\fa}{\fb}(1,2)\right\rangle + \mathcal O(\ep)
,\nn
\end{align}
with $Q^2 = p_V^2$ and $\as = \as(Q)$. Results for generic $\mu$ can easily be obtained from
renormalization group evolution (RGE) arguments. 

\subsection{Finite remainder: Higgs}
In this section, we report the finite remainders for the Higgs process. More
precisely, we consider a theory where the Higgs is coupled directly to gluons
through the effective interaction Lagrangian
\be
\mathcal L_I = -\lambda_{Hgg} H G_{\mu\nu}^{(a)} G^{\mu\nu,(a)},
\ee
where the (bare) $Hgg$ coupling is given by 
\be
\lambda_{Hgg,b} = -\frac{\as}{12\pi v} C(\as) Z_{\rm eff}(\as).
\ee
In this formula, $\as = \as(\mu)$ is the renormalized coupling in a theory
with 5 light flavors, $v$ is the Higgs v.e.v. and the divergent 
($Z_{\rm eff}(\as)$)
and finite ($C(\as)$) parts of the Wilson coefficient renormalization
are given in the
$\overline{\rm MS}$ scheme by
\bes
&
Z_{\rm eff}(\as) = 1- \frac{\beta_0}{\ep}\lp \frac{\as}{2\pi}\rp
+\left[ \frac{\beta_0^2}{\ep^2}-\frac{\beta_1}{\ep}\right]
 \lp\frac{\as}{2\pi}\rp^2
+\mathcal O(\as^3),
\\
&
C(\as) = 1 + \left[\frac{5}{2}\Ca - \frac{3}{2}\Cf\right] 
\lp\frac{\as}{2\pi}\rp
+
\bigg[
\frac{1063}{144}\Ca^2 - \frac{25}{3}\Ca \Cf + \frac{27}{8}\Cf^2
\\
&\quad\quad\quad
-\frac{47}{72}\Ca \nf - \frac{5}{8}\Cf\nf
-\frac{5}{48}\Ca - \frac{\Cf}{6}
\\
&\quad\quad\quad
+\ln\lp\frac{\mu^2}{m_t^2}\rp
\bigg(
\frac{7}{4}\Ca^2 - \frac{11}{4} \Ca \Cf + \Cf\nf
\bigg)
\bigg]\lp\frac{\as}{2\pi}\rp^2 + \mathcal O(\as^3),
\end{split}
\label{eq:wilscoeff}
\ee
see e.g.~\cite{Grigo:2014jma}. In Eq.~\eqref{eq:wilscoeff}, $m_t$ is the 
top-quark mass, $\beta_0$ has been defined in Eq.~\eqref{eq:b0} and
\be
\beta_1 = \frac{17}{6}\Ca^2 - \frac{5}{3}\Ca\tr\nf - \Cf\tr\nf.
\ee

Combining the result for the $Hgg$ form factor in e.g.~\cite{Gehrmann:2005pd}
with the finite part of the Wilson coefficient renormalization, we obtain for
the Higgs finite remainders
\begin{align}
&\left\langle \FLVijfin{g}{g}(1,2)\right\rangle|_{\mu^2=Q^2} = 
\lp\asontwopi\rp\big[5\Ca-3\Cf\big]\left\langle \FLMij{g}{g}(1,2)\right\rangle + \mathcal O(\ep),
\nn\\
&\left\langle F_{LV^2,gg,\rm fin}(1,2)\right\rangle|_{\mu^2=Q^2} = 
\lp\asontwopi\rp^2
\Bigg[\beta_0^2\pi^2 + \lp \frac{5}{2}\Ca - \frac{3}{2}\Cf\rp^2 \Bigg]
\nn\\
&\quad
\times\left\langle\FLMij{g}{g}(1,2)\right\rangle + \mathcal O(\ep),
\nn\\
&\left\langle \FLVVijfin{g}{g}(1,2)\right\rangle|_{\mu^2=Q^2} = 
\lp\asontwopi\rp^2
\Bigg\{
\Ca^2
\lp
\frac{5105}{324}-\frac{17 \pi^2}{3}-\frac{253 \zeta_3}{36}+\frac{\pi^4}{144}
\rp
\\
&
\quad
+\Ca\nf
\lp
-\frac{458}{81}+\frac{481 \pi^2}{216}-\frac{49 \zeta_3}{18}
\rp
+\Cf\nf
\lp
-\frac{67}{12}+4 \zeta_3
\rp
-\frac{23}{108}\nf^2\pi^2
\nn\\
&\quad
+\bigg[
\frac{1063}{72}\Ca^2+\frac{27}{4}\Cf^2
-\frac{50}{3}\Ca\Cf
-\frac{47}{36}\Ca\nf-\frac{5}{4}\Cf\nf
-\frac{5}{24}\Ca-\frac{1}{3}\Cf
\nn\\
&\quad
+\lp
\frac{7}{2}\Ca^2-\frac{11}{2}\Ca\Cf+2 \Cf\nf
\rp\ln\lp\frac{Q^2}{m_t^2}\rp\bigg]\Bigg\}
\left\langle\FLMij{g}{g}(1,2)\right\rangle + \mathcal O(\ep),
\nn
\end{align}
with $Q^2 = p_H^2$ and $\as = \as(Q)$. Results for generic $\mu$ can easily be
obtained from RGE arguments. 

\section{Real-virtual contribution: definitions}
\label{app:realvirtual}
We consider the one-loop amplitude for the process 
\be
f_1+f_2\to V/H+f_4,
\label{eq:procRV}
\ee
where $V/H$ indicates either Drell-Yan or Higgs production, as in Secs.~\ref{sec:dy} and~\ref{sec:h}.
Following Appendix~\ref{app:virtual}, we write it as
\be
\mathcal A(p_1,p_2,p_4;\{p_V\}) = \mathcal A_0(1,2,4) + 
\asontwopimu\mathcal A_1(1,2,4) + ...,
\ee
with
\be
\mathcal A_1(1,2,4) = \mathcal I_1(1,2,4;\ep)\mathcal A_0(1,2,4) + \mathcal A_{1,\rm fin}(1,2,4).
\ee
In the above, $\mathcal I_1(1,2,4;\ep)$ is the equivalent of $\mathcal I_1(\ep)$ of Eq.~\eqref{eq:catani} for
the $f_1+f_2\to V/H+f_4$ kinematic configurations (cf. Ref.~\cite{Catani:1998bh}), and  $\mathcal A_{1,\rm fin}(1,2,4)$ is finite. 
We then define
\be
I_{124}(\ep) = 2{\rm Re}\big[\mathcal I_1(1,2,4;\ep)\big],
\ee
and write the real-virtual contribution to the NNLO cross-section as
\bes
\d\hat\sigma^{\rm NNLO,RV} = 
\asontwopimu\left\langle I_{124}(\ep) \FLM(1,2,4)\right\rangle + 
\left\langle \FLVfin(1,2,4)\right\rangle,
\end{split}
\ee
with $\FLVijfin{i}{j}(1,2,4)$ finite. The explicit form of $I_{124}$ depends on the color
structure of the process. Since the process Eq.~\eqref{eq:procRV}
must involve either a gluon and a $q\qb$ pair or 3 gluons, we can classify the most general
case according to the position of a gluon and write
\bes
 I_{1_i 2_j 4_g}&=\frac{e^{\ep\gamma_E}}{\Gamma(1-\ep)}
\Bigg\{
\cos(\ep\pi) \lp\frac{\mu^2}{s_{12}}\rp^\ep 
\left[\frac{\Ca-2\Ci}{\ep^2} + \frac{\mathcal X_i-2\gamma_i}{\ep}\right]
\\
& -\left[\lp\frac{\mu^2}{s_{14}}\rp^\ep + \lp\frac{\mu^2}{s_{24}}\rp^\ep \right]
\left[\frac{\Ca}{\ep^2} + \frac{\mathcal X_i+\gamma_g}{2\ep}\right] \Bigg\},
\\
 I_{1_g 2_i 4_j}&=\frac{e^{\ep\gamma_E}}{\Gamma(1-\ep)}
\Bigg\{
\lp\frac{\mu^2}{s_{24}}\rp^\ep 
\left[\frac{\Ca-2\Ci}{\ep^2} + \frac{\mathcal X_i-2\gamma_i}{\ep}\right] 
\\
& -\left[\lp\frac{\mu^2}{s_{14}}\rp^\ep + \cos(\ep\pi) \lp\frac{\mu^2}{s_{12}}\rp^\ep \right]
\left[\frac{\Ca}{\ep^2} + \frac{\mathcal X_i+\gamma_g}{2\ep}\right] \Bigg\},
\\
 I_{1_i 2_g 4_j}&=\frac{e^{\ep\gamma_E}}{\Gamma(1-\ep)}
\Bigg\{
\lp\frac{\mu^2}{s_{14}}\rp^\ep 
\left[\frac{\Ca-2\Ci}{\ep^2} + \frac{\mathcal X_i-2\gamma_i}{\ep}\right]
\\
& -\left[\lp\frac{\mu^2}{s_{24}}\rp^\ep + \cos(\ep\pi) \lp\frac{\mu^2}{s_{12}}\rp^\ep \right]
\left[\frac{\Ca}{\ep^2} + \frac{\mathcal X_i+\gamma_g}{2\ep}\right] \Bigg\},
\end{split}
\ee
with
\be
\mathcal X_q = \frac{3}{2}\Ca,~~~~\mathcal X_g = \gamma_g. 
\ee
Here $s_{ij} = 2 E_i E_j \rho_{ij}$, with $E_{i,j}>0$. 

\section{Splitting functions}
\label{app:split}
In this appendix we collect the relevant splitting functions used in the
main text. We write the LO Altarelli-Parisi splitting functions as
\be
\hat P^{(0)}_{ij}(z) = \hat P^{(0)}_{ij,R}(z) + \hat P^{(0)}_{ij,\delta}\;\delta(1-z),
\ee
with
\bes
& \hat P^{(0)}_{qq,R}(z) = \Cf\left[\frac{2}{(1-z)_+} - (1+z)\right],
\\
& \hat P^{(0)}_{qg,R}(z) = \tr \big[ z^2 + (1-z)^2\big],
\\
& \hat P^{(0)}_{gq,R}(z) = \Cf \left[ \frac{1+(1-z)^2}{z}\right],
\\
& \hat P^{(0)}_{gg,R}(z) = 2 \Ca \left[ \frac{1}{(1-z)_+} +
\frac{1}{z} + z(1-z) - 2\right],
\end{split}
\ee
and
\be
\hat P^{(0)}_{qq,\delta} = \gamma_q = \frac{3}{2}\Cf,~~~
\hat P^{(0)}_{gg,\delta} = \gamma_g = \beta_0 = \frac{11}{6}\Ca - \frac{2}{3}\tr\nf,
~~~ \hat P^{(0)}_{qg,\delta}=\hat P^{(0)}_{gq,\delta} = 0.
\ee

The $\mathcal P'_{ij}$ splitting functions are related to the $\mathcal O(\ep)$
part of the LO splitting functions, and read
\bes
&\mathcal P'_{qq}(z) = \Cf \left( 4 \left[\frac{\ln(1-z)}{1-z}\right]_+ 
-2 (1+z) \ln(1-z) + (1-z)\right),
\\
&\mathcal P'_{qg}(z) =   \tr \bigg( 2\big[z^2+(1-z)^2\big] \ln(1-z) + 2 z(1-z)\bigg),
\\
&\mathcal P'_{gq}(z) = \Cf \bigg( 
2\left[\frac{1+(1-z)^2}{z}\right] \ln(1-z) + z\bigg),
\\
&\mathcal P'_{gg}(z) = 2\Ca \bigg( 2\left[\frac{\ln(1-z)}{1-z}\right]_+
+2 \left[ \frac{1}{z} + z(1-z) -2 \right] \ln(1-z) \bigg). 
\end{split}
\ee

\section{Transition functions}
\label{app:transition}
The transition functions $\mathcal T_{ij}$ introduced in the text are generalizations of NLO Altarelli-Parisi splitting functions.
They depend on the scale $\mu$ of the process (though this dependence is of course entirely determined by RGE arguments), 
on an energy fraction $z$ and, in general, on the choice of partition functions
through the factor
\be
\Theta_{ac} \equiv - \bigg\langle \big[\I-C_{41}\big]\left[
\frac{\rho_{12}}{\rho_{41}\rho_{42}}
\right]
\lp
\tilde w^{41,51}_{5||1}\ln\frac{\rho_{41}}{4}
\rp
\bigg\rangle.
\ee
For the choice Eq.~\eqref{eq:partitions}, we obtain
\be
\Theta_{ac} = 1+\ln 2.
\ee
We now show results for a typical transition function, namely the nonsinglet function relevant for Drell-Yan production. It reads. 
\begin{align}
\mathcal T_{qq}^{\rm NS} &= 
\Cf^2\big[4\tilde\DD_1(z) \big]\Theta_{ac} 
+\Cf^2\Bigg[
8 \tilde\DD_3(z) +16 \tilde\DD_0(z) \zeta (3) +
\frac{(z+1)}{12} \ln ^3(z)
\nonumber
\\
&
+\frac{3 \left(2 z^2+2 z-7\right) \ln ^2(z)}{4 (1-z)}
+
\left(4 (z+1) \ln (z+1)+\frac{22 z^2+5 z-17}{2 (1-z)}\right)
   \ln (z)
\nonumber
\\
&
-\frac{\pi ^2 (8-5 z) z}{3 (1-z)}+\frac{(23 z-17)}{2} 
-\ln ^2(1-z) \left(2 (1-z)+\frac{\left(1-7 z^2\right) \ln (z)}{2 (1-z)}\right)
\nonumber
\\
&
+\ln (1-z) \left(\frac{19
   z}{2}+\frac{\pi ^2 \left(3-5 z^2\right)}{6 (1-z)}+\frac{\left(7 z^2-2 z+7\right) \ln (z)}{1-z}-10\right)
\nonumber
\\
&
+4 (z+1) \text{Li}_2(-z)
-\left(\frac{2 \left(2 z^2-5\right)}{1-z}+\frac{\left(3-5 z^2\right) \ln (1-z)}{1-z}\right) \text{Li}_2(z)
+\left[\frac{1+z^2}{1-z}\right]\times
\nonumber
\\
& \bigg(-\frac{5}{2} \ln (1-z) \ln ^2(z)-4 \text{Li}_2(-z) \ln (z)-\text{Li}_2(z) \ln (z)+\frac{2}{3} \pi ^2 \ln(z)
\nonumber
\\
&
+8 \text{Li}_3(-z)
\bigg)
+\frac{\left(9 z^2+1\right) \text{Li}_3(1-z)}{1-z}-\frac{\left(1-3 z^2\right) \text{Li}_3(z)}{1-z}+\frac{\left(3 z^2+7\right) \zeta_3}{1-z}\Bigg]
\nonumber
\\
&
+\Cf\nf \Bigg[\frac{4}{3} \tilde\DD_2(z)-\frac{20}{9} \tilde\DD_1(z)
+\tilde\DD_0(z) \left(\frac{34}{27}-\frac{\pi ^2}{3}+\frac{2 \ln2}{3}\right)
+\frac{(5 z-11)}{18} 
\nonumber
\\
&
-\frac{1}{3} (1-z) \big[2 \ln (1-z)+\ln2\big]
+\frac{\left(5 z^2+6 z-7\right) \ln (z)}{18 (1-z)}+
\left[\frac{1+z^2}{1-z}\right]\times
\nonumber
\\
&
 \left(-\frac{1}{4} \ln ^2(z)+\frac{2 \text{Li}_2(z)}{3}-\frac{\pi ^2}{9}\right)\Bigg] 
+\Ca\Cf\Bigg[
-\frac{22}{3} \tilde\DD_2(z)
+\left(\frac{134}{9}-\frac{2 \pi ^2}{3}\right) \tilde\DD_1(z)
\nonumber
\\
&
-\left(\frac{208}{27}-\frac{11 \pi ^2}{6}+\frac{2 \ln2}{3}-9 \zeta_3\right)\tilde\DD_0(z)+
\frac{\left(7 z^2-12 z+27\right) \ln ^2(z)}{8(1-z)}
\label{eq:transition}
\\
&
-\left(\frac{83 z^2+114 z-109}{18 (1-z)}+2 (z+1) \ln (z+1)\right) \ln (z)-\frac{2}{9} (17z-5)
\nonumber
\\
&
+
\frac{\pi ^2 \left(19 z^2-6 z+31\right)}{36 (1-z)}+\ln (1-z) \left(\frac{58-55 z}{6} -(1-z) \ln (z)\right)
\nonumber
\\
&
-2 (z+1) \text{Li}_2(-z)-\frac{2 \left(4 z^2-3 z+10\right) \text{Li}_2(z)}{3 (1-z)}
+\left[\frac{1+z^2}{1-z}\right] \bigg(\frac{7 \ln ^3(z)}{12}
\nonumber
\\
&
-\ln ^2(1-z) \ln (z)+2 \text{Li}_2(-z) \ln (z)+\frac{\pi ^2 }{3}
   \ln (z)+\frac{\pi ^2}{6}  \ln (1-z)
\nonumber
\\
&
+\big[\ln (z)-\ln (1-z)\big]\text{Li}_2(z)-5 \text{Li}_3(1-z)-4 \text{Li}_3(-z)-4 \text{Li}_3(z)+\zeta_3\bigg)
\nonumber
\\
&
+\frac{1}{3} (1-z) \ln2\Bigg]
+\LMu \Bigg\{\beta_0\Cf \left(4 \tilde\DD_1(z)-\frac{10}{3} \tilde\DD_0(z)-\frac{\left(1+z^2\right) \ln (z)}{1-z}+z-1\right)
\nonumber
\\
&
+\Ca\Cf
   \left[\left(\frac{\pi ^2}{3}-\frac{4}{3}\right) \tilde\DD_0(z)-\frac{\left(1+z^2\right) \ln ^2(z)}{2 (1-z)}-3 (1-z)-(z+1) \ln (z)\right]
\nonumber
\\
&
+\Cf^2 \bigg[-12\tilde\DD_2(z)-12 \tilde\DD_1(z)-2 (z+1) \text{Li}_2(1-z)+\frac{\left(3 z^2+1\right) \ln ^2(z)}{2 (1-z)}
\nonumber
\\
&
-\frac{\left(4 z^2+2 z-3\right) \ln (z)}{1-z}+2 \ln (1-z) \left(\frac{2 z^2 \ln (z)}{1-z}-z+1\right)+2
   (1-z)\bigg]\Bigg\}
\nonumber
\\
&
+\LMusq \left\{\Cf^2 \left(4 \tilde\DD_1(z)+6 \tilde\DD_0(z)-\frac{\left(3 z^2+1\right) \ln (z)}{2 (1-z)}+z-1\right)-\Cf\beta_0
   \tilde\DD_0(z)\right\},
\nonumber
\end{align}     
where we have defined
\be
\tilde \DD_i(z) \equiv \left[\frac{\ln^i(1-z)}{1-z}\right]_+ - \frac{1}{2}(1+z)\ln^i(1-z),
\ee
and as usual $\beta_0 = 11\Ca/6 - \nf/3$. 
Expressions for all the other relevant transition functions have the same form of Eq.~\eqref{eq:transition}, and can be found in the ancillary file.


\newpage 

\begin{thebibliography}{99}

\bibitem{ant} 
  A. Gehrmann-De Ridder, T. Gehrmann and E. W. N. Glover,
  JHEP {\bf 0509} (2005), 056; Phys. Lett. B {\bf 612} (2005), 49; Phys. Lett. B
  {\bf 612} (2005) 36; A. Daleo, T. Gehrmann and D. Maitre, JHEP {\bf 0704}
  (2007), 016; A. Daleo, A. Gehrmann-De Ridder, T. Gehrmann and
  G. Luisoni, JHEP {\bf 1001} (2010), 118; T. Gehrmann and P.F. Monni, JHEP
  {\bf 1112} (2011), 049; R. Boughezal, A. Gehrmann-De Ridder and
  M. Ritzmann, JHEP {\bf 1102} (2011), 098; A. Gehrmann-De Ridder,
  T. Gehrmann and M. Ritzmann, JHEP {\bf 1210} (2012) 047; J. Currie,
  E.W.N. Glover and S. Wells, JHEP {\bf 1304} (2013) 066.

\bibitem{czakonsub}
  M.~Czakon,
  %``A novel subtraction scheme for double-real radiation at NNLO,''
  Phys.\ Lett.\ B {\bf 693} (2010) 259;
  M.~Czakon,
  %``Double-real radiation in hadronic top quark pair production as a proof of a certain concept,''
  Nucl.\ Phys.\ B {\bf 849} (2011) 250.

\bibitem{czakonsub4d}
  M.~Czakon and D.~Heymes,
  %``Four-dimensional formulation of the sector-improved residue subtraction scheme,''
  Nucl.\ Phys.\ B {\bf 890} (2014) 152.

\bibitem{Boughezal:2011jf}
  R.~Boughezal, K.~Melnikov and F.~Petriello,
  %``A subtraction scheme for NNLO computations,''
  Phys.\ Rev.\ D {\bf 85} (2012) 034025.

\bibitem{Cacciari:2015jma}
  M.~Cacciari, F.~A.~Dreyer, A.~Karlberg, G.~P.~Salam and G.~Zanderighi,
  %``Fully Differential Vector-Boson-Fusion Higgs Production at Next-to-Next-to-Leading Order,''
  Phys.\ Rev.\ Lett.\  {\bf 115} (2015) no.8,  082002.

\bibitem{qt1} S.~Catani and M.~Grazzini, 
Phys. Rev. Lett. {\bf 98}, 222002 (2007).
\bibitem{qt2} M.~Grazzini, JHEP {\bf 0802} (2008) 043.

\bibitem{njet1}
  R.~Boughezal, C.~Focke, X.~Liu and F.~Petriello,
  Phys.\ Rev.\ Lett.\  {\bf 115} (2015) no.6,  062002.

\bibitem{njet2} J. Gaunt, M. Stahlhofen, F. J. Tackmann, and J. R.Walsh, JHEP {\bf 09} (2015),  058. 

\bibitem{colorful}
  V.~Del Duca, C.~Duhr, A.~Kardos, G.~Somogyi and Z.~Trocsanyi,
  %``Three-Jet Production in Electron-Positron Collisions at Next-to-Next-to-Leading Order Accuracy,''
  Phys.\ Rev.\ Lett.\  {\bf 117} (2016) no.15,  152004;
  V.~Del Duca, C.~Duhr, A.~Kardos, G.~Somogyi, Z.~Szor, Z.~Trocsanyi and Z.~Tulipant,
  Phys.\ Rev.\ D {\bf 94} (2016) no.7,  074019.

%%%


%\cite{Moult:2016fqy}
\bibitem{tackmann:nj:pow}
  I.~Moult, L.~Rothen, I.~W.~Stewart, F.~J.~Tackmann and H.~X.~Zhu,
  Phys.\ Rev.\ D {\bf 95} (2017) no.7,  074023;
  I.~Moult, L.~Rothen, I.~W.~Stewart, F.~J.~Tackmann and H.~X.~Zhu,
  Phys.\ Rev.\ D {\bf 97} (2018) no.1,  014013;
  M.~A.~Ebert, I.~Moult, I.~W.~Stewart, F.~J.~Tackmann, G.~Vita and H.~X.~Zhu,
  JHEP {\bf 1812} (2018) 084.

\bibitem{frank:nj:pow}
  R.~Boughezal, X.~Liu and F.~Petriello,
  JHEP {\bf 1703} (2017) 160;
  R.~Boughezal, A.~Isgro and F.~Petriello,
  Phys.\ Rev.\ D {\bf 97} (2018) no.7,  076006.

%\cite{Ebert:2018gsn}
\bibitem{Ebert:2018gsn}
  M.~A.~Ebert, I.~Moult, I.~W.~Stewart, F.~J.~Tackmann, G.~Vita and H.~X.~Zhu,
  %``Subleading Power Rapidity Divergences and Power Corrections for $q_T$,''
  arXiv:1812.08189 [hep-ph].
  %%CITATION = ARXIV:1812.08189;%%
  %4 citations counted in INSPIRE as of 28 Jan 2019

\bibitem{magnea}
  L.~Magnea, E.~Maina, G.~Pelliccioli, C.~Signorile-Signorile, P.~Torrielli and S.~Uccirati,
  %``Local Analytic Sector Subtraction at NNLO,''
  JHEP {\bf 1812} (2018) 107;
  L.~Magnea, E.~Maina, G.~Pelliccioli, C.~Signorile-Signorile, P.~Torrielli and S.~Uccirati,
  %``Factorisation and Subtraction beyond NLO,''
  JHEP {\bf 1812} (2018) 062.

\bibitem{herzog}
  F.~Herzog,
  %``Geometric IR subtraction for final state real radiation,''
  JHEP {\bf 1808} (2018) 006.
  %doi:10.1007/JHEP08(2018)006
  %[arXiv:1804.07949 [hep-ph]].
  %%CITATION = doi:10.1007/JHEP08(2018)006;%%
  %8 citations counted in INSPIRE as of 28 Jan 2019

%\cite{Caola:2017dug}
\bibitem{Caola:2017dug}
  F.~Caola, K.~Melnikov and R.~R\"ontsch,
  %``Nested soft-collinear subtractions in NNLO QCD computations,''
  Eur.\ Phys.\ J.\ C {\bf 77} (2017) no.4,  248.
  %doi:10.1140/epjc/s10052-017-4774-0
  %[arXiv:1702.01352 [hep-ph]].
  %%CITATION = doi:10.1140/epjc/s10052-017-4774-0;%%
  %30 citations counted in INSPIRE as of 28 Jan 2019

\bibitem{Caola:2017xuq}
  F.~Caola, G.~Luisoni, K.~Melnikov and R.~R\"ontsch,
  %``NNLO QCD corrections to associated $WH$ production and $H \to b \bar b$ decay,''
  Phys.\ Rev.\ D {\bf 97} (2018) no.7,  074022.
  %doi:10.1103/PhysRevD.97.074022
  %[arXiv:1712.06954 [hep-ph]].
  %%CITATION = doi:10.1103/PhysRevD.97.074022;%%
  %12 citations counted in INSPIRE as of 28 Jan 2019

\bibitem{max}
  F.~Caola, M.~Delto, H.~Frellesvig and K.~Melnikov,
  %``The double-soft integral for an arbitrary angle between hard radiators,''
  Eur.\ Phys.\ J.\ C {\bf 78} (2018) no.8,  687.
  %doi:10.1140/epjc/s10052-018-6180-7
  %[arXiv:1807.05835 [hep-ph]].
  %%CITATION = doi:10.1140/epjc/s10052-018-6180-7;%%
  %3 citations counted in INSPIRE as of 28 Jan 2019

\bibitem{maxtc}
  M.~Delto and K.~Melnikov,
  %``Integrated triple-collinear counter-terms for the nested soft-collinear subtraction scheme,''
  arXiv:1901.05213 [hep-ph].
  %%CITATION = ARXIV:1901.05213;%%

\bibitem{Hamberg:1990np}
  R.~Hamberg, W.~L.~van Neerven and T.~Matsuura,
  %``A complete calculation of the order $\alpha-s^{2}$ correction to the Drell-Yan $K$ factor,''
  Nucl.\ Phys.\ B {\bf 359} (1991) 343
   Erratum: [Nucl.\ Phys.\ B {\bf 644} (2002) 403].
   %doi:10.1016/S0550-3213(02)00814-3, 10.1016/0550-3213(91)90064-5
  %%CITATION = doi:10.1016/S0550-3213(02)00814-3, 10.1016/0550-3213(91)90064-5;%%
  %1121 citations counted in INSPIRE as of 28 Jan 2019

\bibitem{Anastasiou:2002yz}
  C.~Anastasiou and K.~Melnikov,
  %``Higgs boson production at hadron colliders in NNLO QCD,''
  Nucl.\ Phys.\ B {\bf 646} (2002) 220.
  %doi:10.1016/S0550-3213(02)00837-4
  %[hep-ph/0207004].
  %%CITATION = doi:10.1016/S0550-3213(02)00837-4;%%
  %1018 citations counted in INSPIRE as of 28 Jan 2019

\bibitem{Frixione:1995ms}
  S.~Frixione, Z.~Kunszt and A.~Signer,
  %``Three jet cross-sections to next-to-leading order,''
  Nucl.\ Phys.\ B {\bf 467} (1996) 399.
  %doi:10.1016/0550-3213(96)00110-1
  %[hep-ph/9512328].
  %%CITATION = doi:10.1016/0550-3213(96)00110-1;%%

\bibitem{Frixione:1997np}
  S.~Frixione,
  %``A General approach to jet cross-sections in QCD,''
  Nucl.\ Phys.\ B {\bf 507} (1997) 295.
  %doi:10.1016/S0550-3213(97)00574-9
  %[hep-ph/9706545].
  %%CITATION = doi:10.1016/S0550-3213(97)00574-9;%%
  %289 citations counted in INSPIRE as of 28 Jan 2019

%\cite{Ellis:1991qj}
\bibitem{Ellis:1991qj}
  R.~K.~Ellis, W.~J.~Stirling and B.~R.~Webber,
  %``QCD and collider physics,''
  Camb.\ Monogr.\ Part.\ Phys.\ Nucl.\ Phys.\ Cosmol.\  {\bf 8} (1996) 1.
  %%CITATION = CMPCE,8,1;%%
  %595 citations counted in INSPIRE as of 04 Feb 2019

\bibitem{Catani:1999ss}
  S.~Catani and M.~Grazzini,
  %``Infrared factorization of tree level QCD amplitudes at the next-to-next-to-leading order and beyond,''
  Nucl.\ Phys.\ B {\bf 570} (2000) 287.
  %doi:10.1016/S0550-3213(99)00778-6
  %[hep-ph/9908523].
  %%CITATION = doi:10.1016/S0550-3213(99)00778-6;%%
  %224 citations counted in INSPIRE as of 28 Jan 2019
 
%\cite{Ball:2014uwa}
\bibitem{nnpdf}
  R.~D.~Ball {\it et al.} [NNPDF Collaboration],
  %``Parton distributions for the LHC Run II,''
  JHEP {\bf 1504} (2015) 040.
  %doi:10.1007/JHEP04(2015)040
  %[arXiv:1410.8849 [hep-ph]].
  %%CITATION = doi:10.1007/JHEP04(2015)040;%%
  %1439 citations counted in INSPIRE as of 01 Feb 2019

%\cite{Salam:2008qg}
\bibitem{Salam:2008qg}
  G.~P.~Salam and J.~Rojo,
  %``A Higher Order Perturbative Parton Evolution Toolkit (HOPPET),''
  Comput.\ Phys.\ Commun.\  {\bf 180} (2009) 120.
  %doi:10.1016/j.cpc.2008.08.010
  %[arXiv:0804.3755 [hep-ph]].
  %%CITATION = doi:10.1016/j.cpc.2008.08.010;%%
  %141 citations counted in INSPIRE as of 30 Jan 2019

\bibitem{Grazzini:2017mhc}
  M.~Grazzini, S.~Kallweit and M.~Wiesemann,
  %``Fully differential NNLO computations with MATRIX,''
  Eur.\ Phys.\ J.\ C {\bf 78} (2018) no.7,  537.
  %doi:10.1140/epjc/s10052-018-5771-7
  %[arXiv:1711.06631 [hep-ph]].
  %%CITATION = doi:10.1140/epjc/s10052-018-5771-7;%%
  %30 citations counted in INSPIRE as of 30 Jan 2019

\bibitem{Catani:1998bh}
  S.~Catani,
  %``The Singular behavior of QCD amplitudes at two loop order,''
  Phys.\ Lett.\ B {\bf 427} (1998) 161.
  %doi:10.1016/S0370-2693(98)00332-3
  %[hep-ph/9802439].
  %%CITATION = doi:10.1016/S0370-2693(98)00332-3;%%
  %429 citations counted in INSPIRE as of 28 Jan 2019

\bibitem{Gehrmann:2005pd}
  T.~Gehrmann, T.~Huber and D.~Maitre,
  %``Two-loop quark and gluon form-factors in dimensional regularisation,''
  Phys.\ Lett.\ B {\bf 622} (2005) 295.
  %doi:10.1016/j.physletb.2005.07.019
  %[hep-ph/0507061].
  %%CITATION = doi:10.1016/j.physletb.2005.07.019;%%
  %113 citations counted in INSPIRE as of 28 Jan 2019

\bibitem{Grigo:2014jma}
  J.~Grigo, K.~Melnikov and M.~Steinhauser,
  %``Virtual corrections to Higgs boson pair production in the large top quark mass limit,''
  Nucl.\ Phys.\ B {\bf 888} (2014) 17.
  %doi:10.1016/j.nuclphysb.2014.09.003
  %[arXiv:1408.2422 [hep-ph]].
  %%CITATION = doi:10.1016/j.nuclphysb.2014.09.003;%%
  %83 citations counted in INSPIRE as of 28 Jan 2019

\end{thebibliography}
\end{document}